\PassOptionsToPackage{table}{xcolor} %

\documentclass[acmsmall,screen,nonacm]{acmart}

\setcopyright{none}

\acmDOI{}
\acmISBN{}
\acmPrice{}
\acmVolume{}
\acmNumber{}
\acmArticle{}

\usepackage{wrapfig}        %
\usepackage{placeins}       %
\usepackage{tabularx}       %
\usepackage{multirow}       %
\usepackage{makecell}       %
\usepackage{xurl}
\usepackage{nicefrac}
\usepackage{enumitem}
\usepackage{array}
\usepackage{subcaption}
\usepackage[normalem]{ulem} %
\usepackage{soul}           %
\usepackage{tcolorbox}
\graphicspath{{./}{figures/}}

\definecolor{todoyellow}{RGB}{204,153,0}

\definecolor{crimsonglory}{rgb}{0.75, 0.0, 0.2}

\title{DOCKET: A Pluralistic Benchmark for Evaluating AI-Assisted Policy Feedback in U.S. Rulemaking}

\author{Alexander Spangher}
\authornote{Corresponding author.}
\email{alexspan@stanford.edu}
\orcid{0000-0001-6655-9593}
\author{Charlotte Peale}
\email{cpeale@stanford.edu}
\orcid{0000-0002-9959-857X}
\author{Sílvia Casacuberta}
\email{scasac@stanford.edu}
\orcid{0000-0001-5684-4585}
\author{Victor Y. Wu}
\email{victorywu@stanford.edu}
\orcid{0000-0003-4548-9826}
\author{Mirac Suzgun}
\email{msuzgun@stanford.edu}
\orcid{0009-0001-1635-5784}
\author{Olivia H. Martin}
\email{omartin@stanford.edu}
\orcid{0009-0006-5807-862X}
\affiliation{\institution{Stanford University}\city{Stanford}\state{California}\country{USA}}
\author{Dan Bateyko}
\email{drb348@cornell.edu}
\orcid{0000-0002-4829-5464}
\affiliation{\institution{Cornell University}\city{Ithaca}\state{New York}\country{USA}}
\author{Cassandra Handan-Nader}
\email{cassandra.handan-nader@nyu.edu}
\orcid{0000-0003-4868-2509}
\affiliation{\institution{New York University}\city{New York}\state{New York}\country{USA}}
\author{Diyi Yang}
\email{diyiy@cs.stanford.edu}
\orcid{0000-0003-1220-3983}
\author{Sanmi Koyejo}
\email{sanmi@cs.stanford.edu}
\orcid{0000-0002-4023-419X}
\author{Daniel E. Ho}
\email{dho@law.stanford.edu}
\orcid{0000-0002-2195-5469}
\affiliation{\institution{Stanford University}\city{Stanford}\state{California}\country{USA}}

\begin{document}
\addtocontents{toc}{\protect\setcounter{tocdepth}{-1}}

\begin{abstract}

As AI models advance and impact deliberative democracies, there is a growing need and urgency to assess their support of pluralistic viewpoints. Most pluralism benchmarks are insufficient for this: they are not grounded in diversity (i.e., stakeholders across different frames, skills and needs), quality (i.e., rigorous and technical debates), and impact (i.e., invested deliberators and real-world outcomes). %
To address this, we \emph{first} (i) curate DOCKET,\footnote{DOCKET stands for: Dataset of Comments, Knowledge, and Executive Texts.} the most comprehensive dataset of the U.S. Federal Notice \& Comment process to date, a central process in deliberative democracy. DOCKET consists of 42.3M public comments on 22,400 regulations across 44 government agencies between 2016 and 2026 and contains millions of policy debates; we rigorously link public comments to particular policy items proposed by regulators to track the public's attention and the responsiveness of regulators to public input. (ii) \emph{Next}, we demonstrate the pluralistic utility of DOCKET -- we show that public commenters have a wide breadth of perspectives with significant depth in policy expertise and that regulators are often responsive. (iii) \emph{Finally}, we present DOCKETBench, the first systematic evaluation of LLMs' ability to represent human policy perspectives in federal rulemaking.
We find that only 18--30\% %
of LLM-generated feedback aligns with real human comments. LLMs critique proposals broadly but generally, lacking the specific grounding, citations, and lived-experience arguments that humans bring to bear; in addition, most models we test systematically agree with government perspectives (overlapping with up to 68\% of their claims) and agree least with labor unions (as little as 26\% of their claims), demonstrating a fundamental skew in perspectives that even tools like persona modeling do not solve. %
Taken together, we reveal fundamental gaps in how language models navigate real policy issues, and DOCKETBench serves as a testbed for closing those gaps in a crucial real-world domain. %

\end{abstract}


\maketitle

\section{Introduction}

Generative AI has become widespread in government interactions with the public, from official communications \citep{auchincloss2023chatgpt,savoy2024good} to court filings \citep{mata_v_avianca_2023}. This paper focuses on one such consequential setting: notice-and-comment (N\&C) rulemaking, which has been described by legal scholars as \textit{``one of the greatest inventions of modern government''} 
\citep{davis1970supplement}. Through N\&C, agencies translate legislation into regulations by publishing proposed rules, inviting the public to submit ``data, views, or arguments,'' and considering submissions before finalizing them \citep{kerwin2018rulemaking,usc_553}. Language models are beginning to assist --- both by (1) helping citizens draft comments \citep{balla2022responding,hr7528} and by (2) helping agencies analyze submissions, refine proposals, and anticipate public response \citep{cdt2025inventory,hhs2024headstart,icf_commentworks,icf_genai_comments,pnnl2024policyai}\footnote{At least seven federal agencies report using AI tools to support comment analysis, raising familiar questions about which comments receive attention and how public input is translated into regulatory change \citep{cdt2025inventory,hhs2024headstart,icf_commentworks,icf_genai_comments,pnnl2024policyai}.}. Yet scholars and policymakers lack systematic evaluations of this assistance. %

Simultaneously, there is growing awareness that AI models, specifically when attempting tasks in these kinds of settings, must grow and adapt to \textit{pluralistic beliefs, values and preferences} to negotiate 
the needs of a diverse, deliberative public \citep{sorensen2024roadmap,kirk2024prism,conitzer2024position}. Most pluralistic alignment work is evaluated on synthetic disagreements or stated preferences elicited in isolation \citep{bakker2022fine, durmustowards, huang2024collective}, leaving open how models should handle the real-world, high-stakes pluralism.

To address both needs, we introduce an empirical map of the rulemaking process itself: who participates, what claims they raise, which provisions they target, and how agencies respond. We introduce DOCKET: the largest and, to our knowledge, most comprehensive dataset of rulemaking and public comments. DOCKET tracks 44 U.S. federal agencies across 10 years (2016--2026) and consists of 87,005 notices, 17,198 proposed rules, 42.3M public comments from 16 different federal commenting systems, and 22,400 finalized rules.\footnote{DOCKET is U.S.-centric; other rulemaking systems are left to future work.} To understand the role of comments on rulemaking, we identify \textit{specifically} which rule provisions the public commented on, and which comments the agency responded to. We use LLMs to extract provisions from all rules (676K provisions); claims from all comments (13.6M claims); and responses (94K responses) from all finalized rules. We develop a high-performance matching pipeline (claims $\rightarrow$ provisions) and (responses $\rightarrow$ claims). This approach allows us to track what members of the public care about, which comments the government considers deserve a response, and how these comments affect rulemaking. Overall, we find that public commenters engage seriously with regulations ($\sim$30\% bring specialized knowledge), target specific provisions rather than entire rules, and that agencies take up commenter language at meaningful rates (i.e. when rewriting provisions for final rules, agencies directly adopt wording from commenters 11--26\% of the time, and paraphrase another 12--32\% of the time).

For these reasons, DOCKET is an ideal dataset for evaluating the pluralism and policy skills in LLMs. We identify three characteristics driving real-world pluralism: 1) Diversity (or breadth): DOCKET comments come from a large range of human policy perspectives (e.g., industry, advocacy, individuals, law firms, labor unions); 2) Quality (or depth): comments exhibit  deep policy expertise on specific topics, and 3) Impact: comments are consequential -- agencies meaningfully change their policy recommendations directly based on public comments. If pluralistic AI can faithfully represent the public in this setting, it offers a template for AI-mediated deliberation more broadly.

With these insights in hand, we frame our final question: as LLMs become integrated more into our social and political lives, what roles will they play, what feedback will they give, and who will they represent? We design a policy benchmark with the simple goal: ``I am proposing this rule, how should it change?'' We evaluate LLM responses on two primary dimensions: (1) the substance of LLM comments (which provisions they target and what claims they raise), and (2) the style and framing of LLM comments. Although LLM-feedback has been studied in a range of settings (e.g., writing \citep{lee2022coauthor}, voting \citep{argyle2023outofone,santurkar2023whose}, and broader public-opinion simulation \citep{durmus2024globalopinions}), regulatory commentary is particularly challenging because of the technical rigor and high-stakes legal impact of the process. Indeed, in DOCKETBench, we find that LLM responses tend to be less specific,  evidence-based, and technical than human responses, and they do not reflect the broad spectrum of human responses. We find that, across a range of different scaffolding (e.g., persona-driven responses \cite{li2016persona}, RAG \cite{lewis2020retrieval}), most models we test side with government and industry commenters most often and with labor unions least often, and they agree with every group less than human commenters (Table~\ref{tab:bucket_alignment}). Looking forward, we hope DOCKET can serve as a test-bed for civic and deliberative AI.

 \begin{figure}[!tp]
	\centering	
	\includegraphics[width=0.72\textwidth,height=0.9\textheight,keepaspectratio]{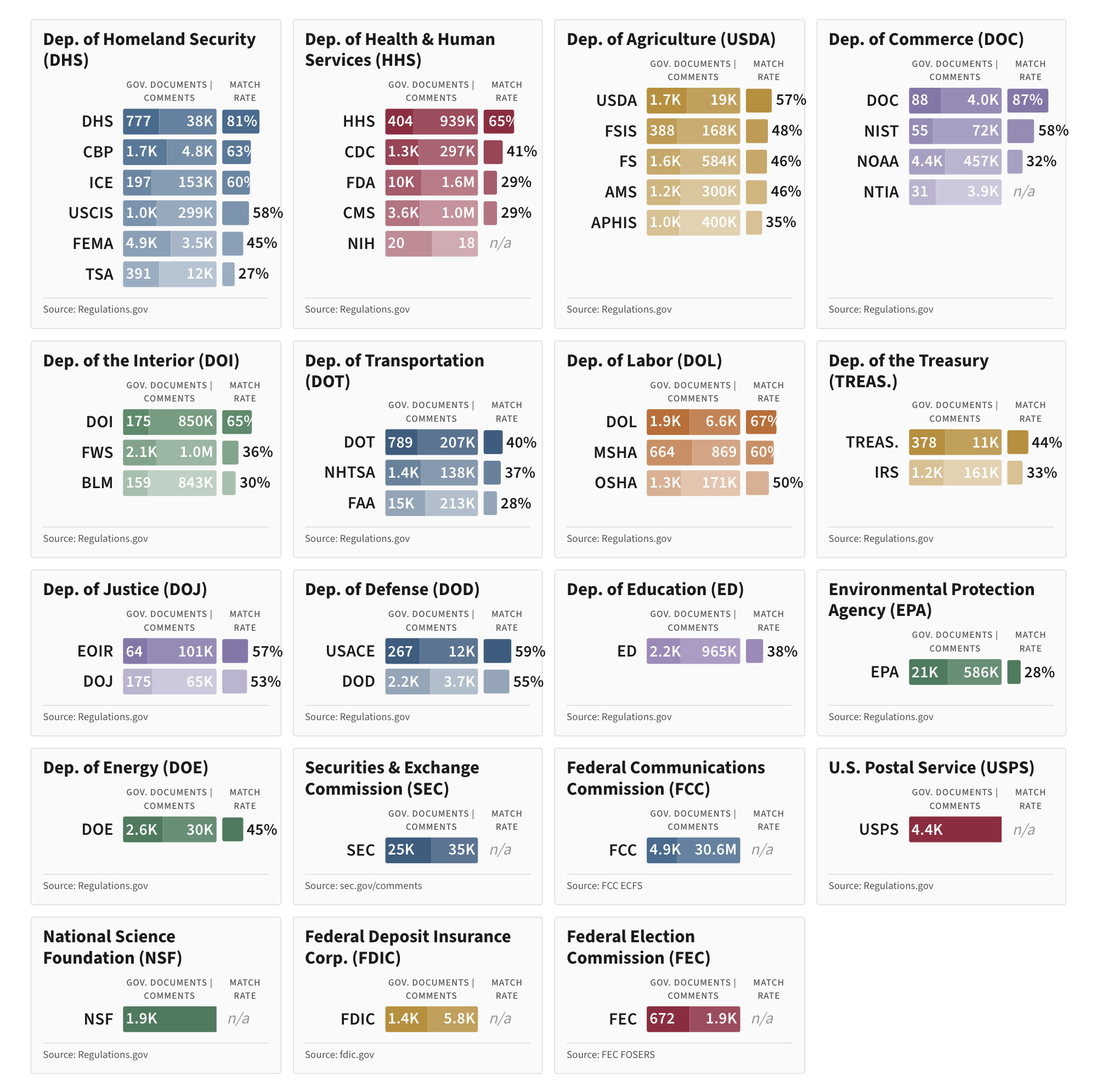}
	\caption{\textbf{Agency-level overview of DOCKET.} DOCKET is composed of proposed rules, comments and finalized rules from 44 agencies, grouped under 19 parent agencies. For each agency we show (i) the number of government documents (notices, proposed rules and finalized rules), (ii) the number of public comments, and (iii) the match rate, or the share of comment clusters that our pipeline links to at least one agency response estimated from a uniform random sample of 1K candidate pairs per agency judged by gpt-5-mini (n/a where we did not sample). Comment volumes range from 18 (NIH) and 869 (MSHA) to over 1M (FDA, CMS, FWS) and 30.6M at the FCC. Differences in match rate across agencies reflect how substantive the comments are and how much response burden the agency takes on under the APA. The numbers are those of Table~\ref{tab:per_agency}.}
	\label{fig:agency-chart}
\end{figure}

\section{Dataset and Methodology}
\label{sec:dataset}

The DOCKET dataset consists of 126,603 regulatory documents (87,005 notices, 17,198 proposed rules, 22,400 finalized rules) and 42.3M public comments across 30,365 dockets.\footnote{A docket is a grouping assigned by the agency to capture one unit of rulemaking: a proposal, comments and final rule.} We collect data from 44 agencies from 2016--2026, enabling analysis of federal regulatory processes spanning four presidential administrations.\footnote{2016-2026 covers four executives: Obama (2016-2017), Trump (2017-2021), Biden (2021-2025), Trump (2025-2026).} Of these, 16,632 are complete rulemaking dockets (containing both a proposed rule and a finalized rule). Figure~\ref{fig:agency-chart} shows aggregate counts for the number of government documents and comments per agency. 
To enable a broad analysis and capture a wide distribution of views, we selected 44 agencies\footnote{The total number of agencies in the U.S. federal government is unknown; only rough counts ($\sim$445 agencies) exist \citep{lewis2018sourcebook}.} sharing the following characteristics: (1) topical similarity: some agencies have similar mandates, enabling future study on finer-grained institutional differences within agencies (e.g., Fish and Wildlife Service (FWS), Bureau of Land Management (BLM) and Environmental Protection Agency (EPA)); (2) diverse cultures: other agencies are notable in having  well-documented institutional cultures (i.e., Immigration and Customs Enforcement (ICE) and the Federal Trade Commission (FTC)); and (3) different subpopulations served: while all agencies bring specialized knowledge to bear to serve the public, agencies interact with different populations (e.g., National Science Foundation (NSF) primarily interacts with scientists while Bureau of Land Management (BLM) interacts more with farmers.).

\subsection{Data Sources}

The primary repository used by the U.S. federal government to collect all the federal records and most public comments is Regulations.gov \citep{regulationsgov}.\footnote{We found regulations.gov, despite having an API, was more efficient to scrape via its bulk-download page. \url{https://www.regulations.gov/bulkdownload}; we script a Playwright browser and query the user for manual captcha-solving.} However, not all agencies send comments to regulations.gov \citep{gao2021rulemaking}, including agencies that typically receive large comment volumes (e.g., FCC and SEC \citep{fcc_ecfs,sec_comments}). As such, in a large-scale effort, we built custom scraping pipelines for 16 distinct federal data sources: regulations.gov and 15 additional agency-specific systems,\footnote{More specifically, we built scrapers and API access scripts for eleven comment systems that expose individual public comments (FCC ECFS, SEC, FERC, Federal Reserve, FDIC, FEC, CFTC, PCAOB, CMS, FS CARA, and NPS PEPC) and four document portals that expose project-level notices or bundled comment archives (BLM ePlanning, NRC ADAMS, USACE Districts, and USITC EDIS). We describe each scraping pipeline in detail in Appendix~\ref{app:scrapers}} which is, to our knowledge, the most comprehensive effort to collect open N\&C data. To validate our pipelines, we built additional scrapers to obtain aggregate counts published by agencies and confirm we retrieve $\sim$100\% of the public comments and documents available (Appendix~\ref{app:agency_list}). Only $\sim$71\% of those comments survive downstream feature extraction in usable form, as some government records were unreachable or PDFs unparsable.\footnote{Ranges from 43\% (NIST) to 95\%+ (TSA, USACE, TREASURY, DOC, NTIA, DOD, FEMA, MSHA, NIH).} DOCKET and DOCKETBench, including the extracted provisions, claims and responses, the validated matches, and the benchmark generations and judgments, are available at \url{https://huggingface.co/datasets/docket-dataset-bench/docket-bench}.

\subsection{Dataset Processing}

We aim to measure changes in rules and the actions of commenters through the N\&C process. To do this, we perform multiple matches, shown in Figure \ref{fig:matching-structure}: (1) \textit{comment $\rightarrow$ proposed rule} tells us which parts of the proposal a comment is discussing; (2) \textit{response $\rightarrow$ comment} tells us which comments the agency is responding to and (3) \textit{comment $\rightarrow$ final rule}: tells us which parts of the final rule changed as a result of the comment. We perform each match as follows:

\begin{figure}[t]
  \centering
  \begin{subfigure}[c]{0.5\textwidth}
    \centering
    \includegraphics[width=\linewidth]{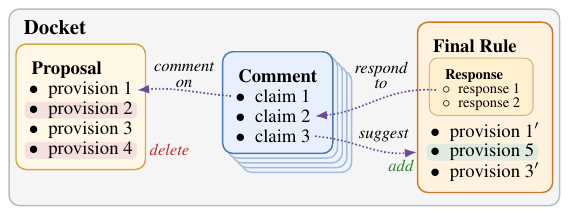}
    \caption{Structure}
    \label{fig:matching-structure-a}
  \end{subfigure}\par\vspace{8pt}
  \begin{subfigure}[c]{\textwidth}
    \centering
    \footnotesize
    \setlength{\tabcolsep}{5pt}
    \renewcommand{\arraystretch}{1.08}
    \begin{tabular}{@{}lr@{\hspace{24pt}}lr@{}}
    \toprule
    \multicolumn{2}{@{}l}{\textbf{Items}} & \multicolumn{2}{l@{}}{\textbf{Links}} \\
    \midrule
    Dockets & 30{,}365 & \textit{comment on}: claim $\to$ proposal provision & 1{,}024{,}001 \\
    Proposals & 17{,}198 & \textit{respond to}: response $\to$ comment & 205{,}646 \\
    \quad provisions & 301{,}892 & \textit{suggest}: claim $\to$ final provision & 494{,}487 \\
    Comments & 42.3M & \textit{delete}: proposal provisions dropped & 70{,}279 \\
    \quad claims & 13.6M & \textit{add}: final provisions that are new & 71{,}857 \\
    Final rules & 22{,}400 & kept as is / modified \textit{(provisions $1'$, $3'$)} & 80{,}246 / 12{,}466 \\
    \quad provisions & 374{,}026 & & \\
    \quad responses & 94{,}692 & & \\
    \bottomrule
    \end{tabular}
    \caption{Counts}
    \label{fig:links}
  \end{subfigure}
  \caption{(a) Structure of provisions, claims and responses in proposals, comments and final rules. \textit{Added} and \textit{deleted} represent provisions removed or added throughout N\&C. (b) Total counts of each in DOCKET. Links are counted after the gpt-5-mini judge (Matching, below). The last three rows align each proposal provision with the final rule in the 6{,}954 dockets that have both documents.}
  \label{fig:matching-structure}
\end{figure}

\paragraph{Parsing:}
We use an LLM\footnote{We use gpt-5-mini to extract provisions, Llama-3.3-70b to extract claims and responses, gpt-5-mini to judge matches, and gpt-4o-mini to judge claim pairs in the benchmark (Appendix~\ref{app:prompts}). This was done purely to choose the most performant model for each task at the lowest cost; all performance considerations were judged by larger models before applying.} to extract provisions from proposed and finalized rules. Specifically, we extract imperative provisions (e.g., ``employers \textit{should} report''; ``brochures \textit{must} contain''; ``\textit{demonstrate} acceptable performance'') because these focus on the \textit{discrete effects} of a rule and how it \textit{impacts} people \cite{hohfeld1913fundamental, leahey2026obligations}.\footnote{Formally, our prompt uses the term ``deontic unit'', a singular permission, obligation, prohibition, condition, or exception made in any imperative in a legal text. We note that this lens does \textit{not} consider \textit{"reasoned explanations"} for rules. Courts have interpreted the APA as requiring rules to have explanations and have found a lack thereof as grounds for overturning the rule; thus, many comments in N\&C address reasoning (e.g., ``rule rests on scientific finding A'' $\rightarrow$ ``commenter challenges A with finding B''. However, we leave a detailed \textit{reasoning-level} analysis of N\&C for future work.}  We use prompts by \citet{leahey2026obligations} to extract 301{,}892 provisions from 17{,}198 proposed rules and 374{,}026 provisions from 22{,}400 finalized rules. Similarly, we use an LLM to extract \textit{responses} from final rules: we extract 94K responses from 22{,}400 finalized rules.\footnote{Courts interpret the APA as requiring agencies to respond to (agree, disagree or simply address) \textit{significant} comments.} Finally, we cluster comments using the MinHash algorithm~\citep{broder1997resemblance} to blunt large-scale comment campaigns \cite{handannader2023fake}, and we use an LLM to extract claims from representative comments from each cluster (we will use ``comments'' and ``clusters'' interchangeably). A claim is a specific statement about an existing or hypothetical provision. %
We extract 13.6M claims from the comments with analyzable text content.\footnote{See Appendix \ref{app:prompts} for all prompts and Appendix~\ref{app:human_validation} for manual validations for these labels.}

\paragraph{Matching:}

With parsed provisions, responses and comments on hand, we construct a matching pipeline.\footnote{Recall, we are performing the following matches: (proposed provision, finalized provision), (proposed provision, comment), (comment, finalized provision), (comment, response).} We design a bi-encoder $\rightarrow$ cross-encoder $\rightarrow$ LLM pipeline \citep{reimers2019sentencebert,karpukhin2020dpr,nogueira2019multistage} to match straightforward pairs with scalable models while using more powerful models for uncertain pairs. See Figure~\ref{fig:links} for full counts of each linking step.
To construct our bi-encoding (BE) match, we embed provisions, responses and claims \citep{babakhin2025nemotron}.\footnote{We embed them using {\tiny \texttt{nvidia/llama-embed-nemotron-8b}} with the retriv library: {\tiny \url{https://github.com/alex2awesome/retriv}.}} Then, we query our retrieval indices and retrieve the top-\emph{k} pairs in each index: for (responses $\rightarrow$ comments), we query our claim indices with each response and retrieve the top-500 claims; for (claims $\rightarrow$ proposed provisions), we query our claim indices with each provision to retrieve claims. \footnote{In earlier experiments, we found that a general CE did not meaningfully improve our performance beyond our BE. For our cross-encoder (CE) architecture, we use ModernBert \citep{warner2024modernbert}.} \footnote{Our cross-encoder design: the CE takes pairs formatted as \texttt{text\_1 <sep> text\_2} and predicts a 1 if there is a match.} We silver-label 500--3,000 pairs per agency with \texttt{gpt-5-mini} (48K pairs in all) and train agency-specific and pooled ModernBERT cross-encoders (CE) on them. Our CE reaches F1 0.69 on 2,982 held-out pairs (0.73 in cross-validation), up from 0.49 for the bi-encoder alone. Per-agency F1 is 0.73--0.81 at FAA, CMS, OSHA and USCIS and below 0.1 at AMS, DOE, DOJ, FDA, FS and NHTSA, which have 1--25 positive validation pairs. We send every pair the CE scores above 0.20 to \texttt{gpt-5-mini} to perform a final validation on, using the opposing-sides prompt in Appendix~\ref{app:prompt_match2}. To validate this labeling, we use a stronger language model, Claude Sonnet, to score 1,000 of these pairs; we find that Sonnet agrees with gpt-5-mini 92.9\% of the time. All experiments are conducted on a cluster of eight b200 Nvidia GPUs. In summary, as shown in Figure~\ref{fig:agency-chart}, we link $\sim$18\% of all comments to agency responses; depending on the agency, about 27--87\% of the comment \textit{clusters} get an agency response. 57\% of provisions from proposed rules survive in final rules (Figure~\ref{fig:links}).

\paragraph{External matching validation:}

To validate our matches, we test how well we retrieve matches identified by \citet{handannader2023fake} in the Federal Communications Commission's (FCC) 2017 ``Restoring Internet Freedom'' proceeding. Because this docket received an extraordinary volume of comments (25M, many from mass campaigns\footnote{A mass comment campaign is when an organized group (e.g. an advocacy organization like \texttt{change.org} or an industry coalition, etc.) distributes template text that many individuals submit verbatim or with minor edits, producing thousands of nearly-identical comments on a single docket.}), the FCC took the uncommon step of directly citing unique comment IDs in their responses. Handan-Nader recovered 1.2k matches between comments and responses. By applying our pipeline, we recall 88\% of these matches and discover 9.2x more responses beyond those cited\footnote{Of our 88\% recall rate with \citet{handannader2023fake}'s gold-corpus, we believe this is a combination of noise stemming from our CE and some weakly tied citations in the gold-corpus (e.g. ``See also Apple Co. filing at p. 42''.) On the 9.2x rate, many comments are not explicitly cited, including group responses (e.g. ``Many commenters
argued X; we disagree because Y,''; our methodology can handle these scenarios.}. \texttt{Gpt-5-mini} as a judge \cite{zheng2023judging} agreed with 96\% of the citation-matched labels. Additionally, we check  LLM labels by hand: the claim-pair judge (52 blind pairs; agreement 0.87, $\kappa$ 0.80) and the direct-uptake classifier (70 blind cases; 80\% confirmed) are both reported in the appendix. Finally, human annotators validated 50 labels from each of eleven pipeline steps, from provision matching to the benchmark's claim-pair judge (Appendix~\ref{app:human_validation}); the model's labels were validated with 95\% accuracy.

\section{Analysis}\label{sec:analysis}

We present an analysis of our dataset DOCKET to motivate the need and design of our policy-feedback benchmark and we establish three properties of pluralism that prior pluralism benchmarks lack \cite{kirk2024prism, durmustowards}: 
(1) \textbf{Diversity (or Breadth):} stakeholder concerns diverge sharply across the corpus; 
(2) \textbf{Quality (or Depth):} commenters tend to focus on specific provisions rather than broad reactions to rules;
and
(3) \textbf{Impact:} agencies respond to commenters and take up their wording when finalizing provisions in rules.
Together, these findings establish that public comment is a structured, impactful artifact rather than generic preference expression. Any LLM tool deployed in this loop must preserve stakeholder heterogeneity to avoid distorting the policy-feedback record.

\paragraph{\ul{Finding \#1: Comments have Impact}. Agencies are broadly responsive to, and take up the wording of, commenters' claims.}

We find that, in their responses to comments (Figure \ref{fig:matching-structure}) agencies express \textit{agreement} with 25\% of comments they respond to, \textit{partial agreement} with 11\% and \textit{disagreement} with 35\% (using LLMs to assess)\footnote{These agreement labels come from our response extractor, which tags each comment--response pair it finds as accepted, partially accepted, disagree, acknowledged, unclear or not relevant; we report accepted as agreement, partially accepted as partial agreement, and disagree as disagreement. For more details, see Appendix~\ref{app:prompt_parsing}.}. Agencies also update rules in ways that directly incorporate literal words, phrases and claims made in comments. Using gpt-5-mini, we annotate matched pairs of comment $\leftrightarrow$ modified provisions, consisting of (proposal, claim, final rule) tuples; and comment $\leftrightarrow$ added provisions consisting of (claim, final rule) tuples with one of the following labels: ``directly taken up'', ``responsive'', ``topic only'', and ``opposed'' (where the agency rejects the suggestion).\footnote{See Appendix \ref{app:uptake} for all prompt details and evaluation details.} As shown in Figure~\ref{fig:uptake_breakdown}, 10.8\% of comments on \textit{modified} provisions and 26\% of comments on \textit{added} provisions have their \textit{wording taken up directly into the rule}; an additional 11.8\% and 32\%, respectively, are classified as \textit{responsive}---where the agency engages the substance of the comment but rephrases it in its own language. Perhaps unsurprisingly, the comments agencies express agreement with (or partial agreement) have their wording taken up 74\% (and 53.1\%) of the time while comments agencies express disagreement with (or simply acknowledge) are taken up 2.6\% (and 2.5\%) of the time. We validated these labels with human annotators: in a blind review of 70 cases (51 labeled direct uptake, 19 distractors), 80\% of the direct-uptake labels held up as language taken from the comment, 86\% if we count close paraphrases (Appendix~\ref{app:uptake}). In sum, comments substantially affect the trajectory of regulations.

\paragraph{\ul{Finding \#2: Comments are Diverse}: Stakeholder concerns diverge sharply. }
\label{sec:stakeholder_diverge}

Different subgroups in our population, defined above, express substantively different concern frames across all agencies and years. Table~\ref{tab:stakeholder_concern} reports the distribution of primary frames, assigned one per comment, across the seven subgroups. The contrasts are large: industry comments tend to be primarily concerned with Economic issues (63\% $\pm$ 18\% per agency); law-firm comments concentrate (67\% $\pm$ 17\%) on Legal concerns; labor unions on Safety (39\% $\pm$ 30\%) and Economic (30\% $\pm$ 28\%); expert organizations on Health (29\% $\pm$ 17\%). Individual commenters are by far the most diffuse. These divergences are not artifacts of topic mix: even within the same docket, stakeholder groups frame their concerns differently. Across 292 dockets where at least two stakeholder types each contributed $\geq$5 claims, the mean within-docket Jensen--Shannon divergence between org-type pairs is 0.18 (1{,}280 pairs), compared to 0.42 for cross-docket comparisons---confirming that roughly half the divergence reflects genuine stakeholder perspective differences rather than docket-topic selection.

These patterns provide large-$N$ empirical support for a long-standing prediction in administrative law. \citet{wagner2010administrative} theorizes that well-resourced participants---industry and law firms---dominate the notice-and-comment process through ``information capture'': by framing their input in the technical and legal registers that agencies are institutionally obligated to address, they shape the informational environment in which rules are made. Our data confirms this at scale: industry concentrates 63\% of its claims on economic cost-benefit framing (the register most directly tied to Executive Order 12866 cost-benefit review), while law firms concentrate 67\% on legal/constitutional framing (the register that triggers agency obligations under the APA). By contrast, individuals---the most numerous commenters---scatter across all frames, diluting their informational influence. \citet{mendelson2011rulemaking} argues that agencies should nonetheless take these diffuse individual comments seriously as expressions of democratic preference, yet acknowledges the structural challenge: individual commenters' concerns are less concentrated and harder for agencies to systematically engage with. The size of these effects matters for DOCKETBench: how accurately a system carries each subgroup's concerns is a natural evaluation axis for any system that aims to summarize, anticipate, or generate public input using LLMs.

\begin{table}[t]
\centering
\tiny
\setlength{\tabcolsep}{2pt}
\renewcommand{\arraystretch}{1.35}
\begin{tabular*}{\textwidth}{@{\extracolsep{\fill}}lccccccccc}
\toprule
 & \makecell{\textbf{Environmental}\\\textbf{impact}\\{\fontsize{4.6}{5.2}\selectfont\itshape pollution, climate}} & \makecell{\textbf{Public}\\\textbf{health}\\{\fontsize{4.6}{5.2}\selectfont\itshape disease, care}} & \makecell{\textbf{Equity \&}\\\textbf{justice}\\{\fontsize{4.6}{5.2}\selectfont\itshape fairness, rights}} & \makecell{\textbf{Economic}\\\textbf{cost-benefit}\\{\fontsize{4.6}{5.2}\selectfont\itshape costs, jobs}} & \makecell{\textbf{Legal \&}\\\textbf{constitutional}\\{\fontsize{4.6}{5.2}\selectfont\itshape authority, APA}} & \makecell{\textbf{Safety}\\\textbf{risks}\\{\fontsize{4.6}{5.2}\selectfont\itshape accidents, workers}} & \makecell{\textbf{Scientific \&}\\\textbf{technical}\\{\fontsize{4.6}{5.2}\selectfont\itshape evidence, methods}} & \makecell{\textbf{Procedural}\\\textbf{objections}\\{\fontsize{4.6}{5.2}\selectfont\itshape process, timing}} & \makecell{\textbf{Other}\\~\\{\fontsize{4.6}{5.2}\selectfont\itshape none of these}} \\
\midrule
\textbf{Law firm}    & \cellcolor[HTML]{F9EDE3}3.1 & \cellcolor[HTML]{FCF4EA}0.8 & \cellcolor[HTML]{FAEFE4}2.7 & \cellcolor[HTML]{EED2C8}12.8 & \cellcolor[HTML]{AC392D}\textbf{\textcolor{white}{67.4}} & \cellcolor[HTML]{FCF4EA}0.7 & \cellcolor[HTML]{F7E8DE}4.8 & \cellcolor[HTML]{F9ECE2}3.7 & \cellcolor[HTML]{F8EBE1}3.9 \\
\textbf{Industry}    & \cellcolor[HTML]{F9ECE2}3.4 & \cellcolor[HTML]{F8EBE1}4.1 & \cellcolor[HTML]{FDF6EC}0.2 & \cellcolor[HTML]{B1463A}\textbf{\textcolor{white}{63.0}} & \cellcolor[HTML]{F5E4DA}6.3 & \cellcolor[HTML]{FBF1E6}2.0 & \cellcolor[HTML]{EBCDC2}14.7 & \cellcolor[HTML]{FCF3E9}1.2 & \cellcolor[HTML]{F7E7DD}5.2 \\
\textbf{Labor union} & \cellcolor[HTML]{FBF2E8}1.3 & \cellcolor[HTML]{F7E7DD}5.3 & \cellcolor[HTML]{F6E5DB}6.1 & \cellcolor[HTML]{D8A196}\textbf{30.4} & \cellcolor[HTML]{F9EDE3}3.2 & \cellcolor[HTML]{CE897E}\textbf{\textcolor{white}{38.9}} & \cellcolor[HTML]{FCF4EA}0.7 & \cellcolor[HTML]{FBF2E8}1.3 & \cellcolor[HTML]{EED2C8}12.7 \\
\textbf{Expert org.} & \cellcolor[HTML]{F5E3D8}6.9 & \cellcolor[HTML]{DAA59A}29.1 & \cellcolor[HTML]{F8EBE1}4.0 & \cellcolor[HTML]{DFAFA5}25.2 & \cellcolor[HTML]{F5E3D9}6.6 & \cellcolor[HTML]{FAF0E6}2.1 & \cellcolor[HTML]{EDD0C5}13.6 & \cellcolor[HTML]{FCF4EA}0.8 & \cellcolor[HTML]{EFD5CA}11.7 \\
\textbf{Government}  & \cellcolor[HTML]{E2B6AC}22.8 & \cellcolor[HTML]{F6E5DA}6.2 & \cellcolor[HTML]{F9EDE3}3.1 & \cellcolor[HTML]{E6C0B5}19.3 & \cellcolor[HTML]{E7C3B9}18.1 & \cellcolor[HTML]{FAEFE5}2.3 & \cellcolor[HTML]{F3DFD4}8.3 & \cellcolor[HTML]{F6E5DB}5.9 & \cellcolor[HTML]{EBCDC2}14.0 \\
\textbf{Advocacy}    & \cellcolor[HTML]{C87B70}\textbf{\textcolor{white}{43.9}} & \cellcolor[HTML]{F0D9CE}10.5 & \cellcolor[HTML]{F4E1D7}7.4 & \cellcolor[HTML]{F1DAD0}10.0 & \cellcolor[HTML]{F3DED4}8.6 & \cellcolor[HTML]{FBF0E6}2.0 & \cellcolor[HTML]{F9EEE3}3.0 & \cellcolor[HTML]{FCF5EB}0.4 & \cellcolor[HTML]{EBCDC2}14.1 \\
\textbf{Individual}  & \cellcolor[HTML]{D3948A}\textbf{34.8} & \cellcolor[HTML]{E5BDB3}20.2 & \cellcolor[HTML]{EFD5CA}11.9 & \cellcolor[HTML]{F4E2D8}7.1 & \cellcolor[HTML]{F7E9DF}4.7 & \cellcolor[HTML]{FBF2E8}1.4 & \cellcolor[HTML]{FDF5EB}0.2 & \cellcolor[HTML]{FDF5EB}0.2 & \cellcolor[HTML]{E6C0B5}19.5 \\
\bottomrule
\end{tabular*}
\vspace{0.8em}
\caption{Stakeholder concerns: the focus of comments across stakeholder groups is shown (in percentages, \%). Stakeholder concerns diverge sharply -- industry concentrates 63\% on Economic, law firms 67\% on Legal/Constitutional, labor unions 39\% on Safety; individuals are most diffuse. See Appendix~\ref{app:concern_taxonomy} for tax. construction, mutual-exclusivity discussion, and examples of each frame. We assigned \textit{one} frame to each claim; frames that do \textit{not} exceed 5\% in any group (e.g. religious/moral, educational and unknown labels) we represent them as ``Other''.}
\label{tab:stakeholder_concern}
\end{table}

\begin{figure}[t]
\centering
\begin{minipage}{0.45\textwidth}
	\includegraphics[width=\linewidth]{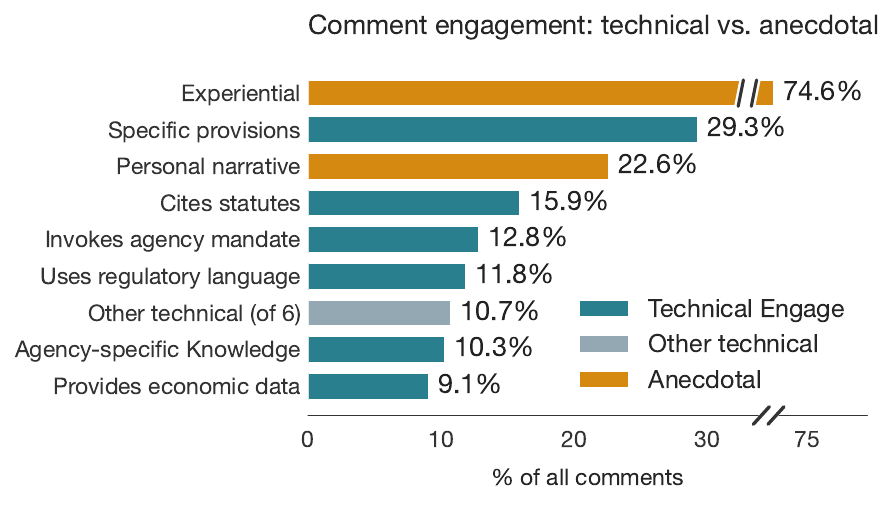}
	\subcaption{How commenters engage with provisions: Arguments used and their technical specificity.}
\end{minipage}\hfill
\begin{minipage}{0.45\textwidth}
	\includegraphics[width=\linewidth]{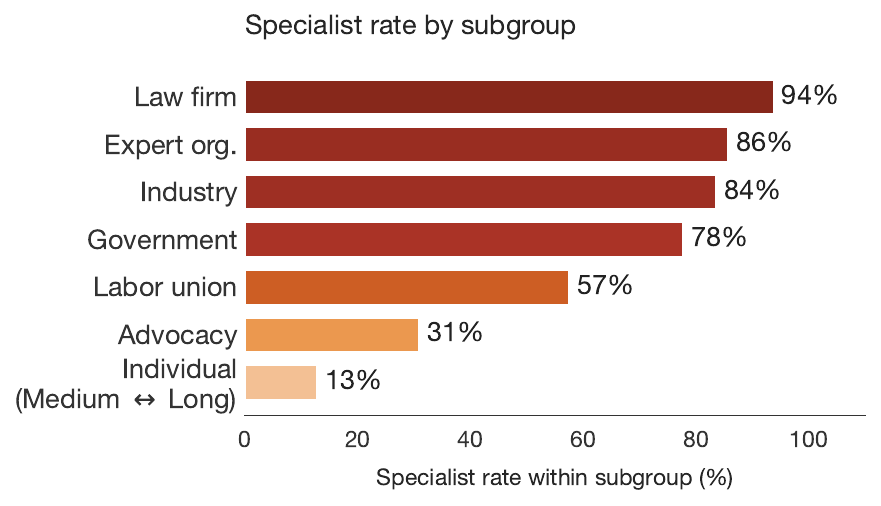}
	\subcaption{Who the commenters are: specialist within commenter pools from different organizations.}
\end{minipage}
\caption{Stakeholder engagement and specialist density.}
\label{fig:engagement_specialist}
\end{figure}

\paragraph{\ul{Finding \#3: Comments are High Quality}: Commenters demonstrate expertise and focus on specific provisions in rules. }

Contrary to scholarly belief that public comment periods are swamped with low-quality comments \citep{cuellar2005rethinking,mendelson2011rulemaking}, our results show a high degree of specialized, evidence-bearing and nuanced responses.\footnote{We do observe a substantial volume of comments originating from campaigns in our dataset, but prior work has indicated that government processing pipelines are capable of filtering such comments out \citep{handannader2023fake,balla2022responding}.} We use an LLM to label each comment as coming from either a specialist or a generalist and to then determine the subgroup of the commenter (i.e.,\ Individual, Law firm, Industry, Labor union, Advocacy NGO, Expert, Government\footnote{See Appendix~\ref{app:subgroup_counts} for full list and definitions.}). As shown in Figure~\ref{fig:engagement_specialist}, 30\% of the comments in our dataset contain technical aspects: addressing specific provisions, citing statutes, etc. Specialists are from every affiliation: nearly 1 in 10 comments from type \texttt{individual} are technical.

Finally, we observe that commenters are often targeted and specific in their input. Each engaged provision is addressed by an average of 7.8 and a median of 2 comments. A small fraction of provisions attract the bulk of attention while the rest receive only one or two distinct claims. General claims (e.g., ``hands off our Medicare'') would match many provisions and produce a much flatter distribution; instead, we observe commenters intentionally naming and responding to particular regulatory provisions. The implication for benchmarking is direct: rules cannot be treated as a single unit of prediction. Any system that summarizes, anticipates, or generates public input must operate at the \emph{provision} level and predict not just \emph{which} provisions attract comment but \emph{how much}.

\section{Can LLMs Reproduce Pluralistic Public Comment?}

As we imagine a future where AI becomes more involved in government, whether it be through informing policymakers or assisting decision-making and critiquing policy, we must ensure AI is \emph{pluralistically aligned}, or accurately represents diverse human perspectives~\cite{sorensen2024roadmap}. Unlike existing pluralism benchmarks \citep{kirk2024prism,durmus2024globalopinions}, which typically measure whether LLMs can represent diverse \textit{general} human perspectives, \textit{legal pluralism} offers a challenging and grounded alignment domain: legal debates can be specific and technical, involving groups and stakeholders that are seldom considered by model designers. To address this, we introduce DOCKETBench, a benchmark that leverages DOCKET to create a \textit{challenging legal pluralism} benchmark measuring the breadth of perspectives expressed in policy deliberations. The challenge of DOCKETBench is: \textit{how closely can an LLM's critiques resemble the breadth of human comments received by a real federal rule?}

\begin{wraptable}{r}{0.47\textwidth}
	\centering
	\scriptsize
	\setlength{\tabcolsep}{3pt}
	\renewcommand{\arraystretch}{1.15}
	\begin{tabularx}{\linewidth}{@{}l>{\raggedleft\arraybackslash}X>{\centering\arraybackslash}X@{}}
		\toprule
		\textbf{Model} & \textbf{Agrees with most} & \textbf{Agrees with least} \\
		\midrule
		gpt-4o-mini              & gov {\tiny\textit{(0.68)}} & labor {\tiny\textit{(0.38)}} \\
		\rowcolor{gray!15}\hspace*{1em}+persona & gov {\tiny\textit{(0.61)}} & labor {\tiny\textit{(0.43)}} \\
		gpt-5-mini               & gov {\tiny\textit{(0.65)}} & labor {\tiny\textit{(0.48)}} \\
		\rowcolor{gray!15}\hspace*{1em}+persona & gov {\tiny\textit{(0.68)}} & labor {\tiny\textit{(0.47)}} \\
		gpt-5                    & industry {\tiny\textit{(0.68)}} & labor {\tiny\textit{(0.61)}} \\
		\rowcolor{gray!15}\hspace*{1em}+persona & gov {\tiny\textit{(0.65)}} & labor {\tiny\textit{(0.56)}} \\
		Llama-3.3-70B            & gov {\tiny\textit{(0.47)}} & labor {\tiny\textit{(0.26)}} \\
		\rowcolor{gray!15}\hspace*{1em}+persona & expert {\tiny\textit{(0.61)}} & advocacy {\tiny\textit{(0.44)}} \\
		Llama-3.1-8B             & gov {\tiny\textit{(0.68)}} & expert {\tiny\textit{(0.40)}} \\
		\rowcolor{gray!15}\hspace*{1em}+persona & gov {\tiny\textit{(0.48)}} & labor {\tiny\textit{(0.36)}} \\
		\midrule
		Human commenters        & individual {\tiny\textit{(0.86)}} & advocacy {\tiny\textit{(0.71)}} \\
		\bottomrule
	\end{tabularx}
	\caption{Whose side LLMs take. For each model, the stakeholder group it agrees with most and least: among judged pairs where an LLM claim and a real claim make a point about the same thing, the share judged \texttt{same}, by the group of the real commenter (commenters counted as the interest they file for). \emph{Human commenters}: randomly sampled real commenters judged against the other comments in their docket, groups with ten or more pairs only. Overall agreement is 0.43--0.66 for the models and 0.75 for human commenters; the full per-group table with confidence intervals is Table~\ref{tab:agreement_full}.}
	\label{tab:bucket_alignment}
\end{wraptable}

DOCKETBench challenges LLMs to align with humans, pluralistically, along four primary dimensions (Table~\ref{tab:benchmark_results}): 
\textbf{(A) Provision Targeting} (i.e., which provisions in proposals get attention) \textbf{(B) Claim Alignment} (i.e., what claims are raised and what stances are taken) \textbf{(C) Frame Alignment} (i.e., how concerns are framed) and \textbf{(D) Style Alignment} (i.e., what style comments are written in). For each of these four dimensions we report precision (the share of what the LLM wrote that some human also wrote) and recall (the share of what humans wrote that the LLM also wrote; specifically, Overton recall as defined by \cite{poole2025benchmarking}). Finally, we challenge models to be \textit{steerable}: a \textit{pluralistic} LLM should, and be able to be induced to, accurately represent viewpoints of particular stakeholder subgroups (e.g., industry representatives, advocacy organizations, affected individuals). 
DOCKETBench is descriptive rather than normative: we measure LLM performance against what different subgroups in our data actually expressed, but make no claims that human comments are correct in legal or normative terms. Our benchmark focuses on the question of whether models can faithfully represent the perspectives held by stakeholders in federal policymaking, a necessary prerequisite for tools that would help (1) agencies understand and anticipate public response and (2) members of the public participate more readily in democratic discourse.

\begin{table}[tp]
\centering
\begin{subtable}{\textwidth}
\centering
\scriptsize
\setlength{\tabcolsep}{5pt}
\renewcommand{\arraystretch}{1.0}
\begin{tabular}{@{}lcccc@{}}
\toprule
& \multicolumn{2}{c}{\textbf{(A) Provision targeting}} & \multicolumn{2}{c}{\textbf{(B) Claim alignment}} \\
\cmidrule(lr){2-3} \cmidrule(lr){4-5}
\textbf{Model} & \makecell{LLM-commented provisions\\humans also hit (\%)} & \makecell{Human-commented provisions\\LLM also hit (\%)} & \makecell{LLM claims\\humans also made (\%)} & \makecell{Human claims\\LLM also made (\%)} \\
\midrule
gpt-4o-mini & \cellcolor{blue!20}58 & \cellcolor{blue!33}92 & \cellcolor{blue!13}20.5 & \cellcolor{blue!8}27.0 \\
\hspace*{1em}+persona & \cellcolor{blue!18}56 & \cellcolor{blue!36}95 & \cellcolor{blue!14}21.0 & \cellcolor{blue!34}53.8 \\
gpt-5-mini & \cellcolor{blue!11}47 & \cellcolor{blue!32}91 & \cellcolor{blue!11}19.6 & \cellcolor{blue!21}40.4 \\
\hspace*{1em}+persona & \cellcolor{blue!13}49 & \cellcolor{blue!39}98 & \cellcolor{blue!12}19.8 & \cellcolor{blue!38}58.7 \\
\hspace*{1em}+fewshot & \cellcolor{blue!11}47 & \cellcolor{blue!38}97 & \cellcolor{blue!12}19.9 & \cellcolor{blue!13}32.6 \\
\hspace*{1em}+RAG & \cellcolor{blue!13}49 & \cellcolor{blue!30}89 & \cellcolor{blue!15}21.2 & \cellcolor{blue!13}31.7 \\
\hspace*{1em}+tool & \cellcolor{blue!10}45 & \cellcolor{blue!30}89 & \cellcolor{blue!14}20.9 & \cellcolor{blue!11}30.5 \\
gpt-5 & \cellcolor{blue!11}47 & \cellcolor{blue!36}95 & \cellcolor{blue!8}18.1 & \cellcolor{blue!18}37.8 \\
\hspace*{1em}+persona & \cellcolor{blue!14}50 & \cellcolor{blue!40}\textbf{99} & \cellcolor{blue!13}20.3 & \cellcolor{blue!40}\textbf{60.4} \\
\hspace*{1em}+fewshot & \cellcolor{blue!11}47 & \cellcolor{blue!40}\textbf{99} & \cellcolor{blue!13}20.3 & \cellcolor{blue!10}29.0 \\
\hspace*{1em}+RAG & \cellcolor{blue!29}69 & \cellcolor{blue!22}82 & \cellcolor{blue!23}24.7 & \cellcolor{blue!10}28.6 \\
\hspace*{1em}+tool & \cellcolor{blue!16}53 & \cellcolor{blue!32}91 & \cellcolor{blue!18}22.5 & \cellcolor{blue!12}31.3 \\
Llama-3.3-70B & \cellcolor{blue!30}71 & \cellcolor{blue!21}81 & \cellcolor{blue!35}30.3 & \cellcolor{blue!14}33.3 \\
\hspace*{1em}+persona & \cellcolor{blue!40}\textbf{83} & \cellcolor{blue!8}68 & \cellcolor{blue!40}\textbf{32.5} & \cellcolor{blue!26}45.6 \\
Llama-3.1-8B & \cellcolor{blue!8}43 & \cellcolor{blue!31}90 & \cellcolor{blue!17}22.1 & \cellcolor{blue!16}35.1 \\
\bottomrule
\end{tabular}
\caption{\textbf{Content.} (A) Does the model comment on the same provisions real commenters did? Precision: share of LLM-commented provisions that a human commenter also commented on. Recall: share of human-commented provisions the LLM also hit. Recall is high because the models spread their claims over nearly the whole proposal (78--99\% of a docket's provisions, against 46\% for human commenters), so for most models precision sits near that 46\% base rate. (B) Does it say the same things about them? Precision: share of LLM claims a judge rates \texttt{same} as some human claim on the docket. Recall: share of human claims with such an LLM match. \emph{+persona} rows pool seven stakeholder personas; persona moves (A) by at most 13 points and claim precision by at most 2.2, but raises claim recall by 12--27 points, mostly because persona runs write three to six times more claims (Llama-3.3-70B excepted; Appendix~\ref{app:persona_aggregation}).}
\label{tab:benchmark_content}
\end{subtable}

\vspace{1pt}

\begin{subtable}{\textwidth}
\centering
\scriptsize
\setlength{\tabcolsep}{5pt}
\renewcommand{\arraystretch}{1.0}
\begin{tabular}{@{}lcccc@{}}
\toprule
& \multicolumn{2}{c}{\textbf{(C) Argumentative frame}} & \multicolumn{2}{c}{\textbf{(D) Writing style}} \\
\cmidrule(lr){2-3} \cmidrule(lr){4-5}
\textbf{Model} & \makecell{LLM frames\\humans also used (\%)} & \makecell{Human frames\\LLM also used (\%)} & \makecell{LLM styles\\humans also used (\%)} & \makecell{Human styles\\LLM also used (\%)} \\
\midrule
gpt-4o-mini & \cellcolor{green!8}71.6 & \cellcolor{green!8}47.1 & \cellcolor{green!8}64.8 & \cellcolor{green!9}58.9 \\
gpt-5-mini & \cellcolor{green!30}78.8 & \cellcolor{green!30}55.8 & \cellcolor{green!31}76.0 & \cellcolor{green!30}61.9 \\
\hspace*{1em}+fewshot & \cellcolor{green!29}78.3 & \cellcolor{green!28}54.8 & \cellcolor{green!27}74.1 & \cellcolor{green!22}60.8 \\
\hspace*{1em}+RAG & \cellcolor{green!24}76.9 & \cellcolor{green!16}50.4 & \cellcolor{green!37}78.8 & \cellcolor{green!36}62.7 \\
\hspace*{1em}+tool & \cellcolor{green!27}77.6 & \cellcolor{green!21}52.1 & \cellcolor{green!33}76.7 & \cellcolor{green!37}62.9 \\
gpt-5 & \cellcolor{green!40}\textbf{81.9} & \cellcolor{green!40}\textbf{59.6} & \cellcolor{green!40}\textbf{80.1} & \cellcolor{green!26}61.3 \\
\hspace*{1em}+fewshot & \cellcolor{green!29}78.4 & \cellcolor{green!28}55.0 & \cellcolor{green!37}78.5 & \cellcolor{green!26}61.3 \\
\hspace*{1em}+RAG & \cellcolor{green!36}80.7 & \cellcolor{green!30}55.5 & \cellcolor{green!34}77.3 & \cellcolor{green!18}60.2 \\
\hspace*{1em}+tool & \cellcolor{green!29}78.5 & \cellcolor{green!28}55.0 & \cellcolor{green!39}79.5 & \cellcolor{green!21}60.6 \\
Llama-3.3-70B & \cellcolor{green!13}73.1 & \cellcolor{green!15}50.0 & \cellcolor{green!16}68.8 & \cellcolor{green!8}58.8 \\
Llama-3.1-8B & \cellcolor{green!12}73.0 & \cellcolor{green!15}50.0 & \cellcolor{green!16}68.5 & \cellcolor{green!40}\textbf{63.3} \\
\bottomrule
\end{tabular}
\caption{\textbf{Style.} (C) Does the model argue the way real commenters argue, and (D) write the way they write? We extract argumentative frames (e.g.\ ``appeal to statutory intent'') and style markers (e.g.\ ``cites current market trends'') from each comment and judge LLM--real pairs on the same docket. Precision: share of LLM frames (styles) that a human one matches. Recall: share of human frames (styles) that an LLM one matches. Persona moves every cell here by under 8 points (largest: gpt-4o-mini style precision, +7.3), so persona rows are omitted.}
\label{tab:benchmark_style}
\end{subtable}
\caption{\textbf{LLMs reproduce the style and framing of public comments far better than their substance.} DOCKETBench results on 558 dockets; each cell is a per-docket mean. Rows below a model are that model with few-shot, RAG, or RAG-plus-tool augmentation. ``LLM \ldots'' columns are precision, ``Human \ldots'' columns recall. Cells are shaded within each column, darker = better; the best value in each column is bold. Claim pairs are judged by gpt-4o-mini; frame and style pairs by Llama-3.3-70B (Appendix~\ref{app:pluralism_methods}). Frame and style precision reaches 65--82\%, while claim precision stays at 18--30\%, and no prompting strategy closes the gap.}
\label{tab:benchmark_results}
\end{table}

\subsection{Benchmark Creation}\label{sec:benchmark_creation}
DOCKETBench is built directly from DOCKET using the provisions, claims and matches described in Section~\ref{sec:dataset}. We describe now the decisions guiding DOCKETBench and goal of our evaluations.  %
\paragraph{Dataset filtering:}
Starting from 17,198 proposed rules,\footnote{A docket is a collection of one or more proposals, comments on these proposals, and finalized rule.} we apply the following filters: (1) there must be $\geq 1$ proposal per docket\footnote{Some dockets only consist of notices or finalized rules, so may lack a proposal.} with 5--100 extracted proposed provisions (i.e., \textit{to exclude very small and very large rules}); (2) there must be $\geq$ 3  (claim, proposed provision) matched pairs (i.e. \textit{to exclude provisions with no, or off-topic comments}). To expand our (claim, proposed provision) matches, we distill an additional cross-encoder to explicitly include opposing comments (Appendix~\ref{app:prompt_match2}), since a comment that takes the opposite position from the proposed text still counts as a valid match. This includes an additional 30\% of true-positive (claim, provision) pairings. These filters yield a final benchmark population of 558 dockets, 22{,}837 proposed provisions, 13{,}824 comments, and 43{,}371 unique real claims. The 558 dockets span 19 federal parent agencies. On average, provisions receive 7.8 claims (median: 2), with a 17\% agency response rate on medium-length and long comments.

\paragraph{Metrics:} We evaluate alignment between LLM-generated and human comments along two dimensions: \textbf{Content-Level Alignment} (metrics A-B) and \textbf{Style-Level Alignment} (metrics C-D). \newline\textit{(A) Provision Targeting (Precision \& Recall):} Using the claim $\rightarrow$ provision mappings described in Section \ref{sec:dataset}, we measure how often models and humans target the same provisions within a proposal.
We report, in Table~\ref{tab:benchmark_results}, the share of LLM-commented provisions that humans also hit (i.e. LLM precision) and the share of human-commented provisions that the LLM also hit (i.e. LLM recall).
\newline 
\textit{(B) Claim Alignment, (C) Argumentative Frame, and (D) Writing Style (Precision \& Recall):} For these metrics, we compare human and LLM-generated comments on the same proposals. We use an LLM to extract claims, argumentative frames, or style keywords from each comment.\footnote{\textbf{Claim examples}: ``rule would gut religious accommodation protections'', ``prevent release of accounting for disclosures''. \textbf{Framing examples}: ``Competitive disadvantage analysis'', ``Appeal to statutory intent''. \textbf{Style examples:} ``Attaches supporting documents'', ``Cites current market trends''. See Tables \ref{tab:comment_to_claims}, \ref{tab:extraction_examples}.}An LLM comment ``matches'' a human comment when another LLM judges the extracted elements to be equivalent.\footnote{For example, an LLM and human who both employ ``cost benefit analysis'' framing would match on metric (C).} We calculate Precision and Recall based on these matches. We report in Table~\ref{tab:benchmark_results} the share of LLM claims (frames, styles) that humans also made or used, and the share of human claims (frames, styles) that the LLM also made or used.
\newline \textit{Agreement by stakeholder group:}
This metric asks whose side the LLM takes. For every judged pair where the LLM claim and the human claim make a point about the same provision (i.e. the judge labels it \texttt{same} or \texttt{opposing}), we compute, for each stakeholder group, $\text{Agreement(group)} = \nicefrac{|\texttt{same}|}{(|\texttt{same}| + |\texttt{opposing}|)}$ over the human claims. We report for each model, in Table~\ref{tab:bucket_alignment}, the group it agrees with most and least; the full per-group table, with confidence intervals and the same number for human commenters, is Table~\ref{tab:agreement_full} in Appendix~\ref{app:bucket_alignment}. %

\paragraph{Benchmark design and testing:}
As stated previously, each task is driven by the \textit{baseline} prompt: \textit{``given \textit{<proposal>}, generate a full comment [that highlights the concerns of stakeholder group X]''} We test the following variants: 

\begin{itemize}[leftmargin=*, itemsep=1pt, parsep=0pt, topsep=2pt]
    \item \textbf{Vanilla:} baseline prompt on: gpt-\{4o-mini, 5-mini, 5\}, Llama-3.3-70B, and Llama-3.1-8B.
    \item \textbf{+persona:} we add persona-specific guidance to the baseline prompt. We run seven passes per task, each conditioning on a different stakeholder persona;\footnote{Full text of persona-guided prompts in Appendices~\ref{app:prompt_gen_vanilla}--\ref{app:subgroup-persona}; examples of persona-generated comments in Appendix~\ref{app:full_comment_examples}. Generation hyperparameters are kept identical across models: temperature 1.0 (default), no top-$p$ truncation. For gpt-5-family models we use \texttt{reasoning\_effort=low} and OpenAI \texttt{service\_tier=flex}; vLLM-served open-weight models use the model's default sampling.} we then pool the outputs with uniform weight before measuring their alignment.
    \item \textbf{+augmentations:} we test several common augmentations on pure language modeling including: \textbf{+fewshot} (randomly selecting $k=2$ human-written comments in the prompt); \textbf{+RAG} (we build a bm25 mock ``fact'' database by using an LLM to extract evidence, studies, cited statutes and other facts from human comments from DOCKET \textit{not} in DOCKETBench and select the top $k=3$ to include in the prompt); \textbf{+tool} (the model has the option to query the bm25 database we constructed up to three times during its generation run).
\end{itemize}

\subsection{Benchmark Results}
\label{sec:benchmark_results}

We now present results from evaluating our baseline strategies on DOCKETBench. We find, overall, that \textbf{alignment on content is much lower than alignment on framing devices and style.} Specifically, we see large \textit{content-level} gaps (i.e. metrics (A) and (B)) between LLMs and human comments (claim precision 18--30\%), but small frame and style gaps (metrics (C) and (D), precision 65--82\%). This suggests current LLMs' stylistic pluralism and content-level pluralism abilities are quite divergent: LLMs can write comments that \textit{sound like} real commenters but fail to capture the opinions each subgroup cares about. We also find that persona conditioning does not have a uniformly positive impact. Persona conditioning helps weaker models (gpt-4o-mini: +7.3pp style prec.) but not stronger ones (gpt-5-mini: -0.1pp style, prec.). Additionally, we find that \textbf{+Persona} changes whose side the model takes but not which groups it covers: pooled over personas, agreement with each group shifts by a few points at most (Table~\ref{tab:bucket_alignment}). In other words, giving a model a persona changes its stance much more than it changes which real arguments it makes (Appendix~\ref{app:bucket_alignment}). These findings diverge from prior work indicating that (a) language models are steerable, and that (b) persona guidance can be effective in inducing more representative models. We hypothesize that, unlike in generalist domains, \textit{technical and legal domain pretraining text put a strong prior over generative performance that overwhelms naive steerings a user might apply}.

One vivid example of this is \emph{whose side} LLMs take (Table~\ref{tab:bucket_alignment}). Only 21--33\% of LLM claims make a point that a real commenter \textit{also} made (i.e. in our data, the cases where a judge labels the pair \texttt{same} or \texttt{opposing}). Another 55--63\% are on the same \textit{topic} as a real claim but make a point no commenter made, and 8--17\% of LLM comments are on topics that no human commenter focused on (Appendix~\ref{app:agreement_anatomy}). On the claims that \textit{do} line up with a real claim, LLMs tend to agree more with government and industry positions and less with labor and advocates. Humans show more inter-commenter agreement than human-LLM agreement. Two reasons can explain this wide spread. Firstly, LLMs hedge: 18\% of the time they argue both sides of the same real point. Secondly their default position follows the procedural and economic framing that industry and government commenters use, so they side against safety, health and environmental claims more often (Appendix~\ref{app:agreement_anatomy}). Three quarters of labor claims, for example, are safety claims, where only 6\% of LLMs claims are. This gap holds after controlling for retrieval similarity, agency and judge \textit{and} prompting LLMs to \textit{specifically} act more like labor advocates. We caveat: labor's positions are unpopular with human commenters too: other commenters \textit{also} agree with labor less than with any other group; however, LLMs still do so even more.

Even when LLMs overlap with human comments, the positions they argue are often more moderate.  LLMs, we find, produce comments that overwhelmingly advocate modifying proposals, rather than adding or deleting them: 85.9\% of gpt-5-mini claims recommend modification, compared with 43.8\% of human claims.
LLMs rarely oppose proposals outright: only 5.1\% of their claims are oppositional, compared with 30.4\% of human claims.
In addition, the \textit{kinds} of arguments that LLMs make to advocate their points skew differently, and LLMs tend to argue from different perspectives. In contrast to human commenters, who mix experiential, economic and technical arguments,
we find that 60.3\% of gpt-5-mini claims are procedural, compared with 15.7\% of human claims, while normative arguments fall from 26.6\% in human comments to 4.5\% in gpt-5-mini comments, and experiential arguments fall from 5.8\% to 0.8\% (Table \ref{tab:argument_dist}). Finally, as shown in Table~\ref{tab:benchmark_results}, models touch 78--99\% of a docket's provisions, humans 46\%. The attention patterns given by LLMs across provisions, independent of model size or persona prompting, are far more uniform than the attention given by humans. Humans concentrate heavily on a small fraction of provisions; models distribute claims uniformly.

\section{Discussion}
\label{sec:discussion}

Notice-and-comment exists as a way for agencies to receive input from the public. The Administrative Procedures Act (APA) asks for ``data, views, or arguments'' from anyone and makes agencies answer the significant ones \citep{usc_553}. As we can see in our data, this process brings in diverse stakeholder views:  Table~\ref{tab:stakeholder_concern} shows industry stakeholders commenting primarily on cost concerns, law firms commenting on statutory authority, labor unions commenting on safety, individuals commenting on a range of concerns. No single group, we see, raises the full set of concerns; the set comes from the mix. If LLMs were to be used in notice\& comment, as suggested by \cite{balla2022responding,hr7528,cdt2025inventory,hhs2024headstart,pnnl2024policyai}, we would expect models to be able to reflect comments for all of these groups. However, our results show that they cannot achieve this aim. These results extend: a tool that summarizes a hundred thousand comments for a rule-writer and lands on the procedural reading each time tells the agency what the median lawyer would have said, which the agency already knew. \citet{wagner2010administrative} worried that well-resourced commenters crowd the record with technical and legal argument. A model that defaults to that register would also achieve this end.

This has been addressed, in computer science, primarily through introducing persona prompts and steering vectors \citep{argyle2023outofone,santurkar2023whose}. In our case, it did very little. Asked to write as a labor union, a model writes more claims (4/5 models wrote 3-6x as many as humans; Llama-3.3-70B wrote half as many), takes a different stance, and still argues from the same procedural and economic frames; its agreement with what real labor commenters said moves by a few points at most for the gpt models (10--18 points for Llama-family models) (Table~\ref{tab:bucket_alignment}, Appendix~\ref{app:bucket_alignment}). The persona changes the stance and style, but not the material. In other words, models can be told whom to speak for, but they do not know what that group actually wishes to comment on. Retrieval, we find, helped gpt-5 more than any persona did (claim precision 24.7 vs.\ 18.1, Table~\ref{tab:benchmark_results}), which points the same way: the missing ingredient is information, not instruction.

Several of our findings are novel in relation to prior work on LLM pluralism. In \citet{argyle2023outofone} a demographic persona was enough to move GPT-3's survey answers toward the group's own, and in \citet{santurkar2023whose} and \citet{durmus2024globalopinions} steering prompts moved opinion distributions toward the named demographic or country. Here, personas impact claims by at \textit{most} 2.2 points and shift agreement towards any one group by at most a few points for gpt models (Llama-family models shift by 10--18 points on labor). To rule out confounders, we tested several, including: the effects of LLM-generated comment-length (i.e., longer comments have more claims, and will thus match more). We matched human units to different numbers of extracted units, $k\in\{5, 10, 15,\ldots\}$. Shown in Appendix~\ref{app:claim_budget} (Figure~\ref{fig:budget_curves_app}), recall levels off around $k$=35--50, claim precision moves by less than 0.4 points across all budgets, and unconstrained generation lands on the human median (46 vs.\ 48 claims per docket) but does not track how much commenting a docket actually drew. Additionally, we found that comment length, topic and agency bias were important signals for comment uptake; our observed results were robust after controlling for these variables.

One explanation for why LLMs reflect such narrow views, here, is that the language of notice\&comment is quite restrictive. A proposed rule is written in a narrow register: section numbers, defined terms, modal verbs, the formulas of the Federal Register. Conditioning on this in the prompt pulls a model toward a specific part of the pretraining distribution: the language of the lawyers and agency staff who wrote the documents it learned from. Everything we measure points to this pull. Sixty percent of LLM claims are procedural against 16\% of human claims; experiential claims nearly vanish (0.8\% vs.\ 5.8\%); the models spread their claims almost evenly over the provisions where humans concentrate on a few; and the groups the models agree with most, government and industry, are the ones that write in that register themselves. In prior settings, which typically ask for responses to surveys, a persona is steering a model in a more general text setting, which might explain why it has a greater steering effect: pretraining has shown the model more realistic responses from these groups in these settings. If this hypothesis is correct, steering should fail most in other formal domains. We did not test this directly; we leave this open for future work.

So far, our findings have focused on low model performance and the shortcomings of current pluralism strategies for mitigating them. Another lens with which to view our results is the \textit{strength} LLMs demonstrate on sections of DOCKETBench. Concerns are emerging in government and technology that systems for democratic deliberation may become overwhelmed with rising AI input; improvements on style and framing can increase the number of comments making past filtering systems \cite{handannader2023fake}. Table \ref{tab:benchmark_results} shows strong baseline model performance in Style and Argumentation, strong +persona impacts, and scaling trends for model size; together, these are indications of increased performance being imminently achievable. We urge a broader conversation around what \textit{kinds} of improvements should be valued in benchmarks: pluralistic alignment, in Style and Argumentation, means \textit{sounding like} a subgroup, not necessarily accurately articulating their concerns. We hope in future work to continue exploring AI detection strategies for optimal human-in-the-loop collaboration. We hope that DOCKET, and its derivative, DOCKETBench can serve as a valuable tool not just for the use-cases outlined here, but other use-cases in the broader community.

\section{Related Work}\label{sec:related-work}

Motivating DOCKET is a large literature studying how to better inform policymakers and promote participation in rulemaking, both theoretically \citep{libgober2020strategic, stephenson2006costly, gailmard2017participation} and empirically. Scholars have used notices, rules, and comments to study influence \citep{yackee2006bias} and conduct detailed analyses of particular regulators---the SEC \citep{nixon2002friends, ban2019presence}, EPA \citep{balla2022lost}, DOT \citep{naughton2009understanding}, FCC \citep{defigueiredo2006erulemaking}, and Dodd-Frank financial regulations \citep{carpenter2024inequality}. But these studies typically cover tens to hundreds of rules over short time windows; ours covers magnitudes more, allowing us to test at scale cross-agency heterogeneity and longitudinal dynamics. Our work also relates to administrative-law scholarship on the democratic promise and limits of notice-and-comment rulemaking. The APA requires agencies to solicit and consider public ``data, views, or arguments'' before issuing final rules \cite{usc_553}, and scholars have long treated this process as a central mechanism of public participation in administrative governance \cite{davis1970supplement,kerwin2018rulemaking,coglianese2006citizen}.
At the same time, prior work has questioned whether rulemaking actually democratizes agency decision-making or instead privileges repeat players, technical experts, and organized interests \cite{cuellar2005rethinking,mendelson2011rulemaking,wagner2010administrative}. DOCKET provides evidence for this debate by linking public comments to specific proposed provisions, agency responses, and final-rules, making it possible to measure not only who participates but what claims they raise, which provisions they target, and how agencies respond.

A second line of related work studies mass participation, comment campaigns, and the integrity of
public-comment systems. Agencies increasingly receive large volumes of duplicative, campaign-driven,
or malattributed submissions, raising questions about how comments should be filtered, summarized,
and weighed \cite{balla2022responding,mendelson2011rulemaking,balla2022lost}. Handan-Nader's
study of the FCC's ``Restoring Internet Freedom'' proceeding shows both the scale of inauthentic
participation and the methodological difficulty of identifying which comments agencies actually
address \cite{handannader2023fake}. Recent policy discussions have extended these concerns to
computer-generated comments and comment-management infrastructure \cite{hr7528}. DOCKET
builds directly on this literature, but shifts the focus from detecting or discounting mass comments to
mapping the substantive structure of public input at scale.

Our benchmark also connects to work on pluralistic alignment and opinion representation in language
models. Recent work argues that AI systems should not collapse diverse human preferences into a
single aggregate objective, but should instead preserve disagreement across individuals, cultures, and
social groups \cite{sorensen2024roadmap,conitzer2024position}. Datasets and benchmarks such as
PRISM and global-opinion evaluations test whose views models represent and whether models can
capture subjective, multicultural, or politically heterogeneous judgments
\cite{kirk2024prism,durmus2024globalopinions,durmustowards}. DOCKETBench differs in evaluating pluralism in a consequential institutional setting: stakeholders are not answering surveys or hypothetical preference questions, but making provision-specific claims about proposed law. This setting requires models to recover not only broad viewpoints, but also the concrete claims, legal frames, evidentiary practices, and stakeholder-specific priorities.

Finally, DOCKET relates to technical work on retrieval, legal-text representation, and model-based
evaluation. Our extraction of provisions is grounded in legal theories of deontic relations
\cite{hohfeld1913fundamental} and recent work that represents regulatory obligations as discrete
units whose survival can be tracked from proposed to final rules \cite{leahey2026obligations}. Our
matching pipeline draws on modern retrieval and reranking methods, including sentence embeddings,
dense retrieval, and cross-encoder reranking \cite{reimers2019sentencebert,karpukhin2020dpr,nogueira2019multistage},
as well as large-scale serving and evaluation tools \cite{kwon2023efficient,zheng2023judging}. The
benchmark interventions we test also build on common LLM adaptation strategies: persona prompting
\cite{li2016persona}, retrieval-augmented generation \cite{lewis2020retrieval}, and tool-mediated
generation. In parallel, agencies and contractors are beginning to use AI systems for public-comment
analysis and regulatory review \cite{cdt2025inventory,hhs2024headstart,icf_commentworks,icf_genai_comments,pnnl2024policyai}.
If LLMs are used to summarize comments, anticipate stakeholder objections, or assist rule drafters, failures of pluralistic representation may directly affect how public input is understood.

\section{Conclusion}

We set out to measure two things: how the public takes part in federal rulemaking, and whether language models can take part in the same way. For the first, DOCKET links 42.3M comments, 17,198 proposed rules and 22,400 final rules at the level of the individual provision, claim and agency response. The record does not match the common belief that comment periods are swamped by low-quality input. Commenters engage specific provisions, bring specialized knowledge about a third of the time, and divide by interest in what they argue about. Agencies answer them, and when they rewrite a provision they adopt commenter wording 11--26\% of the time.

For the second, DOCKETBench asks a model the question an agency would put to a commenter, ``how should this rule change?'', and measures the answer against what real commenters said. The models write comments that sound like public comments; however, they typically represent claims most often that government and industry commenters made. They rarely say what labor unions, advocates or individuals said, and persona prompts do not change that. We release both the dataset and the benchmark so that the tools now being built to draft and read public comments can be judged on whose perspectives they carry rather than on how fluent they sound.

\clearpage

\bibliographystyle{ACM-Reference-Format}
\bibliography{references}

\appendix
\clearpage
\addtocontents{toc}{\protect\setcounter{tocdepth}{2}}
\tableofcontents
\vspace{6pt}

\noindent\textbf{Organization of this appendix.} Appendix~\ref{app:corpus} covers the DOCKET corpus: sources, deduplication, parsing, the stakeholder, concern-taxonomy and uptake analyses, the hand check of the LLM labels (Appendix~\ref{app:human_validation}), and limitations. Appendix~\ref{app:bench} covers DOCKETBench: how we built it, the metrics, the claim-budget and persona choices, the stakeholder-alignment analyses behind Table~\ref{tab:bucket_alignment}, and how LLM claims differ from human claims in direction and argument type (Appendix~\ref{app:direction_argument}). Appendix~\ref{app:prompts} lists every prompt.

\section{The DOCKET corpus: sources, processing, and coverage}
\label{app:corpus}

\subsection{Per-agency corpus statistics}
\label{app:agency_list}

Table~\ref{tab:per_agency} enumerates all 44 federal agencies in DOCKET, with per-agency totals for dockets, public comments, three regulations.gov document classes (notices, proposed rules, finalized rules), three classes of validated cross-document matches, and high-volume comment clusters ($>$100 deduplicated comments per cluster, a proxy for agencies that experience organized commenting campaigns). \emph{Notices+Prop.+Final} is the document total used in Figure~\ref{fig:agency-chart} (for example, DHS: $598+60+119=777$ documents). The three match types are: (\textbf{comment $\to$ proposal}) --- a public-comment claim addresses a specific proposed-rule provision; (\textbf{final rule $\to$ comment}) --- a finalized-rule provision corresponds to a public comment that influenced it; and (\textbf{response $\to$ comment}) --- an explicit agency-response paragraph in the final-rule preamble addresses a specific public comment. Agencies are ordered by comment volume; the bottom 5 rows (FCC, SEC, FDIC, FEC, USPS) are covered through external comment portals (Appendix~\ref{app:scrapers}) --- their proposed/finalized rule counts come from regulations.gov as for the other agencies, but the provision-extraction and matching pipelines have not yet been run on the externally-scraped comment streams. Tabular companion to Figure~\ref{fig:agency-chart}.

\begin{table}[!htbp]
\centering
\tiny
\setlength{\tabcolsep}{2pt}
\renewcommand{\arraystretch}{1.02}
\resizebox{0.92\textwidth}{!}{%
\begin{tabular}{l r r r r r r r r r r r r}
\toprule
\textbf{Agency} & \textbf{Dockets} & \textbf{Comm.\ (analyzable)} & \textbf{Comm.\ (FedReg)} & \textbf{Cov.} & \textbf{Notices} & \textbf{Prop.} & \textbf{Final} & \textbf{(c$\to$p)} & \textbf{(f$\to$c)} & \textbf{(r$\to$c)} & \textbf{Clust.\ $>$100} & \textbf{Match} \\
\midrule
FDA        & 3{,}111 & 1{,}233{,}623 & 1{,}557{,}309 & 79\% & 8{,}874 & 440 & 766 & 72{,}302 & 19{,}691 & 15{,}102 & 234 & 29\% \\
CMS        & 760 & 674{,}012 & 1{,}015{,}135 & 66\% & 2{,}703 & 421 & 455 & 183{,}163 & 94{,}687 & 40{,}345 & 346 & 29\% \\
ED         & 1{,}148 & 627{,}874 & 964{,}550 & 65\% & 1{,}773 & 197 & 249 & 43{,}506 & 26{,}055 & 2{,}505 & 111 & 38\% \\
FWS        & 740 & 617{,}767 & 1{,}008{,}149 & 61\% & 766 & 879 & 482 & 40{,}252 & 24{,}318 & 9{,}815 & 92 & 36\% \\
HHS        & 79 & 601{,}023 & 939{,}404 & 64\% & 37 & 135 & 232 & 25{,}357 & 2{,}252 & 1{,}371 & 104 & 65\% \\
BLM        & 43 & 539{,}683 & 843{,}022 & 64\% & 0 & 43 & 116 & 8{,}289 & 10{,}992 & 665 & 69 & 30\% \\
EPA        & 5{,}268 & 537{,}389 & 585{,}966 & 92\% & 9{,}946 & 5{,}548 & 5{,}255 & 171{,}029 & 78{,}307 & 64{,}463 & 22 & 28\% \\
NOAA       & 1{,}032 & 399{,}645 & 457{,}173 & 87\% & 506 & 1{,}074 & 2{,}818 & 83{,}429 & 37{,}525 & 7{,}053 & 33 & 32\% \\
DOI        & 44 & 388{,}196 & 850{,}023 & 46\% & 84 & 38 & 53 & 1{,}206 & 2{,}791 & 332 & 87 & 65\% \\
FS         & 104 & 353{,}209 & 583{,}657 & 61\% & 1{,}537 & 46 & 47 & 5{,}687 & 1{,}625 & 1{,}101 & 34 & 46\% \\
APHIS      & 469 & 324{,}064 & 399{,}933 & 81\% & 818 & 99 & 115 & 35{,}132 & 11{,}509 & 4{,}861 & 39 & 35\% \\
CDC        & 750 & 264{,}017 & 296{,}643 & 89\% & 1{,}272 & 26 & 45 & 2{,}960 & 3{,}741 & 1{,}844 & 27 & 41\% \\
USCIS      & 188 & 232{,}808 & 299{,}138 & 78\% & 877 & 44 & 104 & 66{,}710 & 27{,}192 & 1{,}371 & 35 & 58\% \\
ICE        & 33 & 231{,}125 & 153{,}310 & 151\% & 174 & 9 & 14 & 5{,}624 & 4{,}017 & 432 & 6 & 60\% \\
DOT        & 442 & 190{,}209 & 207{,}433 & 92\% & 623 & 85 & 81 & 16{,}481 & 9{,}197 & 1{,}409 & 21 & 40\% \\
AMS        & 326 & 180{,}889 & 299{,}749 & 60\% & 411 & 420 & 404 & 22{,}999 & 18{,}949 & 3{,}093 & 63 & 46\% \\
FAA        & 8{,}056 & 144{,}673 & 212{,}593 & 68\% & 3{,}047 & 4{,}848 & 6{,}944 & 93{,}881 & 54{,}692 & 22{,}454 & 22 & 28\% \\
IRS        & 415 & 133{,}532 & 160{,}556 & 83\% & 180 & 514 & 473 & 17{,}866 & 14{,}040 & 3{,}637 & 29 & 33\% \\
OSHA       & 167 & 129{,}483 & 171{,}346 & 76\% & 1{,}124 & 100 & 103 & 21{,}573 & 6{,}029 & 1{,}573 & 14 & 50\% \\
FSIS       & 170 & 84{,}404 & 167{,}744 & 50\% & 293 & 41 & 54 & 7{,}205 & 5{,}402 & 1{,}023 & 29 & 48\% \\
NHTSA      & 454 & 76{,}075 & 137{,}946 & 55\% & 1{,}126 & 132 & 127 & 11{,}412 & 2{,}900 & 7{,}339 & 35 & 37\% \\
EOIR       & 23 & 75{,}777 & 100{,}887 & 75\% & 0 & 18 & 46 & 13{,}459 & 4{,}392 & 182 & 30 & 57\% \\
DOJ        & 68 & 60{,}959 & 64{,}838 & 94\% & 28 & 48 & 99 & 10{,}941 & 293 & 386 & 8 & 53\% \\
NIST       & 28 & 31{,}024 & 72{,}271 & 43\% & 38 & 6 & 11 & 2{,}459 & 215 & 321 & 17 & 58\% \\
DHS        & 171 & 29{,}886 & 38{,}216 & 78\% & 598 & 60 & 119 & 3{,}311 & 9{,}407 & 209 & 2 & 81\% \\
DOE        & 258 & 28{,}208 & 30{,}452 & 93\% & 2{,}369 & 77 & 130 & 5{,}221 & 3{,}612 & 1{,}231 & 2 & 45\% \\
USDA       & 33 & 16{,}427 & 18{,}635 & 88\% & 1{,}631 & 24 & 66 & 3{,}548 & 1{,}381 & 526 & 3 & 57\% \\
TSA        & 28 & 11{,}869 & 11{,}943 & 99\% & 345 & 13 & 33 & 1{,}022 & 741 & 204 & 0 & 27\% \\
USACE      & 39 & 11{,}754 & 12{,}198 & 96\% & 150 & 57 & 60 & 4{,}775 & 4{,}038 & 180 & 0 & 59\% \\
TREASURY   & 140 & 10{,}407 & 10{,}591 & 98\% & 137 & 67 & 174 & 2{,}283 & 1{,}220 & 444 & 0 & 44\% \\
CBP        & 83 & 8{,}667 & 4{,}838 & 179\% & 1{,}383 & 40 & 241 & 23{,}115 & 3{,}177 & 1{,}406 & 0 & 63\% \\
DOL        & 32 & 5{,}739 & 6{,}573 & 87\% & 1{,}751 & 36 & 64 & 2{,}148 & 219 & 5{,}993 & 2 & 67\% \\
DOC        & 21 & 3{,}837 & 4{,}045 & 95\% & 16 & 21 & 51 & 1{,}966 & 974 & 947 & 1 & 87\% \\
NTIA       & 19 & 3{,}644 & 3{,}855 & 95\% & 18 & 2 & 11 & 152 & 0 & 98 & 0 & --- \\
DOD        & 265 & 3{,}548 & 3{,}658 & 97\% & 1{,}912 & 81 & 245 & 3{,}108 & 3{,}135 & 745 & 0 & 55\% \\
FEMA       & 125 & 3{,}343 & 3{,}483 & 96\% & 4{,}697 & 33 & 174 & 3{,}548 & 1{,}903 & 436 & 0 & 45\% \\
MSHA       & 83 & 840 & 869 & 97\% & 585 & 52 & 27 & 6{,}874 & 3{,}869 & 537 & 0 & 60\% \\
NIH        & 4 & 18 & 18 & 100\% & 3 & 6 & 11 & 8 & 0 & 13 & 0 & --- \\
NSF        & 0 & 0 & 0 & n/a & 1{,}933 & 2 & 11 & 0 & 0 & 0 & 0 & --- \\
\midrule
\textbf{Subtotal (regs.gov)} & \textbf{25{,}219} & \textbf{8{,}269{,}647} & \textbf{11{,}698{,}149} & \textbf{71\%} & \textbf{53{,}765} & \textbf{15{,}782} & \textbf{20{,}510} & \textbf{1{,}024{,}001} & \textbf{494{,}487} & \textbf{205{,}646} & \textbf{1{,}517} & \textbf{---} \\
\midrule
\multicolumn{13}{l}{\textit{External-portal agencies (notices/rules counted from regulations.gov; comments from agency-specific portals; matching pipelines not yet run on these comment sources):}} \\
FCC        & 4{,}732 & 30{,}608{,}766 & 30{,}608{,}766 & $\dagger$ & 2{,}936 & 889 & 1{,}121 & --- & --- & --- & --- & --- \\
SEC        & 120 & 34{,}761 & 34{,}761 & $\dagger$ & 24{,}506 & 224 & 351 & --- & --- & --- & --- & --- \\
FDIC       & 202 & 5{,}784 & 5{,}784 & $\dagger$ & 1{,}117 & 146 & 173 & --- & --- & --- & --- & --- \\
FEC        & 92 & 1{,}921 & 1{,}921 & $\dagger$ & 589 & 49 & 34 & --- & --- & --- & --- & --- \\
USPS       & 0 & 0 & 0 & $\dagger$ & 4{,}092 & 108 & 211 & --- & --- & --- & --- & --- \\
\midrule
\textbf{Grand total} & \textbf{30{,}365} & \textbf{38{,}920{,}879} & \textbf{42{,}349{,}381} & \textbf{---} & \textbf{87{,}005} & \textbf{17{,}198} & \textbf{22{,}400} & \textbf{1{,}024{,}001+} & \textbf{494{,}487+} & \textbf{205{,}646+} & \textbf{1{,}517+} & \textbf{---} \\
\bottomrule
\end{tabular}%
}
\vspace{2pt}
\caption{Per-agency aggregate counts for the 44 federal agencies in DOCKET. \textbf{Dockets} = distinct rulemaking dockets containing at least one comment. \textbf{Comm.\ (analyzable)} = comments in our corpus that survive the full text-extraction and feature-extraction pipeline (i.e., have non-empty extracted text on which we run claim-extraction and downstream analysis); the raw scrape from regulations.gov exactly matches the FedReg aggregate for every agency we checked, so the gap below 100\% reflects \emph{stub} or attachment-only records (no OCR-able body text), not missing scrapes. \textbf{Comm.\ (FedReg)} = comment count the agency officially reports to the Federal Register, also equal to the count of records in our raw bulk-download dump. \textbf{Cov.} = analyzable / FedReg; values $>$100\% (ICE, CBP) reflect agencies where the FedReg aggregate undercounts what regulations.gov actually exposes via bulk download. \textbf{Notices / Prop.\ / Final} = distinct regulations.gov documents of each type for all 44 agencies, including the externally-comment-sourced FCC/SEC/FDIC/FEC/USPS; the document totals shown in Figure~\ref{fig:agency-chart} are Notices+Prop.+Final summed (e.g., DHS: $598+60+119=777$). The three match columns are abbreviated by initial: \textbf{(c$\to$p)} = comment $\to$ proposal, \textbf{(f$\to$c)} = final rule $\to$ comment, \textbf{(r$\to$c)} = response $\to$ comment; all three are validated by the calibrated gpt-5-mini judge with opposing-sides framing. \textbf{Clust.\ $>$100} = MinHash-deduplicated comment clusters whose constituent comment count exceeds 100 (high-volume / likely-organized-campaign clusters; see Appendix~\ref{app:campaigns}). The ``+'' on grand-total match counts indicates the regulations.gov subtotal: matching has not yet been run on the externally-scraped comment streams. ``$\dagger$'' marks external-portal agencies, for which the Federal Register publishes no aggregate comment count. \textbf{Match} = share of comment clusters linked to at least one agency response, estimated from a uniform random sample of 1K candidate pairs per agency judged by gpt-5-mini (the match rate shown in Figure~\ref{fig:agency-chart}); ``---'' where the agency was not sampled.}
\label{tab:per_agency}
\end{table}
\clearpage

\subsection{Per-source scraping pipelines}
\label{app:scrapers}

Although regulations.gov is the primary federal portal for public comments, several large-volume agencies maintain their own systems with idiosyncratic formats. We built 15 custom scraping pipelines to bring these into a unified schema. Each pipeline normalizes the source-specific output into the same \texttt{(docket\_id, comment\_id, body\_text, attachment\_paths, posted\_date, agency)} record format used downstream by our parsing and matching code.

Table~\ref{tab:external_sources} lists all 16 data sources. We distinguish two categories: \emph{comment systems}, which expose individual public comments with per-comment metadata (commenter, date, body text); and \emph{document portals}, which expose project-level documents, notices, or bundled comment archives but not individual per-comment records.

\begin{table}[!htbp]
\centering
\small
\caption{Data sources in DOCKET. ``Comment system'' sources expose individual public comments; ``document portal'' sources expose project-level documents or bundled comment archives.}
\label{tab:external_sources}
\setlength{\tabcolsep}{4pt}
\begin{tabular}{lllp{5.5cm}}
\toprule
\textbf{Source} & \textbf{Type} & \textbf{Access} & \textbf{URL} \\
\midrule
Regulations.gov & Comment system & Bulk download & \url{regulations.gov/bulkdownload} \\
FCC ECFS & Comment system & API + bulk & \url{fcc.gov/ecfs} \\
SEC & Comment system & HTML scrape & \url{sec.gov/comments} \\
FERC eLibrary & Comment system & Playwright & \url{elibrary.ferc.gov} \\
Federal Reserve & Comment system & Camoufox & \url{federalreserve.gov/apps/foia/proposedregs.aspx} \\
FDIC & Comment system & HTML scrape & \url{fdic.gov/federal-register-publications} \\
FEC FOSERS & Comment system & HTML scrape & \url{docquery.fec.gov} \\
CFTC & Comment system & Camoufox & \url{comments.cftc.gov} \\
PCAOB & Comment system & HTML scrape & \url{pcaobus.org/about/rules-rulemaking} \\
CMS NCD & Comment system & Camoufox & \url{cms.gov/medicare/coverage} \\
FS CARA & Comment system & HTML scrape & \url{cara.fs2c.usda.gov} \\
NPS PEPC & Comment system & HTML scrape & \url{parkplanning.nps.gov} \\
\midrule
BLM ePlanning & Document portal & HTML scrape & \url{eplanning.blm.gov} \\
NRC ADAMS & Document portal & API & \url{adams.nrc.gov} \\
USACE Districts & Document portal & HTML scrape & \url{usace.army.mil} (per-district) \\
USITC EDIS & Document portal & HTML scrape & \url{edis.usitc.gov} \\
\bottomrule
\end{tabular}
\end{table}
\paragraph{Regulations.gov.} The default federal portal, covering 30+ agencies. Although it has a public API, the API is rate-limited and incomplete; we instead use the bulk-download endpoint via a Playwright browser session that prompts the user to manually solve a captcha when challenged.

\paragraph{FCC ECFS.} The largest external system, with 30.6M comments. We use both the FCC's bulk-download archive and an hourly-poll API mirror (\texttt{publicapi.fcc.gov}) to track comments during high-volume proceedings. The API imposes a 10K offset cap per query; we bypass this by sharding on hourly \texttt{date\_submission} windows and adaptively splitting to 15-minute intervals when a shard exceeds the cap. For the 2017 Restoring Internet Freedom proceeding (Docket~17-108) we additionally use the curated corpus released by \citet{handannader2023fake} as ground truth for our matching-pipeline validation (Section~\ref{sec:dataset}).

\paragraph{SEC.} Comment letters on proposed rules are posted with both inline text and PDF attachments. SEC comment letters are unusually long-form (median $\sim$5 pages) and frequently include legal-counsel signatures.

\paragraph{FERC eLibrary.} The system requires session-based browsing; we use Playwright to paginate through each docket's filings and download the associated PDF attachments.

\paragraph{Federal Reserve.} The site uses a session-stateful Blazor SPA with SignalR-based pagination; our scraper uses Camoufox (an anti-detection browser) to retrieve per-rule comment listings and submitter metadata.

\paragraph{FDIC, FEC, CFTC, PCAOB, CMS.} Each of these agencies publishes comments on proposed rules through agency-specific web portals. We scrape each via purpose-built pipelines: FDIC and FEC use simple HTML directory traversal; CFTC and CMS require Camoufox to bypass Cloudflare protection; PCAOB follows a rulemaking-docket index structure.

\paragraph{FS CARA.} The Forest Service's Comment Analysis and Response Application exposes individual per-letter data. We scrape project pages and resolve per-letter download links from detail pages.

\paragraph{NPS PEPC.} The National Park Service's Planning, Environment, and Public Comment system accepts individual public comments on NPS NEPA projects and makes them publicly available alongside planning documents.

\paragraph{Document portals (BLM, NRC, USACE, USITC).} These four systems expose project-level documents and notices rather than individual comments. BLM ePlanning hosts environmental assessments and associated comment letters; NRC ADAMS is a keyword-searchable document database; USACE operates per-district public-notice pages; USITC EDIS manages investigation documents. Each is scraped via a small Playwright pipeline that follows the agency's docket-listing pattern.

\paragraph{Aggregate-count scrapers.} For agency-level coverage validation (Figure~\ref{fig:agency-chart}), we scrape per-agency, per-year aggregate counts of received comments and published documents from the Federal Register (\url{https://www.federalregister.gov}) and from agency-specific transparency pages where available. These aggregate counts are what we use to estimate that we are collecting $\geq 95\%$ of available documents per agency.

\subsection{Comment-campaign detection and deduplication}
\label{app:campaigns}

A defining feature of the federal public-comment record is the \emph{mass-comment campaign}: an organized effort that mobilizes large numbers of citizens to submit substantively identical (or near-identical) comment text on a single rulemaking. Mass campaigns can dominate raw comment counts to the point where docket-level statistics become essentially uninformative without deduplication. In our corpus, the largest single campaign --- on Department of the Interior docket DOI-2017-0002 (the 2017 ``National Monuments Review'') --- contains approximately \textbf{98{,}000} substantively identical comment submissions in one cluster, and the docket as a whole has multiple clusters in the 25K--100K range. The 2017 FCC ``Restoring Internet Freedom'' net-neutrality docket (handled through FCC ECFS rather than regulations.gov, so not in the per-agency table above) attracted on the order of \textbf{25 million} submissions, the majority of which were attributable to organized campaigns and at least some of which were later identified as automated or fraudulently attributed~\citep{handannader2023fake,balla2022responding}.

\paragraph{Deduplication via MinHash clustering.}
We collapse near-duplicate comments into \emph{clusters} using MinHash similarity~\citep{broder1997resemblance}. Two comments are placed in the same cluster if their MinHash signatures (computed over shingles of the comment text) agree above a Jaccard-similarity threshold tuned to capture standard form-letter campaigns (paraphrase-stable to small surface edits like greeting personalization or postal-address insertion, but distinct on substantive content changes). Each cluster receives a stable \texttt{cluster\_uid} formed as \texttt{<docket\_id>::<cluster\_index>}. All claim extraction, claim-to-provision matching, and downstream metric computation operate at the cluster level rather than the raw-comment level, so a 50{,}000-comment campaign contributes one cluster (and one set of extracted claims) rather than 50{,}000 redundant signal copies.

\paragraph{Cluster-size distribution.}
Across the 10.79M raw regulations.gov comments in our deduplication snapshot (an earlier pull than Table~\ref{tab:per_agency}, which counts 11.7M), MinHash deduplication yields \textbf{7.94M distinct clusters}. The distribution of cluster sizes is highly skewed (Table~\ref{tab:cluster_size_distribution}): \textbf{98.7\% of clusters are singletons} (a single submission with no near-duplicate elsewhere in the corpus), but the 424 clusters with $\geq$1{,}000 members account for 20.3\% of all raw comments. The median cluster size is 1; the mean is 1.4; the maximum (DOI-2017-0002) is 98{,}422. Per-agency counts of high-volume clusters ($>$100 members) appear in the right column of Table~\ref{tab:per_agency} above; CMS, FDA, ED, HHS, and FWS lead, reflecting agencies that regulate health, education, and environmental issues (which tend to attract organized advocacy campaigns).

\begin{table}[!htbp]
\centering
\small
\setlength{\tabcolsep}{8pt}
\renewcommand{\arraystretch}{1.1}
\begin{tabular}{l r r r r}
\toprule
\textbf{Cluster size band} & \textbf{Clusters} & \textbf{\% of clusters} & \textbf{Raw comments} & \textbf{\% of raw} \\
\midrule
1 (singleton)        & 7{,}830{,}939 & 98.7\% & 7{,}830{,}939 & 72.6\% \\
2--5                 &     94{,}235  &  1.2\% &    223{,}950  &  2.1\% \\
6--10                &      5{,}577  &  0.1\% &     40{,}953  &  0.4\% \\
11--100              &      4{,}399  &  0.1\% &    124{,}372  &  1.2\% \\
101--1{,}000         &      1{,}093  & 0.01\% &    379{,}823  &  3.5\% \\
1{,}001--10{,}000    &        376    & 0.00\% & 1{,}158{,}354  & 10.7\% \\
10{,}001--100{,}000  &         48    & 0.00\% & 1{,}034{,}779  &  9.6\% \\
\midrule
\textbf{Total}       & \textbf{7{,}936{,}667} & 100\% & \textbf{10{,}793{,}170} & 100\% \\
\bottomrule
\end{tabular}
\vspace{2pt}
\caption{Distribution of MinHash cluster sizes across the 10.79M raw regulations.gov comments in the deduplication snapshot. The corpus is dominated by singletons (one-off submissions, 72.6\% of raw volume), but the largest 424 clusters ($\geq$1{,}000 members each) account for 20.3\% of raw comments. Without deduplication, single mass campaigns would swamp many docket-level statistics.}
\label{tab:cluster_size_distribution}
\end{table}
\paragraph{Why deduplication matters for our metrics.}
All four benchmark axes operate on cluster-level real claims rather than raw comment text. This has two consequences. First, signal weighting: if 20{,}000 commenters submit the identical critique of a provision, that critique contributes one matched (claim, provision) row, not 20{,}000 --- so a model that paraphrases that single shared critique gets equal credit regardless of campaign size. Second, evaluation feasibility: the LLM-judge cost of axis B is proportional to the number of distinct real claims; without deduplication, a single net-neutrality-scale campaign would dominate the budget and crowd out signal from smaller dockets. The 558-docket benchmark population deliberately excludes dockets where one or two campaigns account for the bulk of cluster mass (we filter to $\geq$3 distinct validated real claims per docket, see Section~\ref{sec:benchmark_creation}); the corpus-wide cluster table in this section is pre-filter, intended to characterize the raw-data terrain.

\paragraph{Limitations of MinHash deduplication.}
MinHash is robust to character-level perturbations (whitespace, capitalization, single-token edits) but defeated by deeper paraphrase. Modern advocacy campaigns increasingly employ template engines that vary sentence ordering or substitute synonyms across submissions, which MinHash treats as distinct clusters. In such cases the corpus over-reports unique-cluster counts and under-reports campaign size. We do not attempt to detect such deeper paraphrase systematically; doing so robustly is an open research problem~\citep{balla2022responding,handannader2023fake}. Similarly, our pipeline does not flag comments that may have been generated by language models (a documented and growing concern in federal rulemaking~\citep{balla2022responding,hr7528}).

\subsection{Parsing examples: proposal $\rightarrow$ provisions, comment $\rightarrow$ claims, response section $\rightarrow$ responses}
\label{app:parsing_examples}

To make our extraction pipeline concrete, this subsection shows three real worked examples drawn from the corpus.

\paragraph{Proposal $\rightarrow$ provisions.} Proposed rules contain regulatory text that establishes obligations, permissions, prohibitions, conditions, and exceptions. We extract these \emph{provisions} using gpt-5-mini (Appendix~\ref{app:prompts}). Table~\ref{tab:proposal_to_units} shows two short proposals and their extracted units, illustrating the range of provisions and how a single regulatory section can contain multiple distinct units.

\begin{table}[!htbp]
\centering
\footnotesize
\noindent\fbox{\parbox{0.96\textwidth}{\textbf{Proposal 1: COE-2021-0006} --- \textit{Restricted Area: Eagle River, Richardson Training Area, AK}

\vspace{2pt}\scriptsize
\texttt{\S~334.1305~~Eagle River from Bravo Bridge to its mouth at Eagle Bay in Knik Arm, Richardson Training Area on Joint Base Elmendorf-Richardson, Alaska; restricted area.}

\texttt{(a) Restricted area.} The restricted area consists of navigable waters within an area defined as beginning a point on shore at latitude 61\textdegree19'40.1'' N, longitude 149\textdegree44'20.336'' W; thence easterly\ldots\ 3.06 nautical miles southerly along the river\ldots

\texttt{(b) The regulation.} The restricted area is permanently closed for public use at all times. No persons, watercraft, or vessels shall enter, or remain, in the area except for those authorized by the enforcing agency.

\texttt{(c) Enforcement.} This regulation will be enforced by the Commander, United States Army--Alaska.
}}

\vspace{4pt}

\noindent\fbox{\parbox{0.96\textwidth}{\textbf{Proposal 2: NOAA-NMFS-2023-0044} --- \textit{Reef Fish Resources of the Gulf of Mexico; Commercial Trip Limit for Gray Triggerfish}

\vspace{2pt}\scriptsize
\texttt{\S~622.43~~Commercial trip limits.}

\texttt{(b)} Gray triggerfish---25 fish. The commercial trip limit applies until the commercial quota specified in Sec.~622.39(a)(1)(vi) is reached, which is equal to the commercial ACT. See Sec.~622.39(b) for the limitations regarding gray triggerfish after the commercial quota is reached.
}}

\vspace{6pt}

\setlength{\tabcolsep}{3pt}
\renewcommand{\arraystretch}{1.0}
\begin{tabular}{p{2.2cm} c p{1.3cm} p{7.5cm}}
\toprule
\textbf{Proposal} & \textbf{Unit} & \textbf{Type} & \textbf{Extracted provision text} \\
\midrule
\multirow{4}{2.2cm}{\raggedright Proposal 1}
& 1 & \cellcolor[HTML]{FDEAE6}prohibition & No persons, watercraft, or vessels shall enter, or remain, in the area. \\
& 2 & \cellcolor[HTML]{FDEAE6}prohibition & The restricted area is permanently closed for public use at all times. \\
& 3 & \cellcolor[HTML]{E8F5E9}exception & Except for those authorized by the enforcing agency. \\
& 4 & \cellcolor[HTML]{E3F2FD}obligation & This regulation will be enforced by the Commander, United States Army--Alaska. \\
\midrule
\multirow{3}{2.2cm}{\raggedright Proposal 2}
& 1 & \cellcolor[HTML]{E3F2FD}obligation & Gray triggerfish---25 fish [commercial trip limit]. \\
& 2 & \cellcolor[HTML]{FFF3E0}condition & The commercial trip limit applies until the commercial quota specified in \S~622.39(a)(1)(vi) is reached. \\
& 3 & \cellcolor[HTML]{E8F5E9}exception & See \S~622.39(b) for the limitations regarding gray triggerfish after the commercial quota is reached. \\
\bottomrule
\end{tabular}
\vspace{2pt}
\caption{Two proposed rules shown in raw regulatory text (top) and their extracted provisions (bottom). Units are classified by type: \colorbox[HTML]{E3F2FD}{obligation}, \colorbox[HTML]{FDEAE6}{prohibition}, \colorbox[HTML]{FFF3E0}{condition}, \colorbox[HTML]{E8F5E9}{exception}.}
\label{tab:proposal_to_units}
\end{table}
\FloatBarrier

To make the claim- and response-extraction pipeline concrete, this subsection shows two real worked examples drawn from the corpus. Each example takes a single underlying piece of text (one public comment in Table~\ref{tab:comment_to_claims}; one final-rule preamble section in Table~\ref{tab:response_to_responses}) and shows the structured rows the LLM extractor produced from it.

\paragraph{Comment $\rightarrow$ claims.} Public comments are typically multi-paragraph documents that bundle several distinct positions. We use Llama-3.3-70B (Appendix~\ref{app:prompts}) to break each comment into its constituent \emph{claims}: short, self-contained statements that each address one specific provision. The LLM is instructed to (a) restate the claim with provision-level specificity, (b) avoid pronouns and references to surrounding paragraphs, and (c) produce one claim per discrete point. Table~\ref{tab:comment_to_claims} shows the five claims the extractor produced from a single ACLU comment on the Department of Homeland Security's 2019 ALL-038 ``Insider Threat Program System of Records'' Privacy Act exemption rule.

\begin{table}[!htbp]
\centering
\scriptsize

\noindent\fbox{\parbox{0.96\textwidth}{\textbf{Raw comment (DHS-2019-0034, cluster 4):} \textit{``The ACLU opposes DHS's proposed rule to exempt the ALL-038 Insider Threat Program System of Records from certain provisions of the Privacy Act of 1974. The exemptions risk violating the Fourth Amendment protections of DHS employees. Specifically, exempting the system from 5 U.S.C.\ 552a(c)(3) and (c)(4) could prevent the release of accounting for disclosures\ldots\ Exempting from 552a(d) could prevent individuals from accessing or amending their records\ldots\ Exempting from 552a(e)(1) could compromise the accuracy of investigative information\ldots\ This rule would further expand DHS's disregard for the constitutional rights of its employees.''}}}
\vspace{4pt}

\setlength{\tabcolsep}{4pt}
\renewcommand{\arraystretch}{1.05}
\begin{tabular}{c p{11.5cm}}
\toprule
\textbf{Claim} & \textbf{Extracted claim text} \\
\midrule
\texttt{0} & The proposed rule to exempt the ALL-038 Insider Threat Program System of Records from certain Privacy Act provisions risks violating the Fourth Amendment. \\
\texttt{1} & Exempting the system from 5 U.S.C.\ 552a(c)(3)/(c)(4) could prevent release of accounting for disclosures and alert individuals wrongfully targeted. \\
\texttt{2} & Exempting from 5 U.S.C.\ 552a(d) could prevent access or amendment to records, concealing investigations into misconduct by DHS employees. \\
\texttt{3} & Exempting from 5 U.S.C.\ 552a(e)(1) could compromise accuracy of information obtained during investigations into potential violations of federal law. \\
\texttt{4} & The proposed exemption would further expand DHS's disregard for the constitutional rights of its employees. \\
\bottomrule
\end{tabular}
\vspace{1pt}
\caption{One public comment on the DHS Insider-Threat Privacy Act exemption, shown in raw form (top) and broken into five self-contained claims by the claim extractor (bottom).}
\label{tab:comment_to_claims}
\end{table}
\FloatBarrier

\paragraph{Response section $\rightarrow$ responses.} Final-rule preambles bundle the agency's discussion of every comment it received into a few long ``response-to-comments'' sections. We use Llama-3.3-70B (Appendix~\ref{app:prompts}) to break each response section into discrete (\emph{comment-summary}, \emph{agency-response}) pairs. Each row is the agency's discussion of one issue raised by a commenter, separated from the surrounding discussion of other issues. Table~\ref{tab:response_to_responses} shows five of the eight rows extracted from EPA's response to comments on the 2017 Renewable Fuel Standard ``Point of Obligation'' rulemaking (EPA-HQ-OAR-2016-0544-0449).

\begin{table}[!htbp]
\centering
\scriptsize

\noindent\fbox{\parbox{0.96\textwidth}{\textbf{Raw preamble (EPA-HQ-OAR-2016-0544-0449):} \textit{``EPA received several petitions requesting that the Agency change the `point of obligation' under the RFS program from refiners and importers to downstream parties such as fuel blenders or position holders. Petitioners argued this change would improve program effectiveness\ldots\ EPA denies the requests\ldots''}}}
\vspace{3pt}

\tiny
\setlength{\tabcolsep}{3pt}
\renewcommand{\arraystretch}{1.0}
\begin{tabular}{p{5.5cm} p{6.5cm}}
\toprule
\textbf{Comment summary (extracted)} & \textbf{Agency response (extracted)} \\
\midrule
Petitioners asked EPA to shift the point of obligation downstream, arguing it would improve the program. & EPA denies, finding petitioners did not demonstrate improved effectiveness. \\
Commenters favored designating position holders as obligated parties. & EPA rejects: position holders are not refiners/importers/blenders, not aligned with statutory authority. \\
Commenters alleged EPA must annually reconsider the obligated parties. & EPA disagrees the statute requires annual reconsideration. \\
Commenters supported designating position holders as obligated parties. & EPA believes this is not well aligned with the authority to place the obligation on ``refineries, importers, or blenders.'' \\
\bottomrule
\end{tabular}
\vspace{1pt}
\caption{One final-rule response section (EPA 2017 RFS), shown in raw form (top) and broken into four (comment-summary, agency-response) pairs by the response extractor (bottom).}
\label{tab:response_to_responses}
\end{table}
\FloatBarrier

\subsection{Comment counts by stakeholder subgroup}
\label{app:subgroup_counts}

\begin{figure}[h]
\centering
\includegraphics[width=0.78\textwidth]{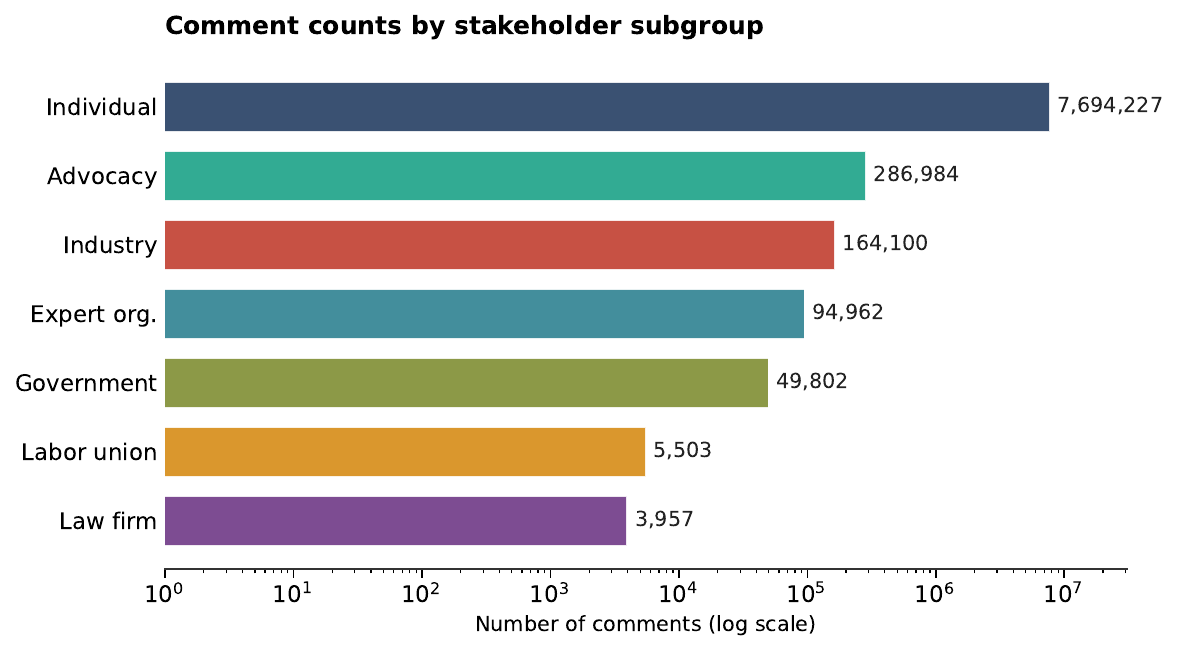}
\caption{Comment counts per stakeholder subgroup across the full DOCKET corpus (8.27M comments). Log-scaled x-axis. Subgroups are derived from \texttt{org\_type} via the seven-bucket taxonomy.}
\label{fig:subgroup_counts}
\end{figure}

Figure~\ref{fig:subgroup_counts} shows the per-subgroup comment counts across the full DOCKET corpus, after mapping the raw \texttt{org\_type} field through our seven-bucket taxonomy (Section~\ref{sec:stakeholder_diverge}). Counts span more than three orders of magnitude: individual commenters dominate raw volume ($\sim$7.7M comments), followed by advocacy NGOs ($\sim$287K), industry ($\sim$164K), expert organizations ($\sim$95K), government ($\sim$50K), and labor unions and law firms in the low thousands ($\sim$5.5K and $\sim$4K respectively). Per-subgroup absolute volume is highly skewed even after de-duplicating form-letter campaigns.

\subsection{Concern taxonomy: construction and examples}
\label{app:concern_taxonomy}

Table~\ref{tab:stakeholder_concern} reports the distribution of primary concern frames across stakeholder groups. The 12-frame taxonomy was developed iteratively through a combination of top-down and bottom-up analysis:

\paragraph{Taxonomy construction.} We began with broad categories grounded in the administrative law literature on public commenting~\citep{yackee2006bias,carpenter2024inequality}: economic, legal, and procedural concerns are canonical axes of regulatory debate. We augmented these with domain-specific categories surfaced by manual inspection of 500 randomly sampled claims across 20 agencies: health, safety, environmental, scientific/technical, and equity/justice emerged as recurrent and distinctive frames that cut across agencies. Religious/moral and national security were added to capture low-frequency but substantively distinct argument types. The ``other'' and ``unknown'' categories absorb residual and ambiguous cases; three low-frequency categories (religious/moral, national security, unknown) are collapsed into ``Other'' in the main-text table.

\paragraph{Mutual exclusivity and overlap.} The taxonomy assigns exactly one primary frame per claim via forced single-label classification. We acknowledge that real claims often invoke multiple frames simultaneously (e.g., a safety concern grounded in scientific evidence, or an economic argument with equity implications). The forced single-label design trades granularity for interpretability: it produces clean distributional comparisons across stakeholder groups (Table~\ref{tab:stakeholder_concern}) at the cost of suppressing secondary frames. The LLM classifier is instructed to select the \textit{dominant} frame---the single framing that best characterizes the claim's core argument.

\paragraph{Example claims by concern frame.} Table~\ref{tab:concern_examples} provides illustrative examples of real public-comment claims classified under each frame.

\begin{table}[h]
\centering
\small
\caption{Example real public-comment claims for each concern frame in our taxonomy. Each claim is the dominant-frame classification assigned by \texttt{gpt-5-mini}; many claims contain secondary frames not reflected in the single-label assignment.}
\label{tab:concern_examples}
\setlength{\tabcolsep}{3pt}
\renewcommand{\arraystretch}{1.3}
\begin{tabular}{p{2.2cm}p{10.5cm}}
\toprule
\textbf{Frame} & \textbf{Example claim} \\
\midrule
Health & ``Restricting plan changes to once per year for dual-eligible, LIS-eligible, and other at-risk beneficiaries will lead to improved health outcomes, such as better medication adherence and completion of preventative screenings.'' \\
Safety & ``The FDA's proposed ban on electrical stimulation devices used to treat self-injurious or aggressive behavior is necessary due to the device's unreasonable risk of serious injury to disturbed, disabled, and vulnerable people.'' \\
Economic & ``The proposed rule requires hospitals to report median payer-specific negotiated charges by MS-DRG, which will require significant additional effort, with an estimated burden at least 10 times greater than CMS' estimate.'' \\
Legal & ``FDA's proposed rule would override existing protections for small-scale farmers by requiring `first receivers' to keep detailed records on products that are exempt from traceability requirements under FSMA.'' \\
Scientific & ``The use of non-FTIR spectroscopy methods, such as NCASI methods 98.01 or 105.01, is requested as an alternative to the specified method for formaldehyde emissions measurement.'' \\
Environmental & ``An extension of the implementation deadline would allow over 1 million tons of additional carbon dioxide to be released.'' \\
Equity & ``The proposed fee increases are designed to target and increase fear among low-income people of color and force immigrant communities to live without protection from deportation.'' \\
Religious & ``The proposed regulation allows religious organizations to require attendance at religious activities as a condition for receiving lifesaving or sustaining federal services.'' \\
Nat'l Security & ``Relaxing I-9 documentation requirements could facilitate the employment of non-citizens or individuals with criminal records, detrimental to national security.'' \\
Procedural & ``The recommended two-year sell-through period for affected new hydronic heaters should be extended indefinitely.'' \\
\bottomrule
\end{tabular}
\end{table}
\subsection{Comment language uptake in final rules}
\label{app:uptake}

\begin{figure}[ht]
\centering
\includegraphics[width=0.62\textwidth]{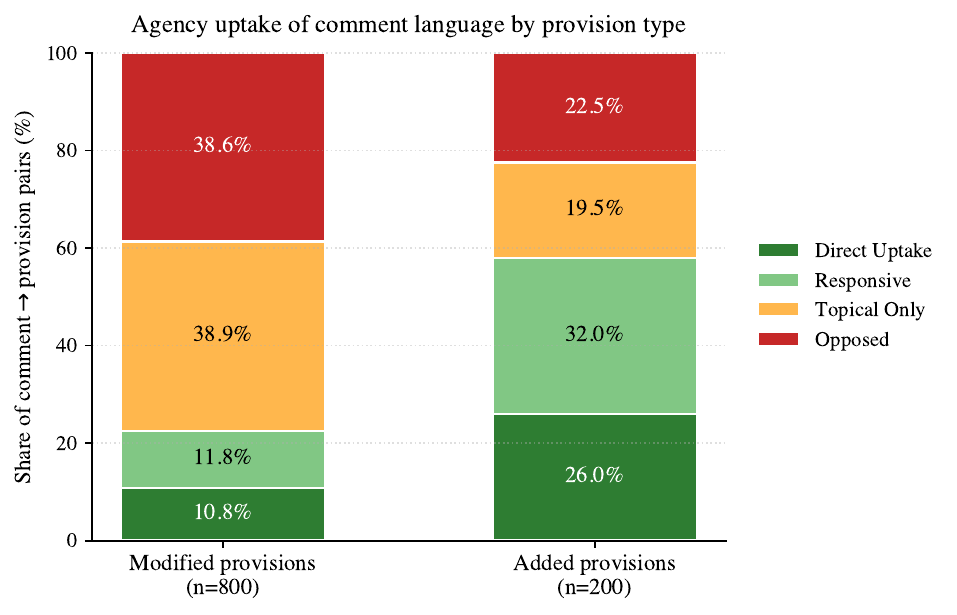}
\caption{Distribution of agency-uptake labels for 1{,}000 sampled comment$\to$provision pairs, split by whether the provision was \textit{modified} from its proposed form ($n=800$) or newly \textit{added} between proposal and final rule ($n=200$). Direct Uptake = final rule adopts specific phrasing from the comment that was absent in the proposed rule; Responsive = same concern addressed in the agency's own words; Topical Only = same topic but no evidence of influence; Opposed = agency rejected the suggestion. Added provisions show notably higher direct-uptake and responsive rates ($26.0\%$ and $32.0\%$, respectively) than modified provisions ($10.8\%$ and $11.8\%$), consistent with newly-introduced rule text tracking commenter language more closely than pre-existing language being amended.}
\label{fig:uptake_breakdown}
\end{figure}

To assess how directly public comments influence final rule \emph{language} (as opposed to policy direction), we sampled 1{,}000 (comment, proposed-rule provision, final-rule provision) tuples from our Match~3 data and classified each with \texttt{gpt-5-mini} into four categories: \textsc{Direct Uptake} (final rule adopts specific phrasing from the comment that was absent in the proposed rule), \textsc{Responsive} (final rule addresses the comment's concern but in the agency's own language), \textsc{Topical Only} (same topic, no evidence of influence), and \textsc{Opposed} (agency rejected the comment's suggestion).
Results: 13.8\% \textsc{Direct Uptake}, 15.8\% \textsc{Responsive}, 35.0\% \textsc{Topical Only}, 35.4\% \textsc{Opposed}.
We checked these labels by hand. We drew 70 cases from the classifier output (51 labeled \textsc{Direct Uptake} and 19 labeled \textsc{Responsive}, \textsc{Topical Only} or \textsc{Opposed} as distractors), shuffled them, hid the classifier label, and highlighted language shared between the comment and the final rule. The annotator confirmed 80\% of the \textsc{Direct Uptake} labels as language taken from the comment (86\% counting close paraphrases, 88\% counting substantive influence without shared language), and found language-level uptake in most of the distractor cases too. On this sample the classifier does not inflate the uptake rate.
Figure~\ref{fig:uptake_breakdown} shows the distribution split by whether the provision was modified from its proposed form or newly added after the proposal stage, and Table~\ref{tab:uptake} shows the eight strongest direct-uptake candidates from the sample.

\begin{table}[ht]
\scriptsize
\setlength{\tabcolsep}{3pt}
\begin{tabular}{@{}p{1.2cm}p{3.8cm}p{3.8cm}p{4.2cm}@{}}
\toprule
\textbf{Docket} & \textbf{Proposed Rule} & \textbf{Public Comment} & \textbf{Final Rule} \\
\midrule

\raggedright FAA-2019-0479
& Install an elbow restrictor, P/N~46610--1, in accordance with the \emph{Accomplishment Instructions} of Bombardier Service Bulletin 84--32--69, Revision~C, dated January~20, 2011.
& The FAA should \textbf{only mandate section~3.B Procedure} of Bombardier Service Bulletin 84--32--69, Revision~C, dated January~20, 2011, as it directly corrects the unsafe condition, and not the entire Accomplishment Instructions.
& Install an elbow restrictor, P/N~46610--1, in accordance with \textbf{paragraph~3.B., ``Procedure,''} of the Accomplishment Instructions of Bombardier Service Bulletin 84--32--69, Revision~C, dated January~20, 2011. \\
\addlinespace

\raggedright FAA-2021-0650
& Time of Designation.\ Intermittent, 0700--1700 local time, Monday--Friday, by NOTAM 7~days in advance.
& The proposed R--2511 restricted area would be active \textbf{less than 36~days a year} and only for 2~hours per active day, making a TFR a more suitable alternative to minimize impact on VFR traffic.
& \textbf{Activation limited to no more than 36~days per year.} \\
\addlinespace

\raggedright EPA-HQ-OAR-2018-0417
& \emph{[No transition timeline for CEDRI electronic reporting.]}
& Dow recommends a transition period of at least \textbf{1~year after the reporting template becomes available in CEDRI} or 2~years after publication of the amendments for the transition to new reporting templates.
& Beginning on April~16, 2021 or \textbf{1~year after the appropriate electronic reporting template becomes available on the CEDRI website}, whichever is later, submit all subsequent reports following the procedure specified in paragraph~(l). \\
\addlinespace

\raggedright CMS-2019-0039
& \textbf{Standardized encounter data}, no later than one~(1) business day after data concerning the encounter is received by the MA organization.
& The proposed rule's requirement for plans to make information accessible through the APIs, including adjudicated claims, \textbf{encounters with capitated providers}, and clinical data, starting on January~1, 2020, will allow patients to exercise their HIPAA right of access.
& \textbf{Encounter data from capitated providers}, no later than one~(1) business day after data concerning the encounter is received by the MA organization. \\
\addlinespace

\raggedright DOT-OST-2022-0144
& You must provide timely notifications to passengers when their wheelchairs and scooters \emph{have been loaded and stowed} in the cargo compartment and \emph{unloaded and retrieved} from the cargo compartment.
& Carriers should be required to \textbf{notify passengers when their wheelchair or scooter is not loaded on the aircraft}, but not necessarily when it is loaded or unloaded, due to potential technological restrictions and implementation costs.
& You must \textbf{notify passengers with disabilities, before the aircraft cabin door closes, whether their wheelchairs or scooters have been loaded} in the cargo compartments of their flights, including whether their device could not fit on the passenger's scheduled flight because of its size or weight. \\
\addlinespace

\raggedright ED-2015-OSERS-0132
& \emph{[No specific criteria framework for measuring significant disproportionality.]}
& States should be allowed \textbf{flexibility} to add additional criteria beyond the \textbf{risk ratio} and \textbf{minimum n~size} thresholds to avoid identifying significant disproportionality that is an artifact of calculations using small numbers.
& (A)~Reasonable \textbf{risk ratio} threshold; (B)~Reasonable minimum cell size; (C)~Reasonable minimum \textbf{n-size}; and (D)~Standard for measuring reasonable progress if a State uses the \textbf{flexibility} described in paragraph~(d)(2) of this section. \\
\addlinespace

\raggedright FAA-2023-2234
& \emph{[No requirement for manufacturer data release.]}
& The proposed AD should include a \textbf{defined date for Boeing to release all service data} to operators, such as December~31, 2024, to ensure sufficient time for modifications prior to the AD's completion due date.
& \textbf{Boeing will release all service data} to allow retrofit of hardware updates to the CFM56--7B nacelle prior to that date. \\
\addlinespace

\raggedright APHIS-2020-0068
& \emph{[No enrichment provision for birds.]}
& \textbf{Environmental enrichment} for all birds is necessary to meet the AWA's goal of ensuring humane care and treatment, and should include \textbf{species-specific} physical and psychological needs.
& The physical environment in the primary enclosures must be \textbf{enriched} by materials or activities that would provide the birds with the means to express noninjurious \textbf{species-typical} activities. \\

\bottomrule
\end{tabular}
\vspace{2pt}
\caption{Top candidates for direct comment-to-rule language uptake. For each case, the final rule contains a specific parameter, term, or formulation that appears in the public comment but was absent from the proposed rule. Boldface highlights the adopted language.}
\label{tab:uptake}
\end{table}
\subsection{Human check of the LLM labels}
\label{app:human_validation}

Every label in the pipeline comes from a model, so one of us checked a sample by hand before submission. For each of eleven steps we drew ten items, stratified across the label values the model produced, showed the annotator the same inputs the model saw together with the model\textquotesingle{}s label, and asked whether the label was right (and, if not, which was). This is a verification check, not a blind re-annotation: the annotator saw the label. Table~\ref{tab:human_validation} gives the counts. The model\textquotesingle{}s label held up on 97 of the 102 items the annotator could decide; two stakeholder-type items were marked unsure because the comment text gave no cue about who wrote it (one was ``please see the attached file''). The five disagreements were borderline rather than gross: a response that mentioned a comment without addressing it, a response the model called ``substantive'' that the annotator read as partial acceptance, a vaccine comment the model filed under ``other'' rather than health, and one provision match and one further response match on which reasonable readers differ. Nine further steps (uptake, claim $\to$ final provision, claim direction, argument type and scope, the fourteen engagement tags, legal score, and requested action) were prepared on the same page but are not yet checked.

\begin{table}[h]
\centering
\footnotesize
\setlength{\tabcolsep}{3pt}
\renewcommand{\arraystretch}{1.08}
\begin{tabular}{@{}llrrrr@{}}
\toprule
\textbf{Step} & \textbf{Where it is used} & \textbf{n} & \textbf{agree} & \textbf{wrong} & \textbf{unsure} \\
\midrule
Claim $\to$ proposed provision (Match 2) & Section~\ref{sec:dataset}, Figure~\ref{fig:links} & 10 & 9 & 1 & 0 \\
Response $\to$ comment & Figure~\ref{fig:agency-chart}, Finding 1 & 10 & 8 & 2 & 0 \\
Response stance (accepted / partial / disagree) & Finding 1 & 10 & 9 & 1 & 0 \\
Concern frame of a comment & Table~\ref{tab:stakeholder_concern} & 10 & 9 & 1 & 0 \\
Stakeholder type of a commenter & Tables~\ref{tab:stakeholder_concern}, \ref{tab:bucket_alignment} & 10 & 8 & 0 & 2 \\
Specialist vs.\ generalist & Figure~\ref{fig:engagement_specialist} & 7 & 7 & 0 & 0 \\
Proposed $\to$ final provision pairing (Match 1) & Figure~\ref{fig:links} & 10 & 10 & 0 & 0 \\
Type of modification & Appendix~\ref{app:bench_construction} & 10 & 10 & 0 & 0 \\
Claim-pair judge (same / related / different / opposing) & Tables~\ref{tab:benchmark_results}B, \ref{tab:bucket_alignment} & 7 & 7 & 0 & 0 \\
Frame and style pair judge & Table~\ref{tab:benchmark_results}(C,D) & 10 & 10 & 0 & 0 \\
Concern frame of an LLM claim & Section~\ref{sec:benchmark_results} & 10 & 10 & 0 & 0 \\
\midrule
\textbf{All} & & \textbf{104} & \textbf{97} & \textbf{5} & \textbf{2} \\
\bottomrule
\end{tabular}
\caption{Hand check of the LLM labels, one annotator, ten items per step (seven where the annotator stopped early). ``Agree'' means the model\textquotesingle{}s label was judged right; ``wrong'' that the annotator chose another label.}
\label{tab:human_validation}
\end{table}

\subsection{Coverage and limitations}
\label{app:coverage_limitations}

We document the known coverage gaps and methodological limitations of DOCKET and DOCKETBench so that downstream users can interpret results in context. The dataset is intentionally large and broad, but it is not exhaustive: every step of collection, parsing, matching, and labeling carries its own sources of error.

\paragraph{Agency-year asymmetry.} Federal commenting volume varies by orders of magnitude across agencies. Coverage is much thinner for the early years of our window (2016--2017) at several agencies, likely because those agencies lacked robust digital comment infrastructure during that window. Some agencies (e.g., MSHA, USPS, and parts of DOI) maintain very low public-comment volumes throughout the period, so per-agency analyses on those slices have limited statistical power. Where agency-year cells are sparse, we recommend treating per-agency findings as descriptive rather than inferential.

\paragraph{Comment systems outside regulations.gov.} The federal government does not have a single, uniform public-comment portal. While regulations.gov hosts comments for the majority of agencies, several large-volume agencies maintain their own systems with non-standard formats. We built 15 custom scraping pipelines to cover these sources---eleven comment systems (FCC ECFS, SEC, FERC, Federal Reserve, FDIC, FEC, CFTC, PCAOB, CMS, FS CARA, NPS PEPC) and four document portals (BLM ePlanning, NRC ADAMS, USACE Districts, USITC EDIS); see Table~\ref{tab:external_sources} in Appendix~\ref{app:scrapers}. Format heterogeneity means we occasionally lose attachments, comment metadata, or page-numbered exhibits. We estimate the loss at well under 5\% of total volume per affected agency, but the exact rate varies and is reported per-agency in the data release.

\paragraph{Comments only available through FOIA.} A small set of public comments are not posted to any web portal and can be released only in response to Freedom of Information Act (FOIA) requests. These comments tend to concentrate at agencies that exempt parts of their regulatory record from public posting (e.g., national-security-adjacent agencies, certain inspector-general filings) or where comments contain confidential business information. Because FOIA releases are slow, partial, and idiosyncratic, we exclude FOIA-only material from the main release.

\paragraph{State-level and sub-federal rulemaking is excluded.} DOCKET covers U.S. federal rulemaking only. The dataset does not include rules issued by state agencies, U.S. territories, tribal governments, county or municipal regulators, or federal-state cooperative bodies. State-level commenting is substantively different in volume, format, and stakeholder mix; users interested in state regulation should look elsewhere.

\paragraph{Form-letter and mass-campaign deduplication is approximate.} We collapse near-duplicate comments using a MinHash similarity threshold, which captures the typical mass-campaign pattern in which thousands of commenters submit identical or near-identical text. However, modern campaigns increasingly insert minor variations (paraphrase, sentence shuffling, or LLM-generated rewrites) that defeat MinHash. We report both deduplicated and raw comment counts; the deduplicated counts likely still over-represent variant campaigns relative to truly independent submissions. Robust LLM-rewrite detection is an open problem and not solved here.

\paragraph{Matching pipeline carries residual error.} Our three-stage matching cascade (bi-encoder retrieval $\rightarrow$ cross-encoder filter $\rightarrow$ LLM judge) is calibrated against hand-curated ground truth, but residual error remains. Internal audit of the proposed-to-final unit matching surfaced approximately 20\% of pairs where we re-classified the upstream label after a second LLM review. The benchmark's claim-to-provision matches use a calibrated opposing-sides judge that improved agreement with human ground truth from 67\% to 96\% on the FCC-RIF reference set, but no LLM judge is perfect, and downstream metrics inherit this residual error. We list per-stage accuracy estimates in Appendix~\ref{app:human_validation}.

\paragraph{Benchmark population is filtered.} The 558 dockets used in DOCKETBench are the entire population that survives our two filters (5--100 proposed provisions; at least three validated claim-to-provision pairs, which implies at least one commented-on provision). This excludes dockets with no commented provisions and dockets at the extreme ends of the size distribution. Conclusions therefore generalize to ``federal rules that draw at least some substantive comment'' and not to silent or extremely small/large dockets. We report the per-agency distribution of the benchmark population in Appendix~\ref{app:benchmark_descriptive}.

\paragraph{Stakeholder bucketing relies on org\_type metadata.} The seven-bucket stakeholder taxonomy (industry, labor, advocacy, expert organization, government, law firm, individual) comes from an LLM classifier (Llama-3.3-70B) that reads the comment text and any stated affiliation, mapped to seven buckets. A second pass also records who the commenter is filing on behalf of; Table~\ref{tab:bucket_alignment} uses it to count law firms as their clients and trade associations as their members (Appendix~\ref{app:bucket_alignment}). We do not validate commenter identity against outside sources. Approximately 10\% of comments cannot be cleanly assigned to a bucket and are placed in an ``other / unknown'' category that we exclude from per-bucket analyses.

\paragraph{Language and accessibility.} All extraction, matching, and benchmark generation steps assume English-language input. A small fraction of public comments are submitted in other languages or as scanned images; these are handled via the same pipeline (extraction is best-effort) but quality is lower. We do not separately evaluate per-language accuracy.

\paragraph{Temporal scope.} The dataset covers 2016--2026. Earlier rulemaking is not included. Because regulatory practice changes over time (e.g., in response to executive orders, court rulings, and changes in administration), aggregate statistics across this window mix periods with different commenting norms. We provide per-year breakdowns in the data release for users who want to restrict to a specific period.

\paragraph{LLM cost and reproducibility.} Several of our pipeline steps use commercial LLM APIs (gpt-5-mini, gpt-5, gpt-4o-mini) which can change behavior over time as providers update model snapshots. We report the model versions and dates of all calls so future replications can detect drift, but exact reproduction of LLM-judge labels may require re-running with the original model snapshots.

\section{DOCKETBench: construction, metrics, and extended analyses}
\label{app:bench}

\subsection{DOCKETBench Construction}
\label{app:bench_construction}

Here, we give more detail on the construction steps in DOCKETBench.

\paragraph{(1) Provisions.} As stated in Section~\ref{sec:dataset}, we segment every proposed and final rule into atomic \emph{provisions} — single normative statements (obligations, permissions, prohibitions, conditions, exceptions) that can be tracked across drafts. To extract units, we use gpt-5-mini at minimal reasoning effort with a 6-shot prompt, applied via a bulk pipeline that recovered \textbf{301{,}892 proposed-rule units across 17{,}198 proposed rules} and \textbf{374{,}026 final-rule units across 22{,}400 final rules} (676K units total; \$67.60 in extraction cost after a two-pass repair of $\sim$23K silent-zero documents). Units carry a section anchor (\texttt{\S}-citation if present), a 1-sentence \texttt{action\_summary}, and the verbatim \texttt{obligation\_text}; the \texttt{(docket\_id, doc\_id, unit\_id)} triple is the canonical identifier used by every downstream stage.

\paragraph{(2) Comment-claim extraction \& (comment $\to$ proposal) matching.} Each public comment is decomposed into its constituent claims (\texttt{claim\_id} = \texttt{docket}::\texttt{cluster}\_\texttt{idx}); claims are then matched to specific proposed-rule units in a three-stage cascade: bi-encoder retrieval $\rightarrow$ ModernBERT cross-encoder filter $\rightarrow$ gpt-5-mini binary judge with explicit ``opposing-sides'' framing (see (4) below). The (comment $\to$ proposal) corpus contains \textbf{1.99M (claim, proposed-unit) pairs} with YES/NO labels (\$28 LLM cost). The validated YES pairs are the population on which all benchmark metrics are computed: across the 558 benchmark dockets, this yields \textbf{43{,}371 unique real claims} (\textbf{223{,}535} (claim, unit) rows after the one-claim-many-units expansion).

\paragraph{(3) Match~1 + modification typology (proposed $\rightarrow$ final unit).} A parallel Match~1 pass aligns each proposed unit to its surviving final-rule counterpart, producing the \texttt{survival\_status} label per unit (survived / modified / removed). All \textbf{12{,}812 modified} units (12{,}466 after the audit in Appendix~\ref{app:prompt_audit} re-labeled some pairs; Figure~\ref{fig:links} reports the audited count) are then sub-labeled by gpt-5-mini under a \textbf{14-class change-type taxonomy} (definition\_clarified, threshold\_or\_scope\_\{looser, tighter\}, exception\_\{added, removed\}, timeline\_\{extended, shortened\}, enforcement\_\{strengthened, weakened\}, agency\_obligation\_change, cosmetic\_only, topic\_mismatch, etc.; \$2.57). The taxonomy was iterated through two manual-audit rounds: v1 mis-coded 46\% of \texttt{stricter} cases (conflating agency-side burden with stringency), motivating the v2 split. The resulting \texttt{cosmetic\_only}+\texttt{definition\_clarified} share (39\%) and \texttt{topic\_mismatch} share (24\%) provide quantitative bounds on upstream Match-pipeline noise.

\paragraph{(4) LLM-as-judge.} All comment-to-provision and response-to-comment matching uses gpt-5-mini judge prompts: the prompt explicitly states that responses (or matches) ``address'' both supporters and opponents of a position. We find that, without this framing, judge agreement with hand-curated ground truth (Nader-FCC RIF set, $n=274$) was only \textbf{67\%}; with it, agreement rose to \textbf{96\%}. The failure mode without it was systematic: the judge labeled opposite-direction true matches as ``not addressed.'' All Match~2/Match~3 labeling uses this prompt. The Axis~B pair judge in the benchmark is a different prompt, run with gpt-4o-mini in batches of ten pairs (Appendix~\ref{app:prompt_axisb}); it labels each (LLM claim, real claim) pair as \texttt{same}, \texttt{related}, \texttt{different} or \texttt{opposing}. We compare it against four other judges in Appendix~\ref{app:bucket_alignment}.

\paragraph{(5) Comment-anticipation benchmark task.} From the 17,198 proposed rules with extracted proposed units, we filter to dockets meeting all of: (a) 5--100 proposed units (excluding micro- and mega-rules), (b) $\geq$ 3 validated (comment $\to$ proposal) matches with distinct real claims, (c) $\geq$ 1 commented-on provision. This yields the \textbf{558-docket benchmark population} (Appendix Table~\ref{tab:benchmark_descriptive}). For each docket, the proposed units (formatted with their section anchors and \texttt{action\_summary}s) are presented to the model; the model must emit anticipated commenter \emph{claims} prefix-tagged \texttt{[P\#]} to specific provisions. The original gpt-5-mini reference run produced 27,936 LLM claims at \$3; the run judged for Table~\ref{tab:benchmark_results} is a later regeneration with 26,139 claims (47 per docket), and Table~\ref{tab:benchmark_descriptive} describes the reference run.

The 558-docket population spans 19 federal parent agencies; with a mean of 25 proposed units per proposal document (median 19; 876 documents, about 40 units per docket); the per-provision real-claim count is heavy-tailed (mean 7.8, median 2). The \textbf{empirical response rate} on (comment $\to$ proposal)-matched claim clusters in the benchmark is \textbf{17\% on medium-/long-comment claim clusters}.

\subsection{Benchmark dataset descriptive statistics}
\label{app:benchmark_descriptive}

We report descriptive statistics on the comment-anticipation benchmark population. The benchmark covers 558 federal rulemaking dockets, selected from DOCKET as those with 5--100 extracted proposed-rule provisions and at least three validated (comment $\to$ proposal) matches. Table~\ref{tab:benchmark_descriptive} summarizes the population.

\begin{table}[!htbp]
\centering
\small
\setlength{\tabcolsep}{8pt}
\renewcommand{\arraystretch}{1.25}
\begin{tabular}{lr}
\toprule
\textbf{Statistic} & \textbf{Value} \\
\midrule
Benchmark dockets                                                & 558 \\
Real claims, matched on provisions (unique)      & 43{,}371 \\
Comments with at least one claim                        & 13{,}824 \\
LLM-generated claims (gpt-5-mini, single run)  & 27{,}936 \\
Mean LLM claims per docket                                & 50.1 \\
\midrule
Provisions with at least one matched comment cluster (corpus-wide) & 62{,}996 \\
Mean comment clusters per matched provision                      & 7.8 \\
Median (95th percentile, max)                                    & 2 (36, 284) \\
\midrule
Empirical response rate ((comment $\to$ proposal)-matched clusters) & 54.1\% \\
Empirical response rate (Match~2 $\cup$ Match~3 clusters)        & 53.1\% \\
Empirical response rate (medium/long, non-form-letter comments)  & 17.0\% \\
Empirical response rate (all comments in benchmark dockets)      & 6.5\% \\
\bottomrule
\end{tabular}
\vspace{2pt}
\caption{Descriptive statistics of the benchmark population. Comment counts are at the cluster level after MinHash deduplication (collapsing form-letter campaigns). Empirical response rate is the fraction of comment clusters whose constituent comment documents were matched to an extracted agency response.}
\label{tab:benchmark_descriptive}
\end{table}
\subsection{Pluralism measurement: methods and evidence extraction}
\label{app:pluralism_methods}

We measure three types of pluralism (inspired by \citep{sorensen2024roadmap}) with two measurements per dimension, precision and recall. This section documents each cell's methodology.

\paragraph{Matching pipeline.} For each dimension (content, frame, style), we compare LLM-generated elements against real human elements from the same docket. The matching proceeds in three stages:

\textit{(1)~Extraction.} For content, we generate claims in \texttt{[P\#]} format (claim-list prompts, Appendices~\ref{app:prompt_gen_vanilla}--\ref{app:subgroup-persona}). For frame and style, we generate full natural-language comments and extract unbounded lists of argument types and writing techniques using Llama-3.3-70B-FP8, producing ${\sim}$10 framings and ${\sim}$14 style elements per comment.

\textit{(2)~Clustering.} Real comments produce ${\sim}$78 unique framings and ${\sim}$118 unique style elements per docket (from ${\sim}$12 comments each), many near-duplicates. We cluster these into canonical groups using Llama-3.3-70B, which assigns each item to a cluster by number reference (to avoid copy errors). This compresses 78$\to$${\sim}$25 canonical framing clusters and 118$\to$${\sim}$23 style clusters per docket.

\textit{(3)~Judging.} For content claims, we embed with all-MiniLM-L6-v2, retrieve top-3 nearest real claims per LLM claim, and judge each pair with gpt-4o-mini. For frame and style, we exhaustively compare each LLM element against all canonical clusters for that docket, judging in batches of 5 pairs per prompt with Llama-3.3-70B-FP8. All judges classify each pair as \texttt{same}, \texttt{related}, \texttt{opposing}, or \texttt{different}; only \texttt{same} counts as a match. 
\paragraph{From matches to metrics.} Given the per-pair judgments, we compute two metrics per dimension, reported in Table~\ref{tab:benchmark_results}.

Let $C_\text{real}^d$ and $C_\text{LLM}^d$ be the real and LLM-generated elements for docket $d$. Let $\text{match}(c, C') = 1$ if any $c' \in C'$ receives a \texttt{same} judgment when paired with $c$.

\textbf{Precision (P)}: mean per-docket precision. Of the LLM's generated elements, what fraction match at least one real element?
$$P = \frac{1}{|D|} \sum_{d \in D} \frac{\sum_{c \in C_\text{LLM}^d} \text{match}(c,\, C_\text{real}^d)}{|C_\text{LLM}^d|}$$

\textbf{Recall (R)}: mean per-docket recall. Of the real human elements, what fraction are covered by at least one LLM element?
$$R = \frac{1}{|D|} \sum_{d \in D} \frac{\sum_{c \in C_\text{real}^d} \text{match}(c,\, C_\text{LLM}^d)}{|C_\text{real}^d|}$$

\paragraph{Extraction and clustering examples.} Table~\ref{tab:extraction_examples} illustrates the three measurement dimensions on a single docket (EPA-HQ-OAR-2017-0091, 2018 Renewable Fuel Standards). Real comments produce ${\sim}$80 unique framings and ${\sim}$139 unique style elements from 12 comments; the LLM generates ${\sim}$15 of each from a single comment. Clustering reduces the 80 real framings to 13 canonical groups.

\begin{table}[h]
\centering
\footnotesize
\caption{Extraction examples from docket EPA-HQ-OAR-2017-0091. \textbf{Left}: real human comments (12 comments $\to$ 80 framings, 139 style elements). \textbf{Right}: LLM-generated comment (1 comment $\to$ 15 framings, 15 style elements). \textbf{Bottom}: canonical clusters after LLM-based clustering of real framings (80 $\to$ 13 clusters).}
\label{tab:extraction_examples}
\setlength{\tabcolsep}{3pt}
\begin{tabular}{p{6cm}p{6cm}}
\toprule
\textbf{Real comment framings} {\scriptsize(sample of 80)} & \textbf{LLM framings} {\scriptsize(gpt-5-mini, 15 total)} \\
\midrule
{\footnotesize
\textbullet~Appeal to economic impact on vulnerable groups \newline
\textbullet~Appeal to environmental benefits of regulation \newline
\textbullet~Appeal to agency duty to protect consumers \newline
\textbullet~Cost-benefit analysis of existing program \newline
\textbullet~Comparison to international regulatory standards \newline
\textbullet~Appeal to personal experience} &
{\footnotesize
\textbullet~Appeal to statutory intent and authority \newline
\textbullet~Analysis of regulatory history and precedent \newline
\textbullet~Economic impact analysis of regulatory choices \newline
\textbullet~Comparison to analogous provisions in related statutes \newline
\textbullet~Invocation of administrative law principles \newline
\textbullet~Reference to agency reporting and transparency} \\
\midrule
\textbf{Real comment styles} {\scriptsize(sample of 139)} & \textbf{LLM styles} {\scriptsize(15 total)} \\
\midrule
{\footnotesize
\textbullet~Addresses specific government official by title \newline
\textbullet~Attaches supporting documents or exhibits \newline
\textbullet~Uses emotional appeal language \newline
\textbullet~Provides personal anecdotal testimony \newline
\textbullet~Cites specific legislation by name} &
{\footnotesize
\textbullet~Cites statutory authority with section references \newline
\textbullet~Proposes specific alternative policy solutions \newline
\textbullet~Uses technical terminology and industry jargon \newline
\textbullet~References historical data and production trends \newline
\textbullet~Uses conditional language for consequences} \\
\bottomrule
\end{tabular}

\vspace{0.5em}
{\footnotesize
\textbf{Canonical clusters} (real framings: 80 $\to$ 13 clusters via Llama-3.3-70B): \textit{Appeal to authority} [4 members], \textit{Appeal to consequences} [7], \textit{Economic appeal} [9], \textit{Environmental appeal} [6], \textit{Personal experience} [5], \textit{Data-driven analysis} [8], \textit{Comparison} [6], \textit{Procedural} [5], \textit{Legal authority} [4], \textit{Public interest} [7], \textit{Industry impact} [6], \textit{Regulatory history} [5], \textit{Transparency} [3].
}
\end{table}
\paragraph{Evidence extraction for RAG baseline.} To build the retrieval-augmented generation baseline, we extract research-level factual evidence from 30,588 real public comments flagged as evidence-bearing by our structural feature classifier. We instruct gpt-4o-mini to extract \emph{only} verifiable, citable facts --- not biographical details, opinions, or vague references. Each extracted item must be a complete, self-contained statement (Table~\ref{tab:evidence_extraction}).

\begin{table}[h]
\centering
\small
\caption{Research-level evidence extraction from real public comments (N=30,588 comments; 27,406 with extractable evidence). The prompt requires each item to be self-contained: a reader must understand what a number measures, what a statute governs, or what a study found without seeing the original comment.}
\label{tab:evidence_extraction}
\setlength{\tabcolsep}{3pt}
\begin{tabular}{p{2.2cm}p{9.5cm}}
\toprule
\textbf{Evidence type} & \textbf{Extracted content (examples)} \\
\midrule
Statutes cited \newline (21.3\%) & 45 C.F.R. \S~164.502(a) establishes the conditions under which protected health information may be disclosed, including limitations on data sharing that could violate privacy laws. \\
\midrule
Published data \newline (48.9\%) & EPA estimated a range of benefits of \$36--\$90 billion from the Mercury and Air Toxics Standards for Power Plants rule, with the reduction of mercury accounting for just \$4--\$6 million of that total. \\
\midrule
Studies cited \newline (20.3\%) & The EPA 2019 Integrated Science Assessment for PM2.5 concluded there is a causal relationship between long-term PM2.5 exposure and cardiovascular mortality. \\
\midrule
Regulatory history \newline (41.5\%) & The 2012 Mercury and Air Toxics Standards required coal-fired plants to reduce mercury emissions by 90\% within 3 years of promulgation. \\
\midrule
Specific alternative \newline (42.4\%) & Incorporate the alternative temperature monitoring option (minimum kiln temperature) into the final rule for additional flexibility. \\
\bottomrule
\end{tabular}
\end{table}
\paragraph{Measurement variables.} We decompose each comment claim into three measurement dimensions, each extracted as an unbounded list of specific elements (Table~\ref{tab:measurement_variables}):

\begin{table}[h]
\centering
\small
\caption{Three measurement variables with example extracted elements. Elements are extracted by Llama-3.3-70B; MiniLM retrieves the three nearest real elements (cosine $> 0.3$) and the LLM judge decides whether each pair matches.}
\label{tab:measurement_variables}
\setlength{\tabcolsep}{3pt}
\begin{tabular}{p{1.8cm}p{3.5cm}p{7cm}}
\toprule
\textbf{Variable} & \textbf{What it captures} & \textbf{Example extracted elements} \\
\midrule
Content & Substantive position taken & ``The proposed 30-day compliance window is insufficient for small manufacturers'' \newline ``The rule should exempt facilities with fewer than 50 employees'' \\
\midrule
Frame & Argumentative approach & ``Cost-burden on small businesses'' \newline ``Environmental justice for frontline communities'' \newline ``APA procedural due-process argument'' \\
\midrule
Style & Evidence and format & ``Cites 42 U.S.C. \S~7411(d) on existing-source emission guidelines'' \newline ``Proposes replacing 30-day deadline with 90 days in \S XXX.XX(a)'' \newline ``References EPA 2019 ISA finding on PM2.5 mortality'' \\
\bottomrule
\end{tabular}
\end{table}
\paragraph{Style-steering prompt variants (a side experiment, not part of Table~\ref{tab:benchmark_results}).} We test five prompt variants that instruct the model to adopt specific structural patterns: (1)~\emph{evidence-rich} (``cite real statute numbers, reference empirical data''); (2)~\emph{experiential} (``write from first-person lived experience''); (3)~\emph{alternative-proposing} (``propose specific replacement regulatory language''); (4)~\emph{legal-grounded} (``cite APA, case law, constitutional provisions''); (5)~\emph{RAG} (retrieve real evidence snippets via BM25 and include as context). The RAG variant uses a BM25 index over 27,406 evidence-bearing extractions; for each benchmark provision, we retrieve the top-3 most similar evidence snippets and include them in the prompt.

\paragraph{Key style finding.} Vanilla LLM claims cite statutes 1.0\% of the time vs.\ 15.9\% in real comments (16$\times$ gap). Prompt steering closes this dramatically --- the legal-grounded variant achieves 85.8\% statute citation and 40.4\% case law citation --- but these citations are likely hallucinated. The RAG variant achieves 26.4\% statute citation with evidence grounded in real prior comments, suggesting that the style gap is partly a knowledge problem addressable with retrieval augmentation.

\subsection{Claim budget selection and calibration}
\label{app:claim_budget}

A key design choice in DOCKETBench is the claim budget $k$: how many claims should the model generate per docket? Without budget control, the model's output volume is fixed regardless of docket complexity. We document the calibration process here.

\paragraph{Human claim volume.} Table~\ref{tab:claim_counts_app} summarizes the distribution of real human claims per docket across the 558 benchmark dockets. The distribution is highly skewed: the median is 48, but the mean is 401 due to mass-comment dockets (max 12,857).

\begin{figure}[h]
\centering
\includegraphics[width=\textwidth]{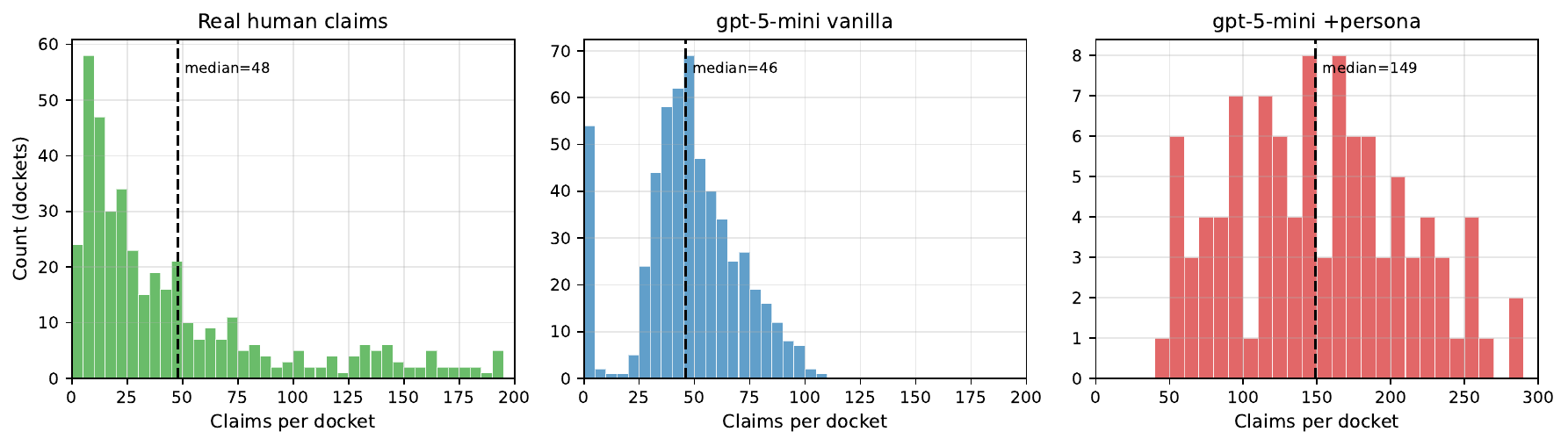}
\caption{Distribution of claims per docket. \textbf{Left}: Real human claims (clipped at 200; long tail extends to 12,857). \textbf{Center}: gpt-5-mini vanilla (concentrated around median 46). \textbf{Right}: gpt-5-mini +persona (7 personas pooled, median 149). Dashed lines show medians.}
\label{fig:claim_distributions_app}
\end{figure}

\begin{figure}[h]
\centering
\includegraphics[width=\textwidth]{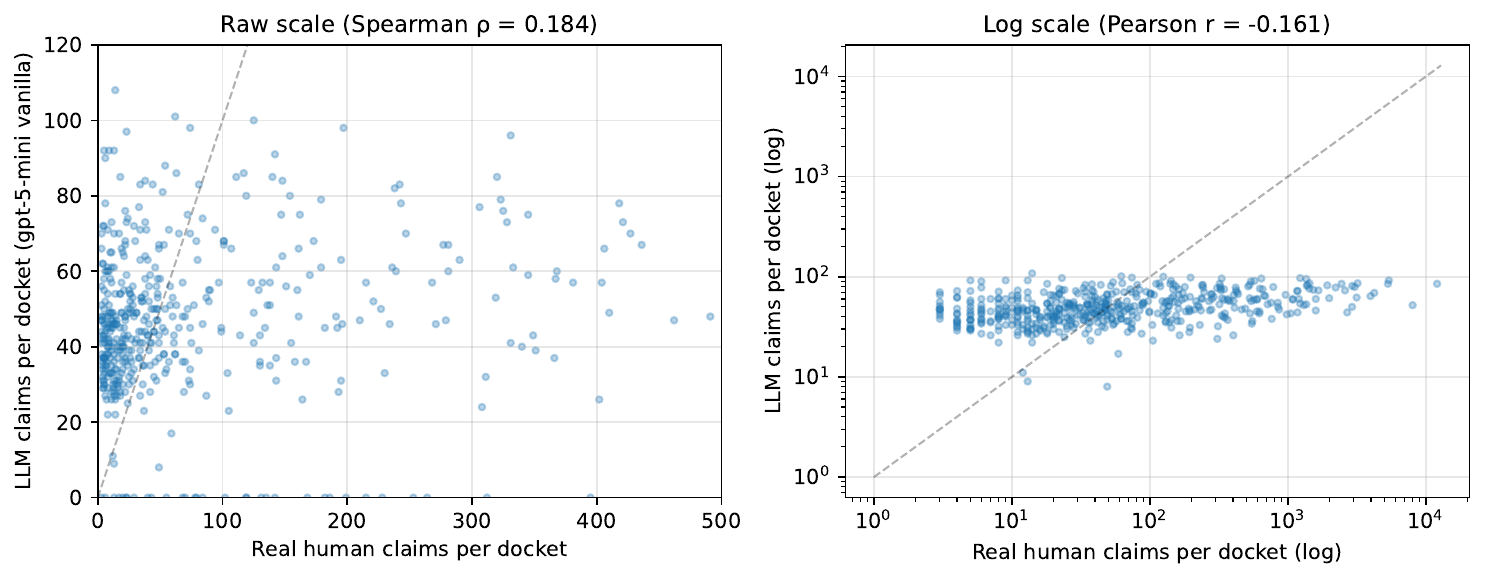}
\caption{Correlation between real human claim count and uncontrolled LLM claim count per docket (gpt-5-mini vanilla, $N$=558). The LLM generates ${\sim}$46 claims regardless of whether the docket received 3 or 12,857 real claims (Spearman $\rho$=0.184, $p<$0.001; Pearson $r$=-0.045, n.s.). This motivates budget control.}
\label{fig:claim_scatter_app}
\end{figure}

\begin{table}[h]
\centering
\small
\caption{Claims per docket in the 558-docket benchmark population.}
\label{tab:claim_counts_app}
\begin{tabular}{l rrrrrr}
\toprule
& Mean & Median & P25 & P75 & Min & Max \\
\midrule
Real human claims & 401 & 48 & 16 & 220 & 3 & 12{,}857 \\
gpt-5-mini vanilla (uncontrolled) & 47 & 46 & 35 & 61 & 0 & 108 \\
gpt-5-mini +persona (uncontrolled) & 151 & 149 & 106 & 190 & 40 & 281 \\
\bottomrule
\end{tabular}
\end{table}
\paragraph{Uncontrolled LLM output volume.} Without a claim budget, gpt-5-mini vanilla generates a median of 46 claims per docket---coincidentally close to the human median (48)---but with a far narrower range (0--108 vs.\ 3--12,857). Figure~\ref{fig:claim_distributions_app} shows the distributions; Figure~\ref{fig:claim_scatter_app} confirms that the LLM does not scale its output to match commenting intensity (Spearman $\rho=0.184$, Pearson $r=-0.045$).

\begin{figure}[h]
\centering
\includegraphics[width=\textwidth]{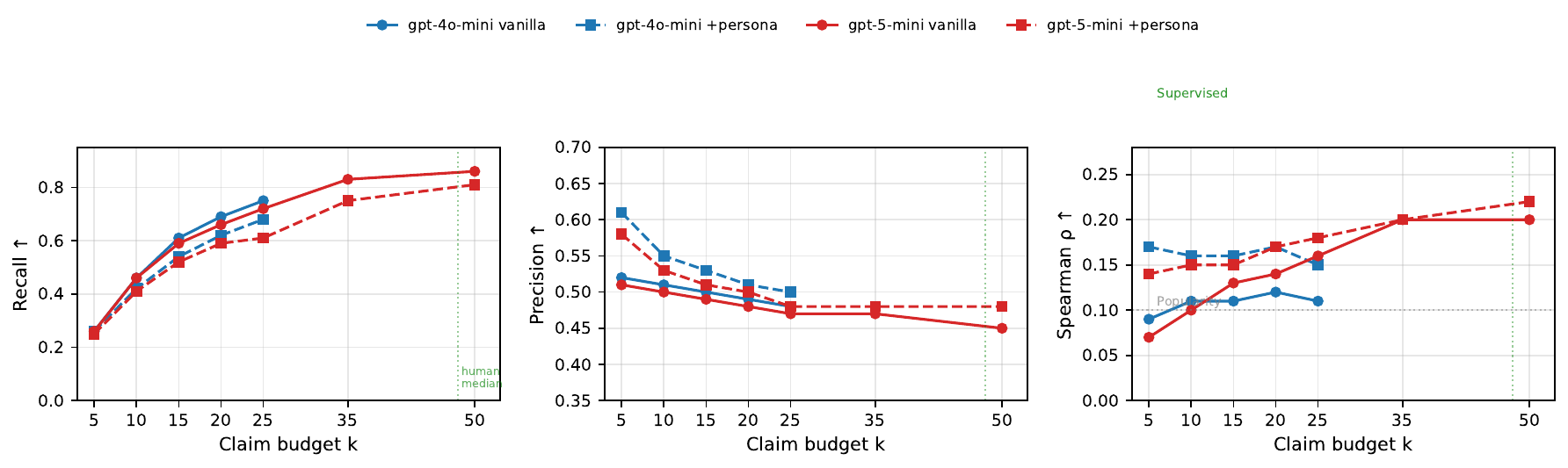}
\caption{Provision targeting as a function of claim budget $k \in \{5, 10, 15, 20, 25, 35, 50\}$. \textbf{Left}: Recall saturates near $k$=35--50 (close to the human median of 48 claims/docket, green dotted line). \textbf{Center}: Precision declines gently; +persona (dashed) maintains higher precision at every $k$. \textbf{Right}: Spearman $\rho$ plateaus at ${\sim}$0.20 for gpt-5-mini. gpt-4o-mini tested at $k$=5--25 only.}
\label{fig:budget_curves_app}
\end{figure}

\paragraph{Budget selection.} We test fixed budgets $k \in \{5, 10, 15, 20, 25\}$ for every model, extended to $k \in \{35, 50\}$ for gpt-4o-mini and gpt-5-mini. Models comply with budget instructions 87--98\% of the time (within $\pm$20\% of requested $k$). gpt-5-mini achieves 98\% compliance; gpt-4o-mini achieves 87\%. Figure~\ref{fig:budget_curves_app} shows provision-targeting metrics as a function of $k$.

\paragraph{Key observations.} (1) Recall keeps rising to $k$=50. (2) Precision decays gently (${\sim}$2pp per 5 claims), so higher budgets help more than they hurt. (3) The real optimum likely lies near $k$=50 (close to the human median) . Table~\ref{tab:benchmark_results} reports uncontrolled generation. We use the budget-controlled runs as a check: claim precision moves by less than 0.4 points across $k$, so the precision gap in the main text is not an artifact of how many claims the models write.

\subsection{Persona aggregation methodology and limitations}
\label{app:persona_aggregation}

We document the design choices and limitations of our persona-conditioned benchmark variant to facilitate correct interpretation of results.

\paragraph{Aggregation.} For the \texttt{+persona} variant, each of 7 personas (industry, labor, advocacy, expert, government, law firm, individual) generates claims independently for the same docket. We \textbf{pool all persona outputs with uniform weight}---i.e., each persona contributes $\frac{1}{7}$ of the generation budget regardless of its prevalence in the true commenter population. In the real data, individual commenters account for the majority of submissions, while law firms and labor unions are relatively rare. We chose uniform weighting for interpretability (each stakeholder voice is represented equally) rather than fidelity to the true population distribution.

\paragraph{Claim count.} The number of claims generated per persona per docket is \textbf{unconstrained}: the model decides how many claims to produce based on the provisions presented. In practice, the ``individual'' and ``advocacy'' personas tend to generate 10--20 claims per docket, while ``law firm'' generates 15--30 (reflecting the model's tendency toward more granular legal arguments). We do not normalize claim counts across personas, which means personas that produce more claims have proportionally more influence on pooled metrics (Axes A--D).

\paragraph{Within-persona diversity.} Each persona is prompted \textbf{exactly once} per docket (temperature 1.0). There is no multi-sample generation, no prompt variation within a persona, and no explicit diversity instructions beyond the persona description itself. Variation across claims within a single persona generation arises solely from the model's stochastic sampling.

\paragraph{Interpretation of Table~\ref{tab:bucket_alignment}.} Table~\ref{tab:bucket_alignment} measures alignment between \emph{all} LLM claims (vanilla or pooled personas) and real claims \emph{grouped by the real commenter's} \texttt{org\_type}. It answers: ``Which types of real commenters does the LLM agree with most?'' It does \emph{not} measure how well each LLM persona matches its corresponding real stakeholder group. The metric is: $\text{Agreement}(X) = \frac{|\texttt{same}|}{|\texttt{same}| + |\texttt{opposing}|}$ for real claims from \texttt{org\_type} $X$. Labor is the smallest group in Table~\ref{tab:bucket_alignment}, with 170--374 pairs per vanilla run. Law firms are counted as their clients in that table.

\paragraph{What this design can and cannot show.} Our persona conditioning is minimal: seven roles, one prompt each, no further steering. This is sufficient to show that (a) role conditioning raises claim recall mostly because four of the five models write three to six times more claims (Llama-3.3-70B wrote half as many), and its effect on precision, frame and style is small and goes both ways (Table~\ref{tab:benchmark_results}), and (b) even with role conditioning, structural deficits persist (procedural tone, uniform attention, modification-oriented stance). However, our setup does \emph{not} constitute a rigorous test of whether LLMs \emph{can} represent diverse stakeholders under more elaborate prompting (e.g., multi-turn elicitation, retrieval-augmented generation, few-shot exemplars drawn from real comments). The structural findings---that LLMs default to procedural, modification-oriented, uniformly-distributed commentary---are robust to persona conditioning but may be addressable with more sophisticated prompting strategies that we do not explore here. We therefore frame our benchmark as measuring ``what LLMs produce under straightforward prompting'' rather than ``what LLMs are capable of producing under optimal conditions.''

\paragraph{Alternative weighting schemes.} Three natural alternatives to uniform pooling are: (1) population-weighted (weight each persona by its share of real commenters), (2) provision-weighted (weight by per-provision org\_type distribution), or (3) claim-count-normalized (constrain each persona to a fixed number of claims). Under population-weighting, the ``individual'' persona would dominate ($\sim$60\% weight), amplifying claims written as citizen testimony. Under claim-count normalization, personas that naturally produce fewer arguments (labor, individual) would receive equal representation alongside more prolific personas (law firm). We report uniform-weight results as the simplest interpretable baseline; alternative weightings can be computed from the released per-persona claim data.

\subsection[Stakeholder alignment: agreement, representation, and steering]{ Stakeholder alignment: agreement, representation, and steering}
\label{app:bucket_alignment}

Table~\ref{tab:agreement_full} is the full version of Table~\ref{tab:bucket_alignment}: agreement with every stakeholder group, with confidence intervals, and the human-commenter row.

\begin{table}[t]
\centering
\scriptsize
\setlength{\tabcolsep}{3pt}
\renewcommand{\arraystretch}{1.15}
\begin{tabular}{@{}lccccccc@{}}
\toprule
\textbf{Model} & \textbf{Individual} & \textbf{Industry} & \textbf{Advocacy} & \textbf{Gov} & \textbf{Expert} & \textbf{Labor} & \textbf{All} \\
\midrule
gpt-4o-mini & 0.57 {\tiny[0.49, 0.64]} & 0.59 {\tiny[0.56, 0.61]} & 0.56 {\tiny[0.53, 0.60]} & 0.68 {\tiny[0.57, 0.78]} & 0.49 {\tiny[0.36, 0.65]} & \textbf{0.38 {\tiny[0.30, 0.46]}} & 0.57 {\tiny[0.55, 0.59]} \\
\rowcolor{gray!15}\hspace*{1em}+persona & 0.53 {\tiny[0.49, 0.57]} & 0.53 {\tiny[0.51, 0.54]} & 0.53 {\tiny[0.50, 0.55]} & 0.61 {\tiny[0.55, 0.68]} & 0.57 {\tiny[0.48, 0.64]} & \textbf{0.43 {\tiny[0.38, 0.48]}} & 0.53 {\tiny[0.51, 0.54]} \\
gpt-5-mini & 0.59 {\tiny[0.54, 0.64]} & 0.60 {\tiny[0.58, 0.62]} & 0.55 {\tiny[0.52, 0.58]} & 0.65 {\tiny[0.57, 0.74]} & 0.51 {\tiny[0.43, 0.60]} & \textbf{0.48 {\tiny[0.41, 0.55]}} & 0.58 {\tiny[0.57, 0.60]} \\
\rowcolor{gray!15}\hspace*{1em}+persona & 0.55 {\tiny[0.50, 0.59]} & 0.56 {\tiny[0.55, 0.58]} & 0.56 {\tiny[0.54, 0.59]} & 0.68 {\tiny[0.61, 0.74]} & 0.52 {\tiny[0.44, 0.60]} & \textbf{0.47 {\tiny[0.42, 0.52]}} & 0.56 {\tiny[0.55, 0.57]} \\
gpt-5 & 0.63 {\tiny[0.58, 0.69]} & 0.68 {\tiny[0.66, 0.70]} & 0.62 {\tiny[0.59, 0.65]} & 0.66 {\tiny[0.55, 0.76]} & 0.63 {\tiny[0.53, 0.74]} & \textbf{0.61 {\tiny[0.53, 0.71]}} & 0.66 {\tiny[0.64, 0.67]} \\
\rowcolor{gray!15}\hspace*{1em}+persona & 0.58 {\tiny[0.54, 0.62]} & 0.60 {\tiny[0.58, 0.61]} & 0.59 {\tiny[0.56, 0.62]} & 0.65 {\tiny[0.59, 0.73]} & 0.60 {\tiny[0.52, 0.67]} & \textbf{0.56 {\tiny[0.51, 0.61]}} & 0.59 {\tiny[0.58, 0.61]} \\
Llama-3.3-70B & 0.41 {\tiny[0.36, 0.49]} & 0.44 {\tiny[0.41, 0.47]} & 0.44 {\tiny[0.41, 0.48]} & 0.47 {\tiny[0.34, 0.60]} & 0.42 {\tiny[0.35, 0.48]} & \textbf{0.26 {\tiny[0.19, 0.33]}} & 0.43 {\tiny[0.41, 0.46]} \\
\rowcolor{gray!15}\hspace*{1em}+persona & 0.44 {\tiny[0.40, 0.50]} & 0.50 {\tiny[0.47, 0.53]} & \textbf{0.44 {\tiny[0.41, 0.48]}} & 0.57 {\tiny[0.46, 0.68]} & 0.61 {\tiny[0.49, 0.74]} & 0.44 {\tiny[0.35, 0.53]} & 0.48 {\tiny[0.46, 0.50]} \\
Llama-3.1-8B & 0.49 {\tiny[0.42, 0.55]} & 0.55 {\tiny[0.53, 0.58]} & 0.49 {\tiny[0.45, 0.53]} & 0.68 {\tiny[0.58, 0.79]} & \textbf{0.40 {\tiny[0.27, 0.56]}} & 0.46 {\tiny[0.38, 0.57]} & 0.53 {\tiny[0.50, 0.55]} \\
\rowcolor{gray!15}\hspace*{1em}+persona & 0.41 {\tiny[0.38, 0.45]} & 0.45 {\tiny[0.43, 0.46]} & 0.42 {\tiny[0.40, 0.44]} & 0.48 {\tiny[0.43, 0.53]} & 0.42 {\tiny[0.37, 0.48]} & \textbf{0.36 {\tiny[0.34, 0.39]}} & 0.44 {\tiny[0.42, 0.45]} \\
\midrule
Human commenters & 0.86 {\tiny[0.70, 0.96]} & 0.77 {\tiny[0.68, 0.85]} & 0.71 {\tiny[0.60, 0.82]} & -- & -- & -- & 0.75 {\tiny[0.69, 0.81]} \\
\bottomrule
\end{tabular}
\caption{Whose side LLMs take. For each pair where an LLM claim and a real claim make a point about the same thing (judge label \texttt{same} or \texttt{opposing}), we report the fraction judged \texttt{same}, split by the stakeholder group of the real commenter. We count each commenter as the interest it files for (a law firm as its client, a trade association as its members). Brackets are 95\% confidence intervals from a docket bootstrap; bold marks the lowest group in each row. The \emph{Human commenters} row runs the same judge on claims from randomly sampled real commenters against the other comments in their docket; we omit cells with fewer than ten pairs. LLMs agree with labor least in eight of ten rows (tied in a ninth) and less than with industry in every row, and every LLM agrees with real commenters less often than real commenters agree with each other.}
\label{tab:agreement_full}
\end{table}

\paragraph{Representation lift, re-estimated.} The lift we printed in the NeurIPS submission ($\text{Lift} = \%\text{ of matched claims from a group} / \%\text{ of comments from the group}$) does not have a null of 1. Models write roughly the same number of claims per docket no matter how large the docket is, so a claim in a small docket is three to four times more likely to be matched than a claim in a large one. Labor claims sit in small FAA, OSHA and MSHA dockets (46\% of them are FAA). Pooling matches across dockets and dividing by a global population share therefore gives labor a 2--4$\times$ ``lift'' even when matching is random within each docket: shuffling group labels within dockets gives 2.75$\times$, and running real human claims from other dockets through the same pipeline gives 2.9--5.5$\times$. The sign of the printed labor cell followed the judge and the protocol, not the model. Table~\ref{tab:bucket_alignment_lift} fixes this. It compares each group's share of matched claims to the share we would expect if each docket's matches were a random draw from that docket's own commenters, standardizes across agencies with weights that do not depend on match rate, and counts each commenter as the interest it files for (all 60 law-firm clusters in the benchmark file for clients; 54 of those are industry). Under this estimator the ten-run consensus for every group is within a few percent of proportional (0.97--1.04$\times$); individual cells range from 0.85 to 1.41$\times$, the widest resting on fewer than three agencies. Dropping FAA dockets moves the pooled within-docket labor estimate from 1.19 to 1.01. Cells marked $\dagger$ rest on fewer than three agencies and carry the FAA problem. The result is not a ceiling effect: in dockets with match rates below 10\%, where a model could have put two to eleven times its share of matches on one group, the estimates are still 0.93--1.05. Matching noise that does not depend on group does push these estimates toward 1, so we can rule out large representation biases, not biases of a few percent.

\paragraph{Persona steering.} We joined each judged pair to the persona that wrote the LLM claim and asked whether persona-$p$ matches land on real group $g$ more often than chance. They do not: the diagonal (persona matches its own group) is 1.04--1.15 and the off-diagonal is 1.03--1.07. Writing as a group does not make a model reproduce that group's actual claims. The same personas move the agreement rate in Table~\ref{tab:bucket_alignment} by 0.2--0.4 (for gpt-5, the government persona agrees with labor 76\% of the time and the individual persona 51\%). Personas change the model's stance. They do not change its substance.

\paragraph{Judge robustness of Table~\ref{tab:bucket_alignment}.} The \texttt{opposing} label is the least stable output of the pair judge. On a 1,987-pair stratified sample, four other judges confirm gpt-4o-mini's \texttt{opposing} calls as \texttt{opposing} 18--33\% of the time (most of the rest become \texttt{related}), but when two judges both give a directional label they agree on the direction 71--84\% of the time. We recomputed Table~\ref{tab:bucket_alignment} over all 1.13M pairs with two open-weights judges. Both reproduce the group pattern: labor is 0.29--0.51 below the other groups under a distilled Gemma-3-12B judge and 0.07--0.27 below under Gemma-3-27B. They do not reproduce the ranking of models by overall agreement, so we do not interpret that ranking. A blind human check of 52 judged pairs, balanced over judge labels, generating models and 21 agencies, agreed with the gpt-4o-mini judge on 86.5\% of pairs ($\kappa$ = 0.80).

\begin{table}[t]
\centering
\scriptsize
\setlength{\tabcolsep}{3pt}
\renewcommand{\arraystretch}{1.15}
\begin{tabular}{@{}lcccccc@{}}
\toprule
\textbf{Model} & \textbf{Individual} & \textbf{Industry} & \textbf{Advocacy} & \textbf{Gov} & \textbf{Expert} & \textbf{Labor} \\
\midrule
gpt-4o-mini & 1.10 {\tiny[0.93, 1.27]} & 1.00 {\tiny[0.97, 1.04]} & 0.96 {\tiny[0.90, 1.01]} & 1.15 {\tiny[0.95, 1.41]} & 0.85 {\tiny[0.60, 1.08]} & 0.94 {\tiny[0.63, 1.25]} \\
\rowcolor{gray!15}\hspace*{1em}+persona & 0.99 {\tiny[0.89, 1.08]} & 1.01 {\tiny[0.98, 1.03]} & 0.99 {\tiny[0.97, 1.03]} & 1.03 {\tiny[0.85, 1.21]} & 0.92 {\tiny[0.76, 1.06]} & 0.92 {\tiny[0.74, 1.20]} \\
gpt-5-mini & 0.94 {\tiny[0.84, 1.04]} & 1.02 {\tiny[0.99, 1.04]} & 1.00 {\tiny[0.96, 1.03]} & 1.03 {\tiny[0.79, 1.26]} & 0.93 {\tiny[0.71, 1.11]} & 1.05 {\tiny[0.87, 1.25]} \\
\rowcolor{gray!15}\hspace*{1em}+persona & 0.97 {\tiny[0.91, 1.03]} & \textbf{1.02 {\tiny[1.00, 1.04]}} & 0.98 {\tiny[0.94, 1.01]} & 1.04 {\tiny[0.87, 1.23]} & 0.92 {\tiny[0.78, 1.05]} & 1.00 {\tiny[0.81, 1.15]} \\
gpt-5 & 0.94 {\tiny[0.83, 1.04]} & 1.03 {\tiny[0.99, 1.06]} & 0.98 {\tiny[0.94, 1.02]} & 0.98 {\tiny[0.78, 1.19]} & 0.86 {\tiny[0.67, 1.07]} & 0.88 {\tiny[0.66, 1.10]} \\
\rowcolor{gray!15}\hspace*{1em}+persona & 0.98 {\tiny[0.92, 1.05]} & 1.01 {\tiny[1.00, 1.03]} & \textbf{0.97 {\tiny[0.94, 0.99]}} & 1.02 {\tiny[0.89, 1.21]} & \textbf{1.10 {\tiny[1.00, 1.22]}} & 1.11 {\tiny[0.93, 1.28]} \\
Llama-3.3-70B & 0.96 {\tiny[0.85, 1.08]} & 1.03 {\tiny[1.00, 1.05]} & 0.98 {\tiny[0.94, 1.02]} & 1.00 {\tiny[0.82, 1.16]} & 0.98 {\tiny[0.78, 1.17]} & 0.91 {\tiny[0.70, 1.14]} \\
\rowcolor{gray!15}\hspace*{1em}+persona & 0.96 {\tiny[0.82, 1.10]} & 1.00 {\tiny[0.97, 1.04]} & 0.98 {\tiny[0.94, 1.02]} & 1.18 {\tiny[0.95, 1.51]} & 1.08 {\tiny[0.84, 1.29]} & 1.13 {\tiny[0.74, 1.30]}$^{\dagger}$ \\
Llama-3.1-8B & 0.95 {\tiny[0.84, 1.05]} & 1.01 {\tiny[0.98, 1.04]} & 0.99 {\tiny[0.95, 1.03]} & 1.09 {\tiny[0.83, 1.39]} & 0.99 {\tiny[0.75, 1.23]} & \textbf{1.41 {\tiny[1.03, 1.74]}}$^{\dagger}$ \\
\rowcolor{gray!15}\hspace*{1em}+persona & 0.96 {\tiny[0.91, 1.02]} & 1.01 {\tiny[0.99, 1.02]} & 0.99 {\tiny[0.97, 1.02]} & 1.02 {\tiny[0.88, 1.13]} & 1.04 {\tiny[0.87, 1.22]} & 0.94 {\tiny[0.71, 1.14]} \\
\midrule
Consensus (10 runs) & 0.97 {\tiny[0.94, 1.00]} & 1.01 {\tiny[1.01, 1.02]} & 0.98 {\tiny[0.97, 0.99]} & 1.04 {\tiny[0.98, 1.10]} & 0.99 {\tiny[0.94, 1.05]} & 1.03 {\tiny[0.95, 1.11]} \\
\bottomrule
\end{tabular}
\caption{Stakeholder representation among matched claims, per model. Each cell is the share of matched real claims written by a group, divided by the share we would expect if each docket\textquotesingle s matches were a random draw from that docket\textquotesingle s commenters. We standardize across agencies (each agency weighted by the group\textquotesingle s claim count there) and count each commenter as the interest it files for (law firms as their clients, trade associations as their members). Brackets are 95\% confidence intervals from a docket bootstrap; bold marks intervals that exclude 1; $\dagger$ marks cells where fewer than three agencies contribute. Those cells are dominated by FAA dockets, where union form letters recur across dockets and inflate labor matching for every model. No group is more than a few percent over- or under-represented; the consensus across runs is 0.97--1.04$\times$.}
\label{tab:bucket_alignment_lift}
\end{table}

\subsection{Why LLMs agree with commenters less often than commenters agree with each other}
\label{app:agreement_anatomy}

\paragraph{What the agreement rate covers.} Table~\ref{tab:bucket_alignment} only counts pairs where the LLM claim and the real claim make a point about the same thing. Table~\ref{tab:unmatched_shares} shows how much of each model's output that is. Between 83\% and 92\% of generated claims retrieve at least one real claim above the cosine threshold. Only 21--33\% get a directional label (\texttt{same} or \texttt{opposing}) against any real claim, and 15--20\% get at least one \texttt{same}. Another 55--63\% are judged \texttt{related} or \texttt{different} to everything they retrieve: they are on the topic of real comments but make a point no commenter made. The last 8--17\% retrieve nothing. So the agreement rate describes the fifth to the third of LLM output that lines up with a real claim. The rest is the substance gap in Table~\ref{tab:benchmark_results}.

\begin{table}[h]
\centering
\scriptsize
\setlength{\tabcolsep}{4pt}
\begin{tabular}{@{}lrrrrrr@{}}
\toprule
\textbf{Model} & \textbf{Generated} & \textbf{Any pair} & \textbf{Directional} & \textbf{$\geq$1 same} & \textbf{Topic only} & \textbf{No counterpart} \\
\midrule
gpt-4o-mini & 14{,}584 & 87.5 & 25.9 & 17.9 & 61.6 & 12.5 \\
\rowcolor{gray!15}\hspace*{1em}+persona & 55{,}358 & 90.2 & 29.1 & 19.1 & 61.1 & 9.8 \\
gpt-5-mini & 26{,}139 & 85.2 & 24.4 & 17.1 & 60.8 & 14.8 \\
\rowcolor{gray!15}\hspace*{1em}+persona & 84{,}490 & 89.1 & 26.2 & 17.9 & 62.9 & 10.9 \\
gpt-5 & 29{,}183 & 82.8 & 20.6 & 15.4 & 62.2 & 17.2 \\
\rowcolor{gray!15}\hspace*{1em}+persona & 92{,}485 & 87.4 & 25.4 & 18.0 & 62.0 & 12.6 \\
Llama-3.3-70B & 30{,}112 & 85.2 & 29.8 & 17.2 & 55.3 & 14.8 \\
\rowcolor{gray!15}\hspace*{1em}+persona & 16{,}653 & 88.1 & 32.5 & 20.0 & 55.5 & 11.9 \\
Llama-3.1-8B & 17{,}962 & 88.4 & 28.4 & 19.1 & 60.0 & 11.6 \\
\rowcolor{gray!15}\hspace*{1em}+persona & 97{,}384 & 91.7 & 32.6 & 19.3 & 59.1 & 8.3 \\
\bottomrule
\end{tabular}
\caption{Where generated claims end up (\% of generated claims per run). \emph{Any pair}: at least one real claim retrieved at cosine $>0.3$. \emph{Directional}: at least one pair judged \texttt{same} or \texttt{opposing}; this is the denominator of Table~\ref{tab:bucket_alignment}. \emph{$\geq$1 same}: at least one \texttt{same}, the numerator of claim precision (micro-averaged here). \emph{Topic only}: retrieved real claims, but every pair judged \texttt{related} or \texttt{different}. \emph{No counterpart}: nothing retrieved.}
\label{tab:unmatched_shares}
\end{table}
\paragraph{Hedging.} For 18\% of the real claims that get any directional label, the same run produces both an agreeing and an opposing claim. 50\% get only agreeing claims and 32\% only opposing ones. Models argue both sides twice as often on labor claims (33\%) as on the rest (17\%). If we count each real claim once, by its majority label, instead of counting pairs, overall agreement rises from 0.53 to 0.59. Hedging explains part of the gap to the human rate but not most of it.

\paragraph{A default stance.} Pooled over the five vanilla runs, agreement is highest for procedural (0.62), scientific--technical (0.56) and economic (0.54) arguments and lowest for environmental (0.43), safety (0.45) and health (0.46) arguments. The models' default positions follow the registers industry and government commenters use (Table~\ref{tab:stakeholder_concern}) and go against the registers advocacy, individual and labor commenters use. Three quarters of labor claims are safety claims; 14\% of other groups' claims are. If we reweight labor to the other groups' frame mix, its agreement rises from 0.44 to 0.49 (the other groups are at 0.54), so framing explains about half the labor gap. Within the safety frame, labor is still below the others (0.41 vs.\ 0.46).

\paragraph{Controls.} The labor gap is not a retrieval artifact. Agreement falls with retrieval cosine for every group, but labor is lower in every cosine bin, and reweighting labor to the other groups' cosine distribution leaves it at 0.40. It is not an agency-mix artifact either: labor is below the other groups inside FAA (0.43 vs.\ 0.53), OSHA (0.42 vs.\ 0.53) and MSHA (0.32 vs.\ 0.55). And it is not specific to the gpt-4o-mini judge (Appendix~\ref{app:bucket_alignment}).

\paragraph{Human baselines.} Labor's low agreement could just mean that no one agrees with labor. The baselines we have say that is largely the case. Inside the same docket, randomly sampled real commenters agree with other commenters 75\% of the time overall and 67\% of the time with labor, though we only have six labor pairs (too few for Table~\ref{tab:agreement_full}, which omits cells under ten pairs). Across dockets, we ran real human claims from other dockets of the same agency through the same retrieval and judging pipeline with an independent open-weights judge: they agree with labor claims 14\% of the time and with every other group 23--36\% of the time (112 labor pairs; 27\% overall). Against these baselines the LLMs are not more opposed to labor than other commenters are. They take the median commenter's side, and the median commenter does not make labor's safety demands. The right test is a within-docket human baseline, judged by the same judge, with enough labor pairs. We did not run it.

\subsection{LLM comment characterization: direction and argument divergence}
\label{app:direction_argument}

To characterize \emph{how} LLM-generated comments differ from human comments, we compare the distribution of comment \textbf{direction} (what the commenter wants the agency to do) and \textbf{argument type} (what reasoning the commenter uses) between LLM-generated and real human claims across the 558-docket benchmark population. Each claim is labeled with a direction from \{modify, oppose, support, inform, exempt\} and an argument type from \{procedural, technical, economic, legal, experiential, normative\} using the same taxonomy prompt applied at generation time (Appendix~\ref{app:prompts}).

\begin{table}[!htbp]
\centering
\small
\begin{subtable}[t]{0.46\textwidth}
\centering
\begin{tabular}{lccc}
\toprule
\textbf{Direction} & \textbf{LLM}  & \textbf{Human} & \textbf{$\Delta$} \\
\midrule
Modify   & \cellcolor[HTML]{B1463A}\textcolor{white}{85.9} & \cellcolor[HTML]{CE897E}\textcolor{white}{43.8} & \cellcolor[HTML]{C04030}\textcolor{white}{\textbf{+42.1}} \\
Oppose   & \cellcolor[HTML]{FBF2E8}5.1 & \cellcolor[HTML]{DAA59A}30.4 & \cellcolor[HTML]{4A7AB5}\textcolor{white}{-25.4} \\
Support  & \cellcolor[HTML]{FDF6EC}2.2 & \cellcolor[HTML]{EDCFC4}12.2 & \cellcolor[HTML]{7BA4CC}\textcolor{white}{-10.0} \\
Inform   & \cellcolor[HTML]{FAF0E5}5.7 & \cellcolor[HTML]{F0D7CD}10.9 & \cellcolor[HTML]{A3C1DE}-5.2 \\
Exempt   & \cellcolor[HTML]{FDF6EC}1.1 & \cellcolor[HTML]{FDF5EB}2.7 & \cellcolor[HTML]{C5D8EC}-1.5 \\
\bottomrule
\end{tabular}
\caption{Comment direction (\%).}
\label{tab:direction_dist}
\end{subtable}
\hfill
\begin{subtable}[t]{0.52\textwidth}
\centering
\begin{tabular}{lccc}
\toprule
\textbf{Argument} & \textbf{LLM} & \textbf{Human} & \textbf{$\Delta$} \\
\midrule
Procedural    & \cellcolor[HTML]{B74B3E}\textcolor{white}{60.3} & \cellcolor[HTML]{EDCFC4}15.7 & \cellcolor[HTML]{B03828}\textcolor{white}{\textbf{+44.6}} \\
Technical     & \cellcolor[HTML]{EDD0C5}16.9 & \cellcolor[HTML]{E3B8AE}22.1 & \cellcolor[HTML]{A3C1DE}-5.2 \\
Legal         & \cellcolor[HTML]{F0D7CD}12.4 & \cellcolor[HTML]{EAC9BD}16.2 & \cellcolor[HTML]{B5CCDF}-3.8 \\
Economic      & \cellcolor[HTML]{FBF2E8}5.1 & \cellcolor[HTML]{EDCFC4}13.6 & \cellcolor[HTML]{7BA4CC}\textcolor{white}{-8.4} \\
Normative     & \cellcolor[HTML]{FCF4EA}4.5 & \cellcolor[HTML]{DFAFA5}26.6 & \cellcolor[HTML]{4A7AB5}\textcolor{white}{-22.1} \\
Experiential  & \cellcolor[HTML]{FDF6EC}0.8 & \cellcolor[HTML]{FAF0E5}5.8 & \cellcolor[HTML]{9DBED9}-5.0 \\
\bottomrule
\end{tabular}
\caption{Argument type (\%).}
\label{tab:argument_dist}
\end{subtable}
\caption{LLM (\texttt{gpt-5-mini}, vanilla, 558 dockets) vs.\ human comment distributions. $\Delta$ in pp. LLMs default to \textit{modify} (+42pp) and \textit{procedural} (+45pp) while under-representing \textit{oppose}, \textit{normative}, and \textit{experiential} modes. Red $\Delta$ = LLM overrepresents; blue = underrepresents.}
\end{table}
The divergence is striking along two axes. First, LLMs adopt a narrow \emph{modify-everything} stance (85.9\% vs.\ 43.8\% human), almost never opposing or supporting provisions outright. Real commenters take diverse directional stances: 30\% oppose provisions entirely, 12\% express support, and only 44\% suggest modifications. Second, LLMs default to \emph{procedural} objections (60.3\% vs.\ 15.7\%) --- ``this provision lacks clarity,'' ``implementation details are insufficient'' --- while dramatically under-representing \emph{normative} arguments (4.5\% vs.\ 26.6\%) that invoke values, fairness, environmental protection, or public health, and \emph{experiential} arguments (0.8\% vs.\ 5.8\%) grounded in lived experience. The mean per-docket Jensen--Shannon divergence is 0.252 for direction and 0.336 for argument type, confirming that these gaps are systematic rather than driven by outlier dockets. In short, LLMs sound procedurally sophisticated but lack the substantive diversity --- the economic data, technical specifics, normative convictions, and personal experience --- that characterizes real public engagement.

\section{All LLM prompts used in the paper}
\label{app:prompts}

We list every LLM prompt directly underpinning a number reported in this paper. We exclude exploratory or downstream-only prompts (e.g., causal-analysis labelers, the legal-framework 10-class classifier, claim-taxonomy scope/argument/direction labels, which follow the frame-label prompt\textquotesingle{}s pattern). Prompts are grouped by the pipeline stage they support (document parsing, matching, comment features, and benchmark generation and judging) and reported with the model used and an estimate of how many calls were made. Total LLM API spend across all stages reported here is approximately \$340 (extraction + matching + audit + benchmark generation across 8 models + Axis~B/C judging).

\subsection{Document parsing}\label{app:prompt_parsing}

\paragraph{Provision extraction.} Extracts the atomic obligations / permissions / prohibitions / conditions / exceptions from each proposed and final rule (Section~\ref{sec:dataset}). Model: \texttt{gpt-5-mini} (\texttt{reasoning\_effort=low}, flex tier). Calls: $\sim$40K rule documents (17{,}198 proposed + 22{,}400 final), \$67.60 total including a two-pass repair. In our prompt, we use the term \textit{deontic units} which refers to an imperative established by a law or regulation.

{\scriptsize
\begin{verbatim}
SYSTEM:
You are a legal analyst specializing in U.S. federal regulatory text. Your task is to extract all
deontic units — obligations, permissions, prohibitions, and requirements — from regulatory text.

A "deontic unit" is any statement that imposes, permits, or prohibits an action on a specific actor.
These are signaled by modal verbs (shall, must, may, may not, should, is required to, is prohibited
from, etc.) or by language that functionally creates a requirement even without a modal verb.

IMPORTANT DISTINCTIONS:
- Extract deontic units ONLY from the actual regulatory text (the part that will become law).
- Do NOT extract from the preamble discussion, background, or DATES/ADDRESSES sections.
- The regulatory text typically appears after "List of Subjects" or starts with amendments to
  specific CFR parts (e.g., "PART 80—...", "§ 123.45").

OUTPUT FORMAT (JSON array; each unit has):
{ "id": <int>, "section": "<CFR ref or empty>", "obligation_text": "<verbatim text>",
  "actor": "<who is obligated>", "modal": "<shall|must|may|may not|...>",
  "action_summary": "<plain-language summary>",
  "type": "<obligation|permission|prohibition|condition|exception>",
  "parent_id": <id or null>, "sub_unit_type": "<threshold|condition|exception|...|null>" }

HIERARCHY: A parent states a general obligation; child sub-units specify thresholds, conditions,
exceptions, alternatives. Not every (a)/(b) creates a sub-unit.

[6 worked examples follow in the actual prompt.]

USER:
Extract all deontic units from the following regulatory document.
Document title: {title}   Agency: {agency}   Docket: {docket_id}
<regulatory_text>{text}</regulatory_text>
\end{verbatim}
}

\paragraph{Response extraction.} Llama-3.3-70B reads each final rule and returns the comment--response pairs it finds as JSON, one object per pair, with the fields \path{content_of_comment}, \path{summarized_content_of_comment}, \path{response_to_comment}, \path{quoted_or_paraphrased}, \path{type_of_response}, \path{reference_scope}, \path{commenter_identifiers_text}, \path{response_engagement_type} and \path{rule_change_outcome}. The stance field \path{type_of_response} takes the values \texttt{accepted}, \path{partially_accepted}, \texttt{disagree}, \texttt{acknowledged}, \texttt{unclear} and \path{not_relevant}; the prompt conveys these through five worked examples rather than a definition list. Finding~1 reports \texttt{accepted} as agreement, \path{partially_accepted} as partial agreement and \texttt{disagree} as disagreement. The prompt, with its examples, is in the released code (\path{extract_responses_from_gov.py}).

\subsection{Match 1 (proposed $\rightarrow$ final)}\label{app:prompt_match1} Aligns each proposed obligation to its surviving final-rule counterpart and assigns one of \{same, modified, removed, added\} per unit. Model: \texttt{gpt-5-mini} (flex). Calls: 7{,}574 dockets (single multi-pair call per docket; 6{,}954 of them have units on both sides and enter Figure~\ref{fig:links}), \$28 total.

{\scriptsize
\begin{verbatim}
SYSTEM:
You are a legal analyst comparing regulatory obligations between a PROPOSED rule and the
corresponding FINAL rule from U.S. federal rulemaking.

You will be given two numbered lists:
  Proposed obligations (p1, p2, ...)   Final obligations (r1, r2, ...)

Classify every obligation into exactly one of:

SAME: substantively identical (minor wording / renumbering differences ignored).
MODIFIED: clear counterpart but substantive change (numbers, deadlines, scope, conditions).
REMOVED: proposed obligation has no counterpart in the final rule.
ADDED: final obligation has no counterpart in the proposed rule.

Every proposed unit must be SAME, MODIFIED, or REMOVED. Every final unit must be SAME, MODIFIED,
or ADDED. SAME / MODIFIED are 1-to-1; if unsure SAME vs MODIFIED, prefer MODIFIED; if unsure
MODIFIED vs REMOVED/ADDED, prefer MODIFIED.

USER:
PROPOSED RULE OBLIGATIONS: {proposed_units}
FINAL RULE OBLIGATIONS:    {final_units}
Output (no other text):
  same: (p_, r_), (p_, r_)
  modified: (p_, r_), ...
  removed: p_, p_
  added: r_, r_
\end{verbatim}
}

\subsection{Match 2 (claim $\rightarrow$ proposed unit)}\label{app:prompt_match2} Calibrated opposing-sides judge: decides whether a single comment claim addresses a specific proposed-rule provision. This is the prompt that drives the Match~2 corpus (1.99M pairs, \$28); the Axis~B pair judge is a separate prompt (Appendix~\ref{app:prompt_axisb}). Model: \texttt{gpt-5-mini} (flex, \texttt{reasoning\_effort=minimal}).

{\scriptsize
\begin{verbatim}
SYSTEM:
You are a legal analyst determining whether a public comment addresses a specific regulatory
provision. The claim may agree with, disagree with, or take an opposing stance toward the
provision — all of these count as "addressing" it. A claim only fails to address the provision
if it is on a completely different regulatory topic.

USER:
Proposed rule obligation:
  Section: {unit_section}   Type: {unit_type}
  Summary: {unit_summary}

Public comment claim:
{claim_text}

Does this claim specifically address this obligation (whether agreeing or disagreeing)?
Answer YES or NO.
\end{verbatim}
}

\subsection{Match audit}\label{app:prompt_audit} Independent gpt-5 (medium reasoning) review of the Match~1 pipeline's REMOVED+NEW pairs and MODIFIED pairs to flag missed merges and spurious matches. Model: \texttt{gpt-5} (\texttt{reasoning\_effort=medium}, flex). Calls: 7{,}574 dockets, \$57.55. Output: 1{,}966 merges + 1{,}770 unmatches + 785 rewires applied to the v3 dataset.

{\scriptsize
\begin{verbatim}
You are auditing the output of a regulatory text-matching pipeline. The pipeline matched
obligations from a PROPOSED federal rule to obligations in the corresponding FINAL rule,
labeling each as "survived", "modified", "removed", or "new". The pipeline uses bi-encoder +
cross-encoder semantic matching; it works well for textually similar pairs but commonly MISSES
cases where a regulatory concept was renumbered, restructured, split, or moved to a different
CFR section. These mistakes typically appear as a "removed" proposed + "new" final pair that are
actually the same concept. It can also produce SPURIOUS "modified" pairs where two units have
surface similarity but address different obligations.

Identify three kinds of relabel suggestions:
1. MERGE   — a (REMOVED proposed) + (NEW final) pair that should be a single MODIFIED pair.
2. UNMATCH — a current (MODIFIED <-> MODIFIED) pair that should NOT be linked.
3. REWIRE  — a current (MODIFIED <-> MODIFIED) pair where the proposed unit's true counterpart
             is a different unit shown.

PROPOSED-side units: {proposed_list}
FINAL-side units:    {final_list}

Output one line per suggestion (or NO_SUGGESTIONS):
  MERGE P_R<i> <-> F_N<j>: <brief reason>
  UNMATCH P_M<i> <-> F_M<j>: <brief reason>
  REWIRE P_M<i>: was F_M<j>, should be F_N<k>: <brief reason>
Be CONSERVATIVE -- only flag with HIGH confidence.
\end{verbatim}
}

\subsection{Cosmetic-only re-judge}\label{app:prompt_cosmetic} Re-classifies pairs that the v1 14-class taxonomy called \texttt{cosmetic\_only}, asking whether the change is truly cosmetic vs. substantive. Used to refine the dataset's \texttt{cosmetic\_only} share from 25.7\% (raw v1) to a more reliable distribution; produced \texttt{cosmetic\_review\_v3.jsonl} (3{,}485 confirmed cosmetic, 1{,}972 actually substantive). Model: \texttt{gpt-5-mini} (flex, minimal reasoning).

{\scriptsize
\begin{verbatim}
SYSTEM:
You are a legal analyst comparing a regulatory provision in a PROPOSED rule to its counterpart
in the FINAL rule. Both appear to be the "same" provision with different wording. Determine
whether the change is PURELY COSMETIC (no real difference in legal effect on a regulated party)
OR SUBSTANTIVE (the obligation, scope, threshold, condition, penalty, or exception meaningfully
changed).

COSMETIC if it is one or more of:
- Wording reorganization, sentence restructuring, simplification with no change to who is
  obligated, what they must do, when, or how.
- Section/paragraph renumbering or cross-reference updates pointing to the same substantive rule.
- Replacing placeholder text with the same content (e.g., "the date" → an explicit date).
- Updating document titles, edition numbers (e.g., "Bulletin Rev 2" → "Rev 3") when content is
  unchanged.
- Removing redundant clauses without changing the underlying obligation.
- Renaming an agency office or contact.
- Adding examples or clarifying language without changing scope or requirement.

SUBSTANTIVE if it is one or more of:
- A specific number, threshold, percentage, deadline, quantity changed.
- A new obligation added, or an existing one removed.
- The set of regulated parties, applicability, exemptions, or coverage changed.
- Penalties, enforcement mechanisms, or compliance methods changed.
- A new exception/exclusion added, or an existing one removed.
- Required documentation, reporting, recordkeeping changed in substance.

When in doubt prefer SUBSTANTIVE.

USER:
Section/identifier (proposed): {section_proposed}
Section/identifier (final):    {section_final}
Type: {unit_type}
PROPOSED text: {proposed_text}
FINAL text:    {final_text}

Output (exactly two lines):
  COSMETIC or SUBSTANTIVE
  <one short sentence (<25 words) explaining why.>
\end{verbatim}
}

\subsection{Topic-mismatch re-judge}\label{app:prompt_topic} Re-classifies pairs flagged \texttt{topic\_mismatch} to confirm whether they are wrong matches or legitimate same-topic revisions. Produced \texttt{topic\_mismatch\_review\_v3.jsonl} (2{,}468 confirmed mismatch, 1{,}501 actually valid). Model: \texttt{gpt-5-mini} (flex, minimal). Same input format as Appendix~\ref{app:prompt_cosmetic}; output \texttt{TRUE\_MISMATCH} or \texttt{VALID\_MATCH} with one-sentence reason.

\subsection{14-class modification taxonomy}\label{app:prompt_mod14} Classifies each \texttt{modified} (proposed $\rightarrow$ final) unit into one of 14 concrete change types. Drives the modification typology reported in Appendix~\ref{app:bench_construction}. Model: \texttt{gpt-5-mini} (flex, minimal). Calls: 12{,}812 modifications, \$2.57 total.

{\scriptsize
\begin{verbatim}
SYSTEM:
You are comparing a PROPOSED regulatory obligation to its FINAL version. Classify the change
into EXACTLY ONE of these 14 categories (each is a CONCRETE textual change pattern, not an
abstract direction):

THRESHOLD/SCOPE:
  threshold_or_scope_tighter | threshold_or_scope_looser
EXCEPTIONS:
  exception_added | exception_removed
TIMELINE:
  timeline_shortened | timeline_extended
ENFORCEMENT:
  enforcement_strengthened | enforcement_weakened
NO BURDEN CHANGE:
  definition_clarified | cosmetic_only
OFF-AXIS:
  agency_obligation_change | respondent_protection_added
PROBLEM CASES:
  mixed_substantive | topic_mismatch

DECISION RULES:
1. Focus on impact on REGULATED PARTIES (operators, manufacturers, applicants), NOT the agency.
2. Adding states/parties to a "may" provision = threshold_or_scope_looser (more flexibility).
3. If texts are semantically equivalent ("within X" = "before X") → cosmetic_only.
4. If texts are about totally different obligations → topic_mismatch.
5. Adding a respondent-protection clause = respondent_protection_added.
6. Pick the SINGLE MOST INFORMATIVE category; if multiple apply, pick the one with the
   largest effect on regulated parties.

[8 worked examples follow.]

USER:
Proposed:  Section: {p_section}  Summary: {p_summary}  Text: {p_text}
Final:     Section: {f_section}  Summary: {f_summary}  Text: {f_text}

Respond exactly:
  label: <one of 14 labels>
  reason: <one short phrase identifying the textual signal>
\end{verbatim}
}

\subsection{Specialist vs. generalist commenter classification}\label{app:prompt_specialist} Identifies whether a comment author writes from substantive domain expertise (legal/technical/insider/experiential) vs. as an ordinary citizen. Used as the gold-label seed for the corpus-wide TF-IDF + LR distillation that produces the per-comment specialist score reported in Section~\ref{sec:analysis} (Finding~3, ``$\sim$30\% specialist rate''). Model: \texttt{gpt-5-mini} (flex, minimal). Calls: 5{,}987 gpt-5-mini-labeled seed comments seeded the distillation; the LR distillate was then applied to all 8.27M comments at no LLM cost.

{\scriptsize
\begin{verbatim}
SYSTEM:
SPECIALIST: the author brings substantive domain expertise — they cite statutes/case law/
regulations correctly, use technical/scientific vocabulary appropriately, demonstrate insider
knowledge of an industry's operations, OR write from substantive occupational expertise. The
author *contributes information* the agency would not already have.

GENERALIST: the author writes from ordinary citizen perspective — values, opinions, generic
concerns, brief support/oppose statements, mass-campaign language, or general lived experience
without substantive expertise. The author *expresses preference* but does not contribute
specialized information.

Identify which expertise modes (if any) are present:
- legal_expert
- technical_scientific
- industry_insider
- experiential_specialist

A short comment can still be specialist; a long emotional letter can still be generalist.

USER:
COMMENT (truncated if long): {text}
Output JSON only:
  {"is_specialist": true|false, "expertise_modes": ["legal_expert", ...], "reason": "<25 words"}
\end{verbatim}
}

\subsection{Primary-frame label}\label{app:prompt_frame} Assigns each comment claim one of 12 primary frames (health, safety, economic\_cost\_benefit, legal\_constitutional, scientific\_technical, environmental, equity\_justice, religious\_moral, national\_security, procedural, other, unknown). Drives Table~\ref{tab:stakeholder_concern} (stakeholder $\times$ frame matrix) and the frame shares of LLM claims in Section~\ref{sec:benchmark_results}. Model: \texttt{gpt-5-mini} (flex, minimal). Calls: 27{,}936 claims for the original gpt-5-mini reference run; for the per-model benchmark sweep, an additional 43{,}371 unique real claims (one-time cache) plus the LLM claims of each evaluated model (14K--100K each).

{\scriptsize
\begin{verbatim}
USER:
Classify the dominant framing of this public-comment claim. Pick exactly ONE from this enum:
- health: public health, disease, medical outcomes
- safety: accidents, worker safety, product safety
- economic_cost_benefit: costs, jobs, competitiveness, ROI, market effects
- legal_constitutional: statutory authority, constitutional rights, APA, takings, due process
- scientific_technical: evidence, methodology, engineering, measurement
- environmental: pollution, ecosystems, climate, wildlife
- equity_justice: fairness, disparate impact, environmental justice, civil rights
- religious_moral: ethical or values-based appeals
- national_security: defense, foreign threats, critical infrastructure
- procedural: process objections, transparency, comment-period mechanics
- other
- unknown

Claim: {claim_text}

Respond with one word, the enum value.
\end{verbatim}
}

\subsection{Anticipated-comment generation: vanilla}\label{app:prompt_gen_vanilla} Asks the model to produce predicted commenter claims given the docket's pooled provisions. Used per docket per evaluated model. Calls per evaluated model: 558 (one per docket).

{\scriptsize
\begin{verbatim}
USER:
Below are provisions from a proposed federal rule.

Your task: What are public comments members of the public are likely to submit on the
following provisions? For each comment, express it as a specific claim — a concrete point
about a specific provision.

Format each claim on its own line, prefixed with the provision number it addresses:
  [P1] claim about provision 1
  [P3] claim about provision 3
  [P1] another claim about provision 1

Generate as many distinct claims as you think real commenters would submit. Be specific —
reference particular requirements, thresholds, or actors mentioned in the provisions. Consider
perspectives from industry, advocacy groups, individuals, and legal experts.

PROPOSED RULE PROVISIONS:
{provisions}

Generate the anticipated public comments:
\end{verbatim}
}

\subsection{Anticipated-comment generation: subgroup persona}\label{app:subgroup-persona} 
Same task structure as Appendix~\ref{app:prompt_gen_vanilla} but conditioned on one of seven stakeholder personas (industry, labor, advocacy, expert, gov, lawfirm, individual). The seven personas are run as 7 independent prompts per docket per model and the outputs are pooled. Calls per evaluated model: $558 \times 7 = 3{,}906$. Each persona description is a 1-paragraph descriptor naming the stakeholder's typical evidence base and concerns.

{\scriptsize
\begin{verbatim}
USER:
Below are provisions from a proposed federal rule. Write public comments FROM THE
PERSPECTIVE OF: {stakeholder_label}.

{stakeholder_persona}

Generate the public-comment claims this stakeholder would submit. Each claim should
be a concrete point about a specific provision, written in the voice and concerns characteristic
of {stakeholder_label}.

Format each claim on its own line, prefixed with the provision number:
  [P1] claim about provision 1

Be specific — reference particular requirements, thresholds, or actors mentioned in the
provisions, AND write in a way that reflects this stakeholder's typical concerns, vocabulary,
and framings.

PROPOSED RULE PROVISIONS:
{provisions}

{stakeholder_label} comments on this proposal:

==== STAKEHOLDER PERSONA TEXT (one used per call) ====
industry:   "a regulated company, trade association, or industry coalition. You focus on
             compliance cost, operational feasibility, market effects, and ROI. You cite
             economic data and industry-specific operational details."
labor:      "a labor union representing workers in the regulated industry. You focus on worker
             safety, job protection, fair labor standards, and economic impact on workers. You
             cite worker statistics and on-the-job experience."
advocacy:   "an environmental, civil rights, or public-interest advocacy organization. You
             focus on broader public good, environmental protection, equity, and the
             regulation's role in achieving social goals. You cite scientific studies, public-
             health data, and advocacy research."
expert:     "an academic researcher, professional society, or expert organization. You focus on
             technical correctness, scientific accuracy, peer-reviewed evidence, and
             methodological soundness. You cite research, technical specifications, and
             disciplinary expertise."
gov:        "a state or local government agency that interacts with this federal regulation. You
             focus on inter-jurisdictional coordination, implementation feasibility for state
             agencies, and federalism issues. You cite state-level data and operational details."
lawfirm:    "a law firm or bar association representing regulated parties. You focus on
             statutory authority, constitutional issues, APA procedural requirements, due-
             process implications, and legal interpretation. You cite case law, statutes, and
             regulatory history."
individual: "an ordinary individual citizen affected by this rule. You focus on personal
             experience, lived impact on you or your family/community, and concrete harms or
             benefits you would face. You speak in first person about direct experience."
\end{verbatim}
}

\subsection{Axis~B pair-alignment judge}\label{app:prompt_axisb} For each validated real claim (i.e. each (comment $\to$ proposal) match) the judge labels the pair as one of \{same, related, different, opposing\}. Model: \texttt{gpt-4o-mini}, ten pairs per call, with the rubric below. We pair each LLM claim with its three nearest real claims in the same docket (MiniLM cosine $>0.3$). This gives 35K pairs for gpt-4o-mini vanilla up to 263K for Llama-3.1-8B persona, over all 558 dockets (504--558 dockets per run have at least one judged pair).

{\scriptsize
\begin{verbatim}
USER:
You compare two regulatory comment claims (A and B) that target the SAME proposed regulatory
provision. The two claims may come from commenters on opposing sides of the issue: agreeing,
disagreeing, raising distinct concerns, or expressing the same point. Choose ONE label that
best describes the relationship of B to A:

- same:     B states the same point as A (a paraphrase or restatement; same direction).
- related:  B is on the same sub-topic as A but makes a different point (compatible direction).
- different: B addresses a different aspect/concern of the provision (no semantic overlap).
- opposing: B takes a contrary position to A on the same point (e.g., A wants extension, B
            wants no change).

Claim A: {a}
Claim B: {b}

Respond with one word: same, related, different, or opposing.
\end{verbatim}
}

\subsection{Full-comment generation prompts and examples}
\label{app:full_comment_examples}

In addition to claim-list generation (Appendices~\ref{app:prompt_gen_vanilla}--\ref{app:subgroup-persona}), we generate \textbf{full natural-language comments} in paragraph form for each docket. These are used for the frame and style analysis (Section~\ref{sec:benchmark_results}). We evaluate five generation methods: vanilla, persona-conditioned, RAG-augmented, few-shot, and RAG with tool calling. Full-comment prompts follow the same provision input as Appendices~\ref{app:prompt_gen_vanilla}--\ref{app:subgroup-persona} but instruct the model to ``write in full paragraphs'' rather than produce tagged claim lists.

\subsection{Full-comment generation: vanilla}

{\scriptsize
\begin{verbatim}
USER:
Below are provisions from a proposed federal rule.

Write a public comment responding to these provisions, as a real member of
the public would. Write in full paragraphs — cite specific evidence,
reference relevant statutes or data, draw on operational experience where
relevant, and propose specific alternatives where you disagree.

Do not use bullet points or numbered lists. Write naturally, as if
submitting a real comment to the Federal Register.

PROPOSED RULE PROVISIONS:
{provisions}

Public comment:
\end{verbatim}
}

\subsection{Full-comment generation: RAG-augmented} Retrieves evidence from a BM25 index over 30K real evidence-bearing comments (extracted per Appendix~\ref{app:pluralism_methods}) and stuffs retrieved snippets into the prompt context.

\subsection{Full-comment generation: few-shot} Includes two real public comments from non-benchmark dockets as style demonstrations before the generation instruction.

\subsection{Full-comment generation: RAG with tool calling} Provides the model with a retrieval tool; the model decides what queries to issue before writing. Up to 3 tool-calling rounds are permitted.

\paragraph{Example generated comments.} Figures~\ref{fig:example_gpt5_vanilla}--\ref{fig:example_4omini} show excerpts from comments generated by different methods and personas on the same docket (EPA-HQ-OAR-2017-0091, proposing 2018 renewable fuel percentage standards). The full comments range from 3K to 16K characters. Notable differences include: \textbf{gpt-5 vanilla} writes as a ``Midwestern small-business owner,'' cites GREET model and Argonne National Lab data, and references Clean Air Act \S211(o)---a highly informed, policy-analyst style; \textbf{gpt-5-mini RAG} writes as a ``soybean farmer,'' grounds claims in retrieved evidence (``Diamond Green Diesel producing 160M gallons''); \textbf{gpt-5-mini +individual persona} uses emotional first-person testimony (``I saw friends and cousins pick up second jobs or leave town''); and \textbf{gpt-4o-mini vanilla} produces the shortest comment (3.2K chars) with vaguer evidentiary claims.

\begin{figure}[h]
\fbox{\parbox{0.97\textwidth}{\scriptsize
\textbf{gpt-5, vanilla prompt.} Docket EPA-HQ-OAR-2017-0091 (2018 Renewable Fuel Standards). 8,476 characters total.

\medskip
I write as a Midwestern small-business owner who operates a modest delivery fleet and as a long-time credit union member who has helped with my cooperative's community outreach. I appreciate the opportunity to comment on both the renewable fuel percentage standards and the advertising statement requirements for federally insured credit unions. I focus below on the provisions that most affect fuel markets and consumer protection.

On the proposed 2018 renewable fuel percentage standards in 40 CFR \S~80.1405(a)(9), I respectfully urge EPA to revisit the balance among the cellulosic, biomass-based diesel, advanced, and total renewable fuel percentages. Under Clean Air Act section 211(o), EPA must set annual renewable fuel obligations that are achievable based on projected supply while advancing the program's air quality and greenhouse gas goals. The proposed percentages---0.131\% for cellulosic, 1.74\% for biomass-based diesel, 2.34\% for advanced biofuel, and 10.62\% for total renewable fuel---appear conservative relative to actual 2018 market availability and the environmental benefits at stake.

First, the 1.74\% biomass-based diesel (BBD) percentage likely understates feasible supply. EPA has repeatedly found that biomass-based diesel and renewable diesel are among the most readily scalable advanced biofuels, and Argonne National Laboratory's GREET model estimates life-cycle greenhouse gas reductions commonly in the 50--80\% range for biodiesel and renewable diesel compared to petroleum diesel, depending on feedstock and pathway. [\ldots]
}}
\caption{Full-comment generation: \textbf{gpt-5, vanilla}. The model adopts the persona of a ``Midwestern small-business owner,'' cites the GREET model, Argonne National Lab, and specific CFR sections. Comment is highly detailed and policy-analyst in style.}
\label{fig:example_gpt5_vanilla}
\end{figure}

\begin{figure}[h]
\fbox{\parbox{0.97\textwidth}{\scriptsize
\textbf{gpt-5-mini, RAG-augmented.} Same docket. 9,417 characters total. Retrieved evidence from BM25 index over 30K real comments.

\medskip
I submit these comments as a soybean farmer and as a long-time member of a small rural credit union. I appreciate the agencies' efforts to provide regulatory clarity, but I am concerned that the proposed percentages for the 2018 Renewable Fuel Standard (RFS) pools and the proposed NCUA advertising requirements as written will have unintended economic and operational consequences.

With respect to the proposed 2018 RFS values (\S~80.1405(a)(9)(i)--(iv)), my livelihood and local rural economy depend on stable, predictable markets for biodiesel and renewable diesel. The RFS is established by statute (42 U.S.C. \S~7545(o)), which gives EPA discretion but also creates an expectation that annual standards reasonably reflect available renewable fuel production and support ongoing investment in the industry. The retrieved evidence shows biodiesel and renewable diesel add real value to agricultural commodities; biodiesel ``directly adds more than \$0.63 of value to the bushel of soybeans,'' and industry capacity is expanding---for example, Diamond Green Diesel was producing about 160 million gallons annually and projected expansion to 275 million gallons by mid-2018. [\ldots]
}}
\caption{Full-comment generation: \textbf{gpt-5-mini, RAG-augmented}. The model writes as a ``soybean farmer,'' grounds claims in \emph{retrieved} evidence from real prior comments (e.g., specific production figures for Diamond Green Diesel).}
\label{fig:example_rag}
\end{figure}

\begin{figure}[h]
\fbox{\parbox{0.97\textwidth}{\scriptsize
\textbf{gpt-5-mini, +individual persona.} Same docket. 7,384 characters total.

\medskip
I am writing as a lifelong resident of a small rural town that depends on agriculture and a handful of locally owned businesses for jobs and economic stability. My family farms soybeans and we once sold a meaningful share of our crop to a nearby cooperative biodiesel plant that provided good-paying work for neighbors and kept dollars circulating locally. When the market for biodiesel softened a few years ago, that plant cut shifts and hired fewer seasonal workers; I saw friends and cousins pick up second jobs or leave town. The proposed renewable fuel standard values (P1--P4) for 2018---in particular the cellulosic biofuel standard of 0.131 percent (P1) and the biomass-based diesel value of 1.74 percent (P2)---feel disconnected from the real lives of people like me. I recognize EPA must reflect real fuel supply in setting annual percentages under the Clean Air Act (42 U.S.C. \S~7545(o)), but setting cellulosic and advanced biofuel targets at such low levels has concrete consequences here: it signals to investors and lenders that markets will not support new or expanded biorefineries, it reduces demand for crops and waste oils that farmers and local processors rely on, and it makes it harder for my community to attract the few good manufacturing jobs that remain in rural America.

The very small cellulosic percentage (P1) is especially troubling. I understand cellulosic fuels are still scaling up technologically, but by effectively writing a near-zero target into the standard, EPA removes a predictable market that would help entrepreneurs move from pilot plants to commercial scale. In my town, a group of young entrepreneurs tried for years to adapt corn stover and livestock manure into advanced biofuel feedstock; they struggled to find long-term buyers and financing because demand from obligated parties was inconsistent. [\ldots]
}}
\caption{Full-comment generation: \textbf{gpt-5-mini, +individual persona}. The model adopts a first-person voice (``I saw friends and cousins pick up second jobs or leave town''), frames regulatory impact through lived experience, and still references specific statutory provisions.}
\label{fig:example_individual}
\end{figure}

\begin{figure}[h]
\fbox{\parbox{0.97\textwidth}{\scriptsize
\textbf{gpt-4o-mini, vanilla prompt.} Same docket. 3,176 characters total (shortest).

\medskip
I appreciate the opportunity to provide comments on the proposed federal rule pertaining to renewable fuel standards and advertising statements. As a concerned citizen and member of the public, it is essential to express my views, particularly on the provisions regarding the establishment of biofuel standard values and the mandated advertising statements for federally insured credit unions.

Regarding P1 to P4, which establish the cellulosic biofuel, biomass-based diesel, advanced biofuel, and renewable fuel standard values, I am concerned that these values may not adequately reflect the current market conditions and advancements in biofuel technology. Specifically, the proposed 2018 cellulosic biofuel standard value of 0.131 percent appears to underutilize the potential of advancements made in this sector. According to the U.S. Energy Information Administration, the production capacity of cellulosic biofuels has been steadily increasing, and more significant investments are being made in research and development. In light of this, I would suggest that the values be reassessed to encourage higher production and utilization of biofuels, which can contribute to reduced greenhouse gas emissions and energy independence.

Furthermore, when considering the biomass-based diesel standard value of 1.74 percent (P2), it is critical to acknowledge the extensive benefits of biodiesel, particularly in terms of sustainability and economic growth. The National Biodiesel Board has reported that biodiesel production supports a significant number of jobs and offers substantial economic returns. Therefore, it would be prudent to elevate the biomass-based diesel standard value to harness these benefits more effectively.

On the topic of advertising statements (P5 to P13), I recognize the importance of transparency in the financial sector and the necessity for consumers to be well-informed about the insurance status of their credit unions. However, I suggest that the language provided in P7 to P9 may be overly restrictive. While clarity is vital, there should be flexibility in how credit unions convey this essential information without being bound to predetermined phrases. [\ldots]
}}
\caption{Full-comment generation: \textbf{gpt-4o-mini, vanilla prompt}. Notably shorter (3.2K vs 7--9K chars), writes as a generic ``concerned citizen,'' and uses vaguer evidentiary claims (``the National Biodiesel Board has reported\ldots'') without specific figures.}
\label{fig:example_4omini}
\end{figure}

\end{document}